\documentclass[a4paper]{article}
\usepackage{etex}
\usepackage{a4wide}
\usepackage[pdftex]{graphicx} 
\usepackage{url}
\usepackage{amsbsy,amssymb,amsgen,amsfonts}
\usepackage{amsmath,amsfonts,color,latexsym}
\usepackage[breaklinks=true,colorlinks=true,citecolor=blue,
            linkcolor=blue,urlcolor=blue]{hyperref}
\usepackage{geometry}
\usepackage{tocloft} 
\usepackage{calc}
\usepackage{float}
\usepackage{color} 
\usepackage{threeparttable} 
\usepackage{comment}
\usepackage{epstopdf}
\usepackage[authoryear,round,longnamesfirst]{natbib}
\usepackage[toc,title]{appendix}
\newcommand{\revision}[2]{#2}
\newcommand{\revisions}[1]{}

\definecolor{darkgreen}{rgb}{0,.5,0}

\def\drawline#1#2{\raisebox{2.5pt}{\rule{#1pt}{#2pt}}}
\def\spacce#1{\hskip #1pt}
\def\solid{\protect\rule[2.5pt]{24.pt}{.5pt}}
\def\bdash{\protect\hbox{\protect\drawline{4}{.5}\spacce{2}}}
\def\dashed{\bdash\bdash\bdash\bdash }

\def\chndot{\protect\hbox
{\protect\drawline{9.5}{.5}\spacce{2}\protect\drawline{1}{.5}\spacce{2}\protect\drawline{9.5}{.5}}}

\def\trian{\raise 1.25pt\hbox{$\scriptstyle\triangle$}\nobreak}
\def\squar{\raise 1.25pt\hbox{$\scriptstyle\Box$}\nobreak}
\def\dtrian{\raise 1.25pt\hbox
   {$\scriptscriptstyle\bigtriangledown$}\nobreak}
\def\circle{$\circ$\nobreak}

\def\solidcircle{$\bullet$\nobreak}
\def\solidtriangle{$\blacktriangle$\nobreak}
\def\solidsquare{$\scriptstyle\blacksquare$\nobreak}

\newcommand{\rmd}{\mathrm{d}}
\usepackage{sectsty}
\allsectionsfont{\sffamily}
\let\LaTeXmaketitle\maketitle
\renewcommand{\maketitle}{{\sf\LaTeXmaketitle}}

\begin{document}
\epstopdfsetup{suffix=} 
\title{DNS of horizontal open channel flow with finite-size, 
  heavy particles at low solid volume fraction}
\author{Aman G Kidanemariam, 
  Clemens Chan-Braun,\\
  Todor Doychev and 
  Markus Uhlmann
  \footnote{Corresponding author: {\tt markus.uhlmann@kit.edu}}
  \\[1ex]
  {\it\small
    Institute for Hydromechanics,
    Karlsruhe Institute of Technology (KIT),
    76131 Karlsruhe, Germany}
}
\date{
}
\maketitle
\begin{abstract}
We have performed direct numerical simulation of turbulent open
channel flow over a smooth horizontal wall 
in the presence of finite-size, heavy particles.  
The spherical particles have a diameter of approximately
$7$ wall units, a density of $1.7$ times the fluid density and 
a solid volume fraction of $5\cdot10^{-4}$. 
The value of the Galileo number is set to 
\revision{%
  $16.6$
}{%
  $16.5$
}, 
while the Shields
parameter measures approximately $0.2$. 
Under these conditions, the particles are predominantly located
in the vicinity of the bottom wall, where they exhibit strong
preferential concentration which we quantify by means of Voronoi
analysis and by computing the particle-conditioned concentration
field. 
As observed in previous studies with similar parameter values,
the mean streamwise particle velocity is smaller than that of the
fluid. 
We propose a new definition of the fluid velocity ``seen'' by
finite-size particles based on an average over a spherical surface
segment,  
from which we deduce in the present case that the particles are
instantaneously lagging the fluid only by a small amount. 
The particle-conditioned fluid velocity field shows that the particles
preferentially reside in the low-speed streaks, leading to the
observed apparent lag. 
Finally, a vortex eduction study reveals that spanwise particle motion
is significantly correlated with the presence of vortices with the
corresponding sense of rotation which are located in the immediate
vicinity of the near-wall particles. 
\end{abstract}

%
\section{Introduction}\label{sec-introduction}
The transport of solid particles in wall bounded turbulent flows is a common 
occurrence in various natural and man-made systems. 
This transport phenomenon has significant implications in many 
environmental and industrial processes. For instance
one of the key environmental variables in river geomorphology is 
sediment transport, which involves the erosion, movement and deposition 
of sediment particles by the fluid flow.
An improved understanding of the coupled interaction between solid particles 
and turbulence 
is highly desirable as it would pave the way for improvements of
engineering-type formulas.
However the complex structure of wall-bounded shear flows and 
the dependence on multiple governing parameters has made this 
task very challenging to the present date. 


Experimental studies of particle-turbulence interaction in 
horizontal channel flow show that 
near-wall coherent structures play an important role in the 
dynamics of  particle motion
\citep[see e.g.][]
{sumer:1978,yung:1989,rashidi:1990,kaftori:1995a,
kaftori:1995b,nino:1996,kiger:2002,righetti:2004}.
\citet{rashidi:1990} investigated
the coupled interaction between phases in a horizontal 
channel flow laden with heavy spherical particles.
They observed that most of the particles, 
once they settle to the bottom wall, accumulate in the low-speed streaks. 
The particles, depending on their size, their density and 
flow Reynolds number, are observed to get lifted up and entrained 
into the outer flow presumably by the action of the coherent structures.
Particles resettling to the bottom are observed to 
migrate into the low-speed fluid regions, supposedly by the 
action of the eddy structures.
\citet{hetsroni:1994} found similar particle segregation at the wall
when investigating particles with a diameter of 10 wall units 
(i.e.\ equal to 10 times the 
viscous length scale $\nu/u_\tau$ where $\nu$ is the
kinematic viscosity of the fluid and $u_\tau$ is the friction velocity). 
However, larger particles having a diameter of more than 
30 wall units did not accumulate in the low-speed fluid regions; 
rather they formed a random distribution 
at the bottom wall 
of their horizontal flume. 
In similar experiments, 
\citet{kaftori:1995a} and \citet{nino:1996} investigated the effect of the 
coherent structures on particle motion, distribution, entrainment
as well as deposition in the wall region. 
They show that heavy
particles directly interact 
with the flow structures. At the bottom wall, particles were observed to
be non-uniformly distributed and to form streamwise aligned streaks.
The shape, length and persistence with time of these streaks was observed
to vary with the particle size and/or the flow rate.
Formation of particle streaks was also observed by 
\citet{yung:1989} even for particles which are completely 
submerged within the viscous sublayer
(diameter smaller than $1.3$ wall units).
The experimental studies of 
\citet{kaftori:1995b,kiger:2002,righetti:2004} 
were focused on
the characteristics of suspended heavy particles 
and their turbulence modulation effects in open as well as closed
channel flows. These
experiments showed that generally, 
the mean streamwise velocity of the particles is smaller than 
that of the fluid except for those particles located in a layer 
very close to the wall 
where contrarily particle velocity
is reported to be on average higher than that of the fluid.
%
\begin{figure}
\begin{center}
  \includegraphics[width=0.7\textwidth]
  {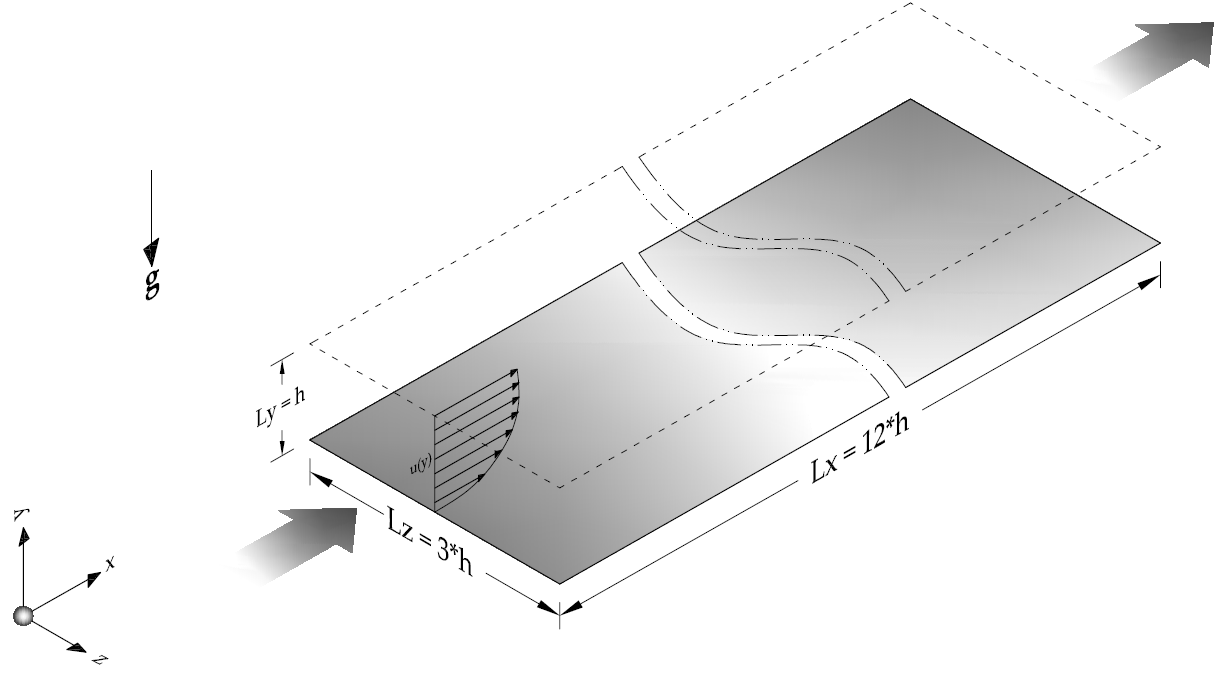}
\end{center}
\caption[Geometrical configuration of the numerical setup.]
{Geometrical 
  configuration of the 
  present simulation. 
  Gravity is directed 
  in the negative \textit{y-} direction. The computational domain is 
  periodic 
  along the \textit{x-} and \textsl{z-}directions.}
\label{fig:geo_setup}
\end{figure}

The majority of available numerical studies on particle-turbulence 
interaction are based on the point-particle approach
\citep[see e.g.][]{rouson:2001,marchioli:2002,narayanan:2003,
picciotto:2005}. Studies based on this approach 
show that there is a strong correlation
between the near-wall coherent structures 
and particle deposition 
and entrainment processes as well as the resulting preferential
particle segregation. The degree of the correlation is 
found 
to 
be dependent on the Stokes number, 
i.e.\ the ratio of the 
particle response time 
to some representative time scale 
of the flow \citep[][]{soldati:2009}. 
However, 
the point-particle approach is 
limited to particle sizes which are much smaller than the 
smallest flow scales.
Therefore, in order to numerically investigate finite-size 
effects, one has to resort to simulations which fully resolve
the particles.

Recently more faithful simulations in which the interface 
between the phases is resolved to accurately compute the flow field 
at the particle scale are starting to emerge both for wall bounded flows 
\citep[][]{pan:1997,uhlmann:2008,shao:2012,
garcia-villalba:2012} 
and unbounded flows 
\citep[][]{kajishima:2002,tencate:2004,lucci:2010,lucci:2011}.
In one of the first simulations of this kind,
\citet{pan:1997} simulated 
horizontal open channel flow 
seeded with a small number of fixed and moving particles of 
diameter 8.5 and 17 wall units
at an average solid volume fraction of $\mathcal{O}(10^{-4})$.
They observed that the presence of particles strongly affected the 
turbulent structures. 
\citet{uhlmann:2008} and \citet{garcia-villalba:2012}
simulated vertically oriented
channel flow seeded with freely moving heavy particles 
with a diameter of
approximately 11 wall units at a solid volume fraction of $0.4\%$. 
These simulations revealed that the particles strongly modified the
flow and 
lead to the formation of very large 
streamwise-elongated 
flow structures. Recently, \citet{shao:2012} simulated  
particle-laden turbulent flow in a horizontal channel. They 
considered heavy particles (with density $1.5$ times that of the
fluid) which are relatively large in size 
(with diameter $10\%$ and $20\%$ of the half channel height), 
at relatively large solid volume fraction values 
(up to 
$7\%$).
They reported that, in cases where settling effects are not 
considered (without gravity), the
particle distribution was homogeneous and the mean 
fluid and particle velocities were equal. 
However, when settling effects are considered,
they reported that most particles settle to the bottom wall 
where they 
accumulate in the low-speed fluid regions. 

\begin{table}\footnotesize
   \centering   
   \begin{tabular}{ccccccccccc}
   \hline\hline\\[-5pt]
   $Re_{\mathrm{b}}$ & $Re_\tau$ & $\rho_\mathrm{p}/\rho_\mathrm{f}$ & 
   $D/h$ & $D^+$ & $St^+$ & $St_\mathrm{b}$ & 
   $\Phi_\mathrm{s}$ & $|\mathbf{g}|h/ u_\mathrm{b}^2$& $Ga$ & $\theta$\\
   \hline\\[-3pt]
   $2870$ & $184.54$ & $1.7$ & 
   $
   10/256
   $ & $7.21$ & 
   $4.91$ & $0.41$ & $0.05\%$ & $0.80$& 
   \revision{%
     $16.58$
   }{%
     $16.49$
   }
   & $0.19$\\[3pt]
   \hline\hline
   \end{tabular}
   \caption{Physical parameters of the simulation.
           $Re_\mathrm{b}$ is the Reynolds number based on the channel depth 
           $h$ and bulk velocity 
           $u_\mathrm{b}$. 
           $Re_\tau$
           is the Reynolds number based on $h$ and the friction velocity
           $u_\tau$. $St^+$ and $St_\mathrm{b}$ are the Stokes numbers based on
           near wall fluid time scale and bulk time scale respectively.
           $D$ is the diameter of the spherical particles.
           $\Phi_\mathrm{s}$ is the 
           global 
           solid volume fraction and $|\mathbf{g}|$
           is the magnitude of acceleration due to gravity.
           $Ga$ and $\theta$ are the Galileo number and Shields
           number, respectively.
           }
    \label{tab:physical_parameters}
\end{table}
\begin{table}[b]
  \footnotesize
   \centering   
   \begin{tabular}{cccccc}
   \hline\hline\\[-5pt]
   $ L_\mathrm{x} \times L_\mathrm{y}\times L_\mathrm{z} $ & 
   $ N_\mathrm{x} \times N_\mathrm{y}\times N_\mathrm{z} $ &
   $D/\Delta x$ & $\Delta x^+$ & $N_\mathrm{p}$ & $T_\mathrm{obs} u_\mathrm{b}/h$\\
   \hline\\[-3pt]
   $12h \times h \times 3h$ & $3072 \times 257 \times 768$ &
   $10$ & $0.72$ & $518$ & 
   \revision{$162$}{$255$}
   \\[3pt]
   \hline\hline
   \end{tabular}
   \caption{Numerical parameters of the simulation.
           $L_i$ and $N_i$ are the computational domain length and number
           of grid points employed in the 
           $i$th coordinate direction, 
           respectively.
           $N_\mathrm{p}$ is the
           number of particles considered. $T_\mathrm{obs}$ is the observation 
           time of the
           simulations excluding 
           the 
           transient period.
         }
   \label{tab:numerical_parameters}
\end{table}
  The general picture that has emerged from previous studies of
  horizontal turbulent channel flow with heavy particles 
  (in the regime where streaky particle patterns are observed) 
  can be summed up briefly as follows. 
  Sweeps of high-speed fluid as well as gravitational settling 
  bring suspended heavy particles towards the wall. 
  Through the action of the quasi-streamwise vortices, particles near
  the wall are forced to move in the spanwise direction into the
  low-speed regions, resulting in particle accumulation in the form of
  persistent streaky particle patterns.  
  The counter-rotating quasi-streamwise vortices flanking the
  low-speed streaks generate ejections of low-momentum fluid away from
  the wall. 
  These ejection events are the predominant sources of upward particle 
  motion from the wall. 
  Concerning the difference between the mean fluid velocity and the
  mean particle velocity (i.e.\ particles apparently lagging the
  fluid), it is believed to be a statistical consequence of particles
  preferentially residing in the low-speed regions, rather than a
  manifestation of particles instantaneously lagging the surrounding
  fluid in a significant manner. 

\begin{figure}
   \centering
        \begin{minipage}{2ex}
        \rotatebox{90}
        {\small
          $\langle k \rangle_\mathrm{xyz}/u_\tau^2$
        }
        \end{minipage}
        \begin{minipage}{.5\linewidth}
        \includegraphics[width=\linewidth]
        {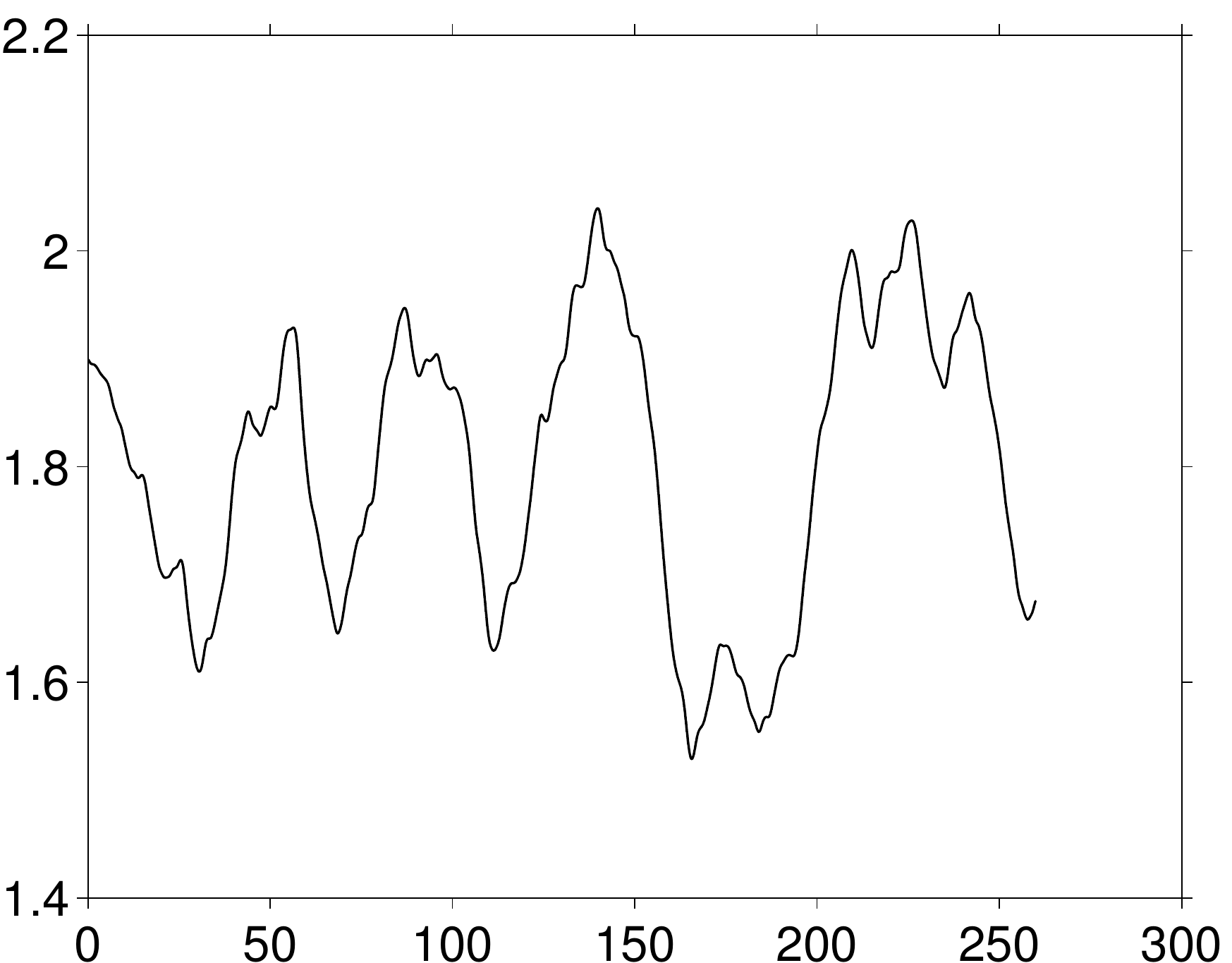}
        \hspace{-0.5\linewidth}\raisebox{.82\linewidth}
        {\small $(a)$}
        \\
        \centerline
        {\small $t/T_\mathrm{b}$}
        \end{minipage}\\[20pt]
        %
        \begin{minipage}{2ex}
        \rotatebox{90}
        {\small $Re_\tau$ }
        \end{minipage}
        \begin{minipage}{.5\linewidth}
        \includegraphics[width=\linewidth]
        {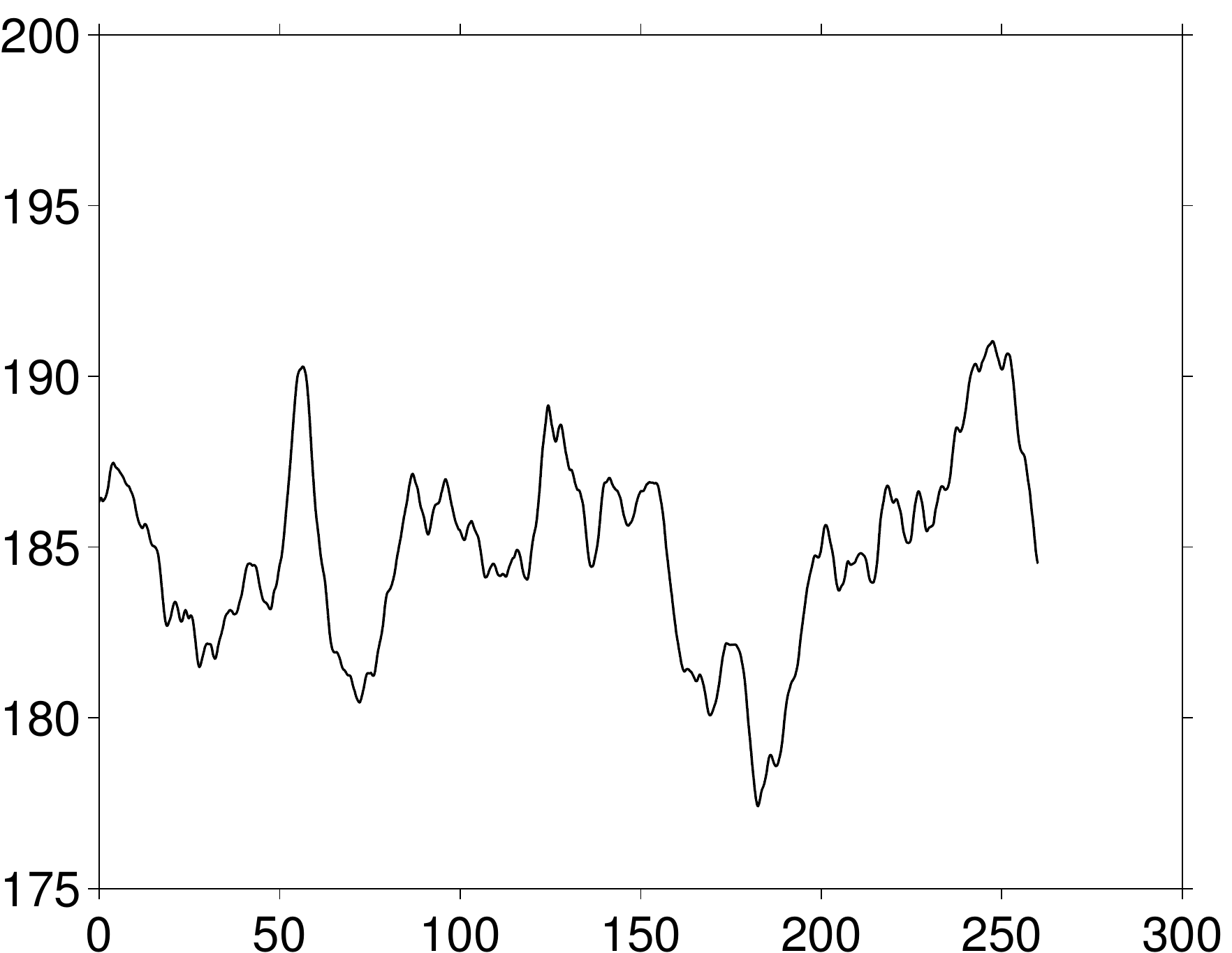}
        \hspace{-0.5\linewidth}\raisebox{.82\linewidth}
        {\small $(b)$}
        \\
        \centerline{\small $t/T_\mathrm{b}$}
        \end{minipage}
         \vspace{2ex}
        \caption{Time evolution of 
                 (\textit{a}) box-averaged turbulent kinetic energy 
                  and
                 (\textit{b}) friction-velocity based Reynolds number.
                 Note that the time coordinate has been shifted to the
                 beginning of the observation interval, excluding initial
                 transients. 
                }
        \label{fig:energy_trend}
\end{figure}
  Despite the great progress achieved through past studies, a number
  of important questions still remain unanswered. The present work is an
  attempt to fill these gaps. 
  First, a detailed characterization of the spatial structure of the
  dispersed phase in horizontal channel flow with finite-size
  particles is still lacking. For this purpose an analysis based upon
  Voronoi tesselation 
  relative to the instantaneous particle
  distribution is an adequate instrument, as has been demonstrated by
  \cite*{monchaux:2010b} and \cite*{monchaux:2012}. 
  Additionally, the particle-pair distribution function yields
  complementary information on the large-scale structure of the
  particle `field'. 
  Second, in order to rigorously confirm the explanation of the
  apparent velocity lag of the particles in terms of preferential
  sampling of the fluid velocity field, the possibility of a
  systematic instantaneous inter-phase velocity lag needs to be
  excluded. This in turn requires the definition of a characteristic
  fluid velocity in the vicinity of the particles (often referred to
  as the velocity `seen' by the particles). 
  However, for finite-size particles a unique definition does not
  \revision{exists}{exist}, 
  and for different reasons previous attempts
  \citep{bagchi:2003,merle:2005,zeng:2008,lucci:2010} do not
  yield the desired results in the present flow configuration. 
  Here we propose a new definition of the fluid velocity
  seen by finite-size particles. 
  A third point of interest concerns the fluid velocity field
  conditioned upon the presence of particles and, in particular, the
  relation between the spanwise motion of near-wall particles (driving
  particles into low-speed streaks) and the presence of coherent
  vortices.  

  In order to investigate these three aspects of the problem,
  highly-resolved flow and particle data (in time and space) is
  required.  
  In the present work we have generated such data by means of  
  interface-resolved direct 
  numerical simulation of particle-laden, horizontal, open 
  channel flow over a smooth wall.
  All relevant scales of the flow problem are resolved by means of a
  finite-difference/\-immersed-boundary technique.  
  The solid volume fraction was set to a relatively low value in order
  to avoid a dominance of inter-particle collisions. 
  Contact between pairs of particles and contact of particles with the
  solid wall is considered as frictionless in the simulations. 
\begin{figure}
  \centering
  \hspace{2ex}
  \begin{minipage}{1ex}
    \rotatebox{90}{\hspace{-6ex}\small $h$}
  \end{minipage}
  \begin{minipage}{.9\linewidth}
    \centerline{\small$(a)$}
    \includegraphics[width=\linewidth]
    {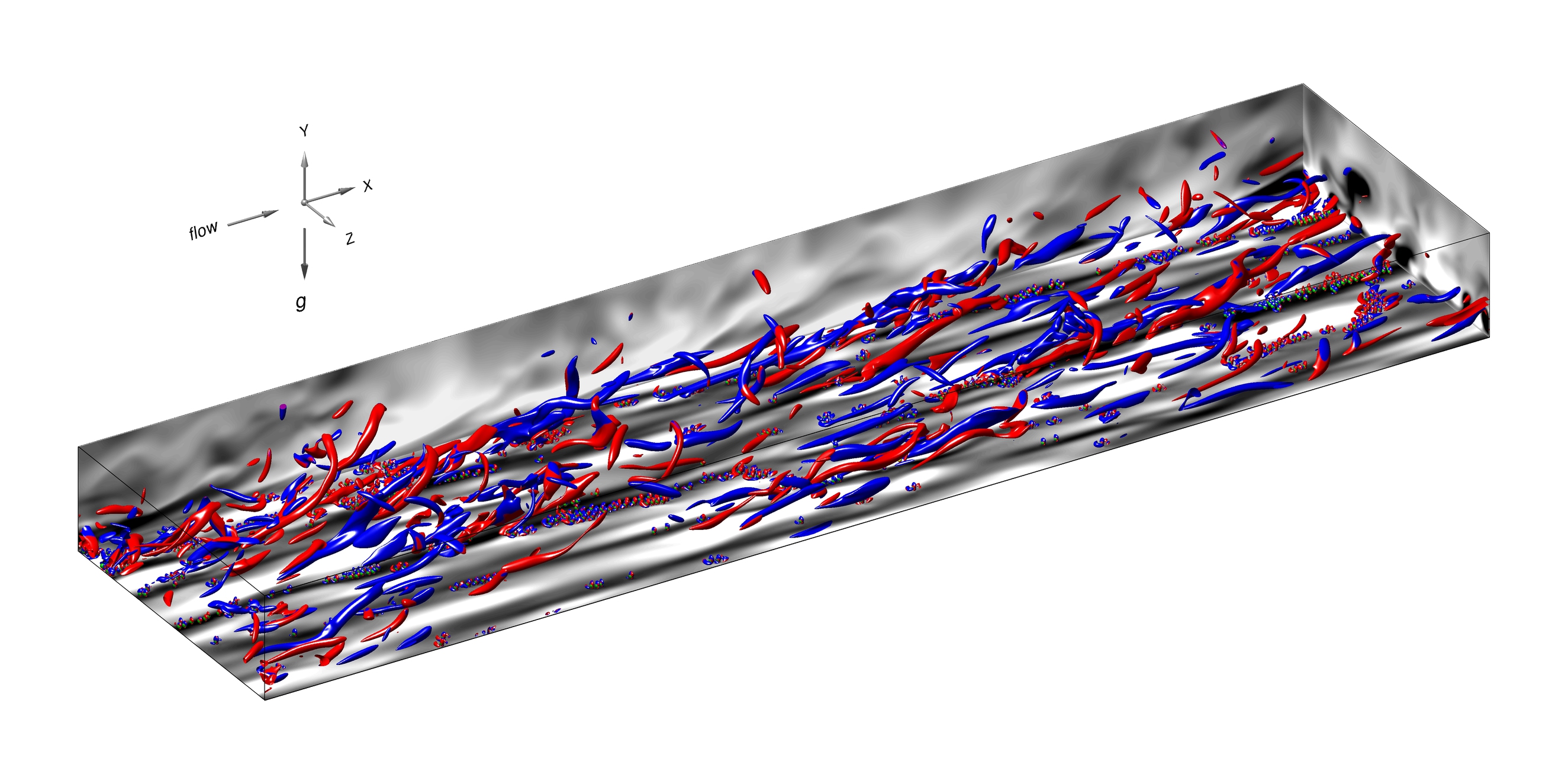}
    \hspace{-.98\linewidth}\raisebox{.04\linewidth}
    {\small $3h$}
    \hspace{.5\linewidth}\raisebox{.09\linewidth}
    {\small $12h$}
  \end{minipage}
  \\[2ex]
  \begin{minipage}{.75\linewidth}
    \centerline{\small$(b)$}
    \includegraphics[width=\linewidth]
    {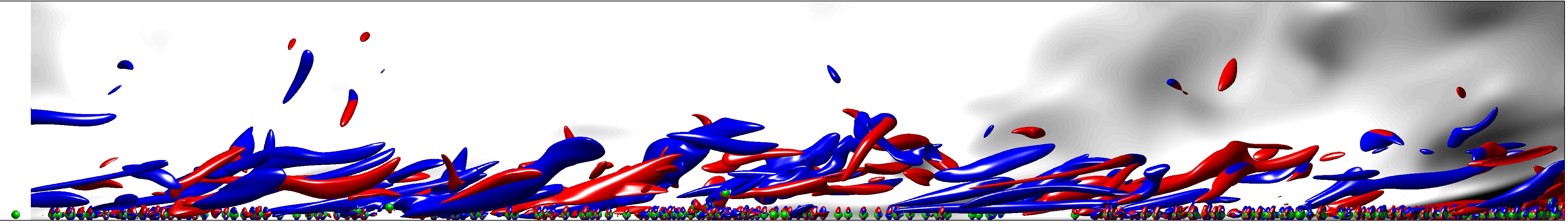}
  \end{minipage}
  \caption{Instantaneous three-dimensional snapshot of
    the flow field
    and the 
    particle positions
    (observed mostly residing at the bottom wall). 
    Strong vortical
    regions are shown by plotting 
    a negative-valued 
    iso-surface of the
    $\lambda_2$ field 
    (cf.\ definition in
    \S~\ref{subsec-role-of-streamwise-vortices-on-distribution-of-particles}).  
    The iso-surfaces are
    colored (red/blue) based on the sign of 
    the streamwise vorticity 
    (positive/negative).
    The greyscales indicate values of the 
    streamwise velocity fluctuation 
    in a 
    wall-parallel plane at 
    $y^+=5$ 
    (projected upon the plane $y=0$ for better visibility) 
    and in wall
    perpendicular planes at $x/h=12$ and $z/h=0$,  
    ranging from black (minimum negative values) to white
    (maximum positive values).
    \revision{}{%
      Note that although the particle surfaces are colored
      in green, particles appear mostly in either red or
      blue color, due to vortical structures located in
      their immediate vicinity.
      A close-up of the same graph looking against the spanwise
      direction is shown in $(b)$ (i.e.\ the flow is from left to right). 
    } 
  }
  \label{fig:3d_flow_field}
\end{figure}
%
\section{Computational methodology and setup}
\label{sec-computational-methodology-and-setup}
\subsection{Numerical method}\label{subsec-numerical-method}
The numerical method employed in the present study is a
formulation of the immersed boundary method for the simulation
of particulate flows developed by \citet{uhlmann:2005a}.
The basic idea of the immersed boundary method is to solve the 
modified incompressible 
Navier-Stokes equations throughout the entire domain $\Omega$
comprising the fluid domain $\Omega_{\mathrm{f}}$ and the domain occupied by
the suspended particles $\Omega_{\mathrm{s}}$ while adding  a 
force term 
which serves to impose the no-slip condition at the fluid-solid
interface. 
The immersed boundary technique is realized in the
framework of a standard fractional step method for
the incompressible Navier-Stokes equations. 
The temporal discretization is semi-implicit,
based on the Crank-Nicolson scheme for the viscous terms and a
low-storage three-step Runge-Kutta procedure for the non-linear part
\citep{verzicco:1996}. The spatial operators are evaluated by central
finite-differences on a staggered grid.
The temporal and spatial accuracy of this scheme are of second order.
In the computation of the forcing term, the necessary
interpolation of variable values from Eulerian grid positions
to particle-related Lagrangian positions
(and the inverse operation of spreading the computed force
terms back to the Eulerian grid) are performed
by means of the regularized delta function approach of
\citet{peskin:2002}. This procedure yields a smooth temporal
variation of the hydrodynamic forces acting on individual
particles while they are in arbitrary motion with respect
to the fixed grid. The employed Cartesian grid is uniform
and isotropic in order to ensure the conservation of
important quantities such as the total force and torque,
during the interpolation and spreading procedure.
The particle motion is determined by the Runge-Kutta-discretized
Newton equations for linear and angular motion of rigid bodies,
driven by buoyancy, hydrodynamic forces/torque and contact forces
(in case of collisions).

%
\begin{figure}
   \centering
        \begin{minipage}{2ex}
        \rotatebox{90}{\small $\quad y/h$}
        \end{minipage}
        \begin{minipage}{.5\linewidth}
          \includegraphics[width=\linewidth]
          {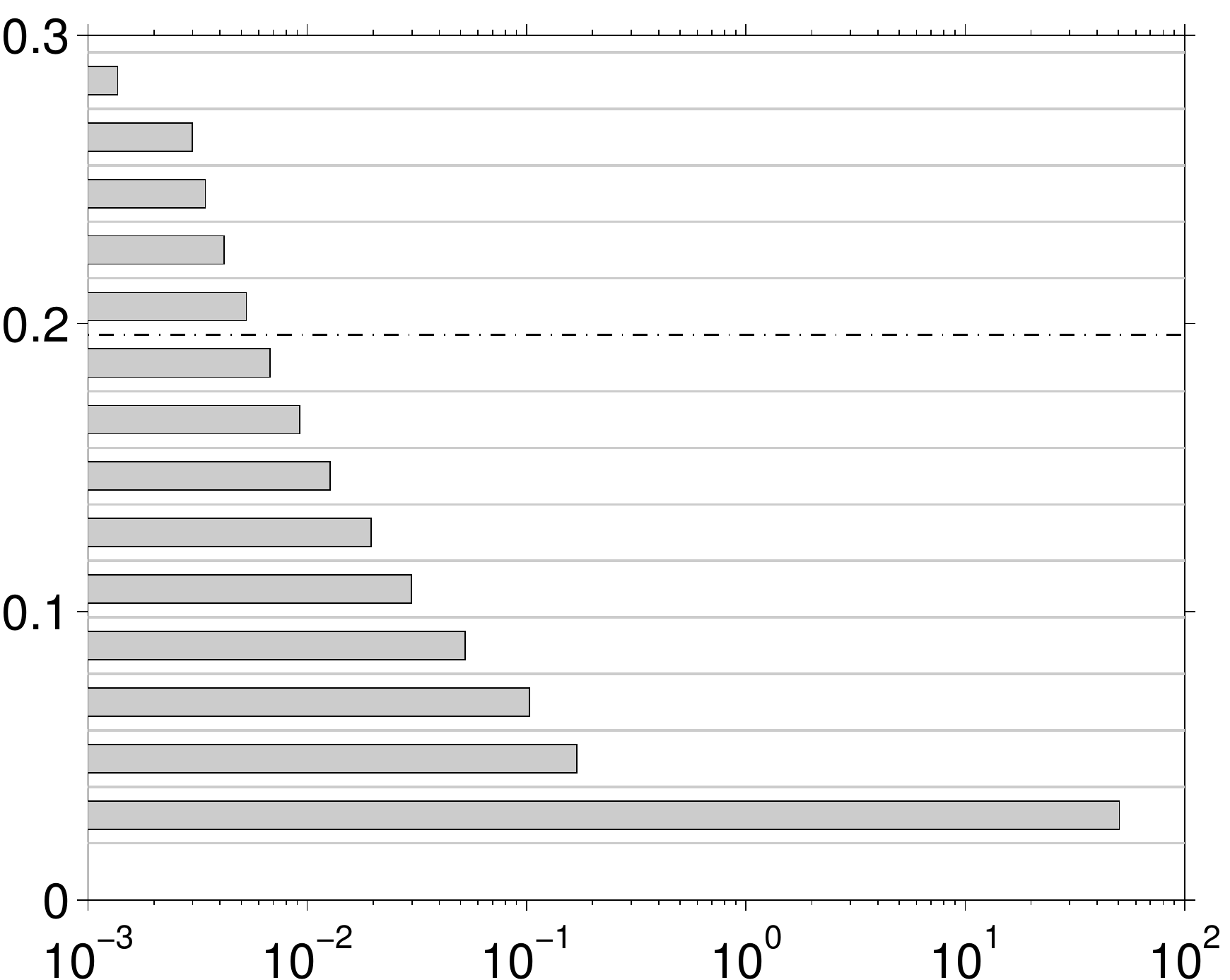}
          \\
          \centerline{\small $\langle\phi_s\rangle/\Phi_s$}
        \end{minipage}      
        \\[-180pt]
        \hspace{100pt}
        \begin{minipage}{.21\linewidth}
          \includegraphics[width=\linewidth]
          {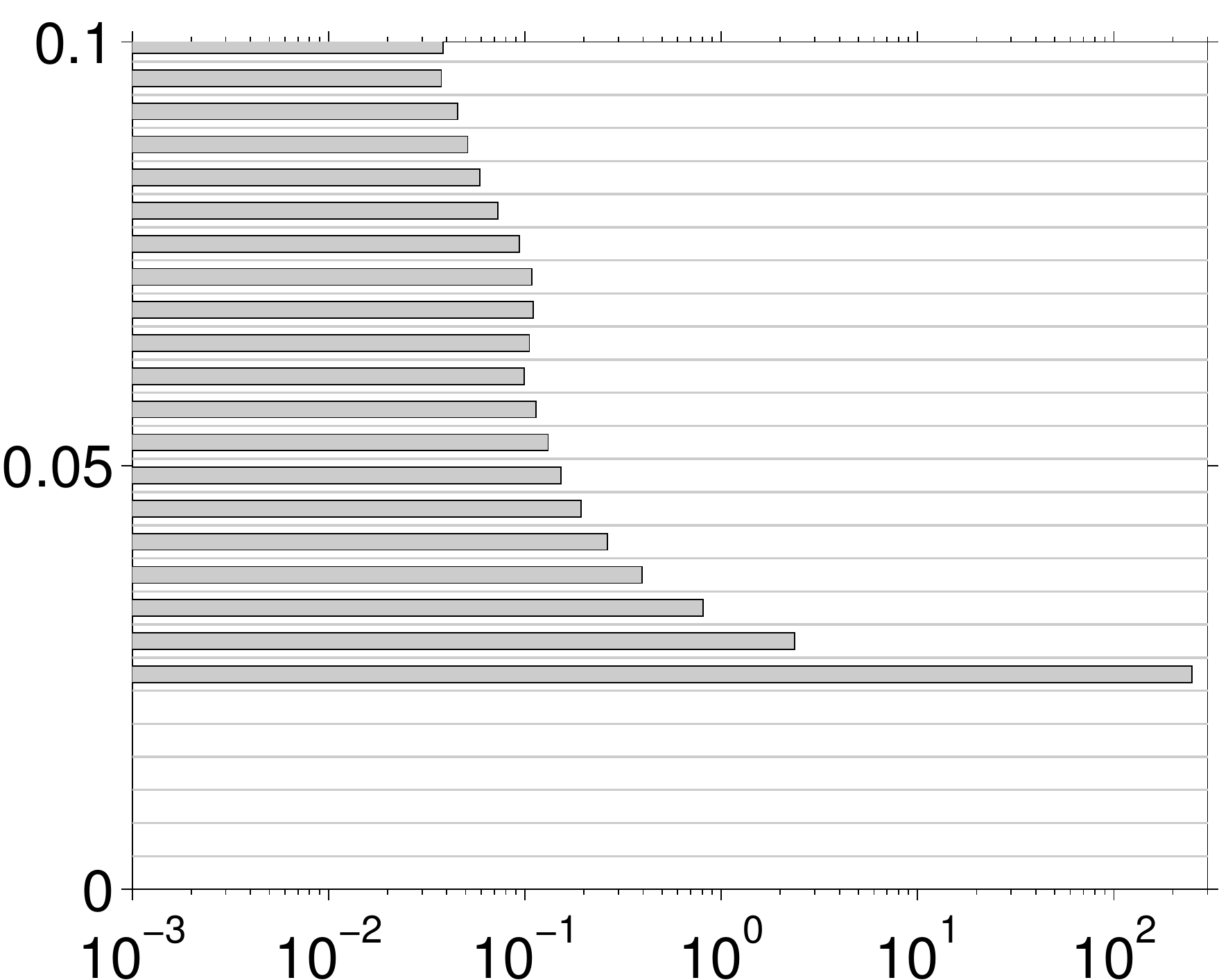}
        \end{minipage}
        \vspace{.24\linewidth}
        \caption{
          Profile of the average solid volume fraction
          $\langle\phi_s\rangle$, normalized by the global value
          $\Phi_s$.
          The width of the averaging
          bins are shown with gray-colored lines. 
          The number of data samples located in bins above the
          chain-dotted line (at $y^+=36$)
          is smaller than 800. 
          Note that the bin adjacent to the wall has zero number of samples
          as there is no particle center in the interval 
          $0\le y_\mathrm{p} < D/2$.
          The 
          insert 
          shows the same quantity for an interval close to the
          wall, evaluated with finer averaging bins $\Delta h=\Delta
          x$. 
        }    
        \label{fig:particle_solid_volume_fraction}
\end{figure}
During the course of a simulation, particles can closely approach
each other. However, very thin 
inter-particle fluid films cannot be 
resolved by a typical grid, and therefore, the correct buildup
of repulsive pressure is not captured, which, in turn, can lead
to possible partial ``overlap'' of the particle positions in the
numerical computation. In practice,
particle-particle contacts are treated  
by a simple repulsive force mechanism \citep{glowinski:1999},
which is active for particle pairs at a distance smaller than 
two grid spacings. The contact force only acts in the normal direction
(along the line connecting the two particles' centers) and frictional 
contact forces are not considered. 
The analogous treatment is applied to particle-wall contact. 
\revision{}{%
  Note that a posteriori evaluation of the particle trajectories
  revealed that the average temporal interval between two collision
  events measures approximately $4.1$ bulk flow time units. 
}

The numerical method employs domain decomposition for
parallelism and has been shown to run on grids of up to $8192^3$,
using up to ${\cal O}(10^5)$ processor cores in scaling tests
\citep[][]{uhlmann:2010}.
The numerical method has been validated on a whole range of
benchmark problems \citep[][]{uhlmann:2004, uhlmann:2005a,
uhlmann:2005b, uhlmann:2006}, and has been previously
employed for the simulation of several flow configurations 
\citep[][]{uhlmann:2008,chan-braun:2011,garcia-villalba:2012}.
\subsection{Flow configuration and parameter values}
\label{subsec-flow-configuration}
Horizontal open channel flow laden with finite-size,
heavy, spherical particles is considered
including the action of gravity. 
As shown in figure \ref{fig:geo_setup},
a Cartesian coordinate system is adopted such that $x$, $y$, and $z$ are 
the streamwise, wall-normal and spanwise directions respectively. 
Mean flow and gravity are directed in the positive $x$ and the 
negative $y$ directions respectively.
The computational domain is
periodic in the homogeneous directions (streamwise and spanwise).
\begin{figure}
   \centering
        \begin{minipage}{2ex}
        \rotatebox{90}{\small $\quad y/h$}
        \end{minipage}
        \begin{minipage}{.5\linewidth}
        \includegraphics[width=\linewidth]
        {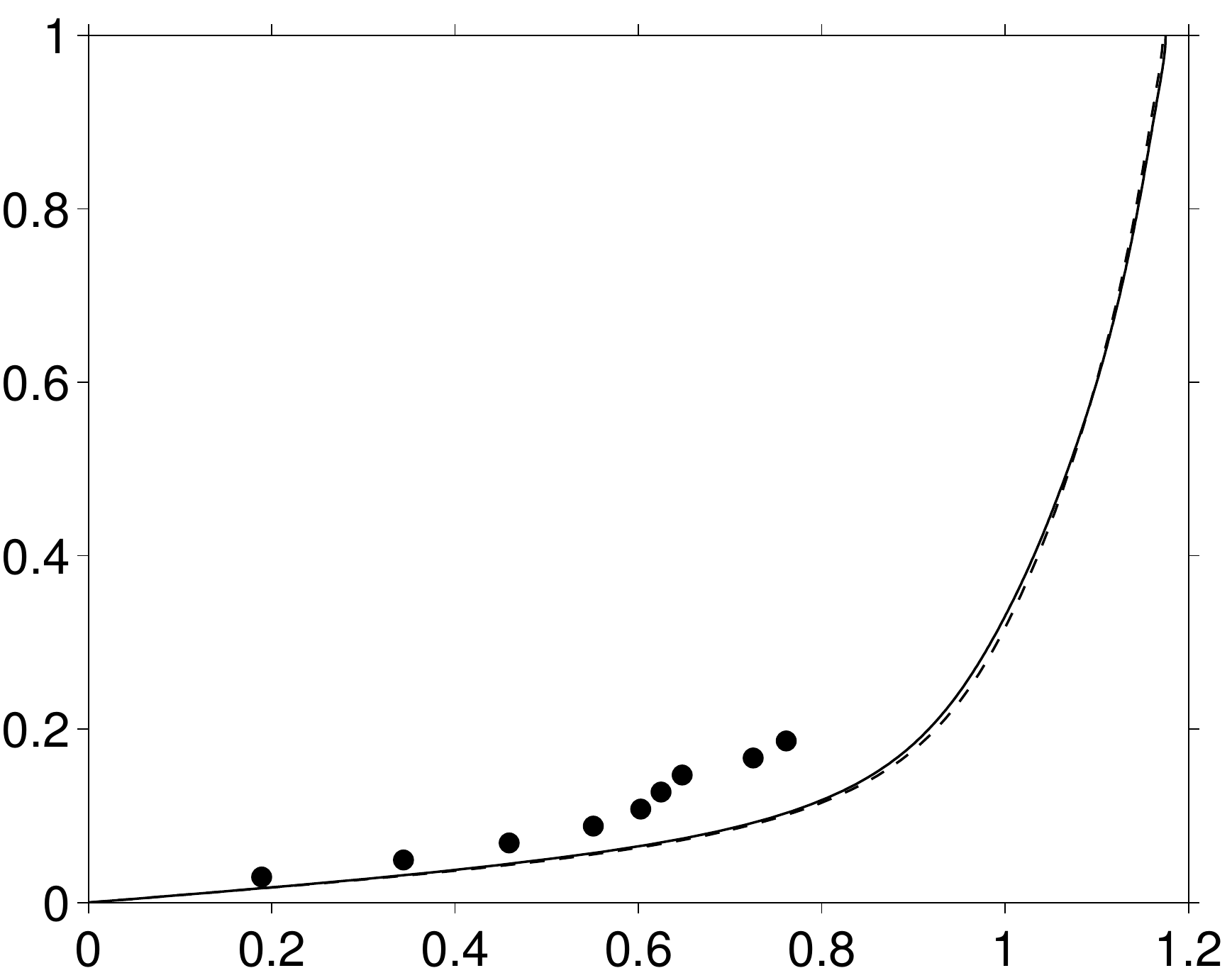}
        \hspace{-0.5\linewidth}\raisebox{.85\linewidth}
        {\small $(a)$}
        \\
        \centerline
        {\small $\langle u_\mathrm{f} \rangle /u_\mathrm{b},
                 \langle u_\mathrm{p} \rangle /u_\mathrm{b}$}
        \end{minipage}\\[20pt]
        %
        \begin{minipage}{2ex}
        \rotatebox{90}
        {\small $\langle u_\mathrm{f} \rangle /u_\tau,
                 \langle u_\mathrm{p} \rangle/u_\tau$ }
        \end{minipage}
        \begin{minipage}{.5\linewidth}
        \includegraphics[width=\linewidth]
        {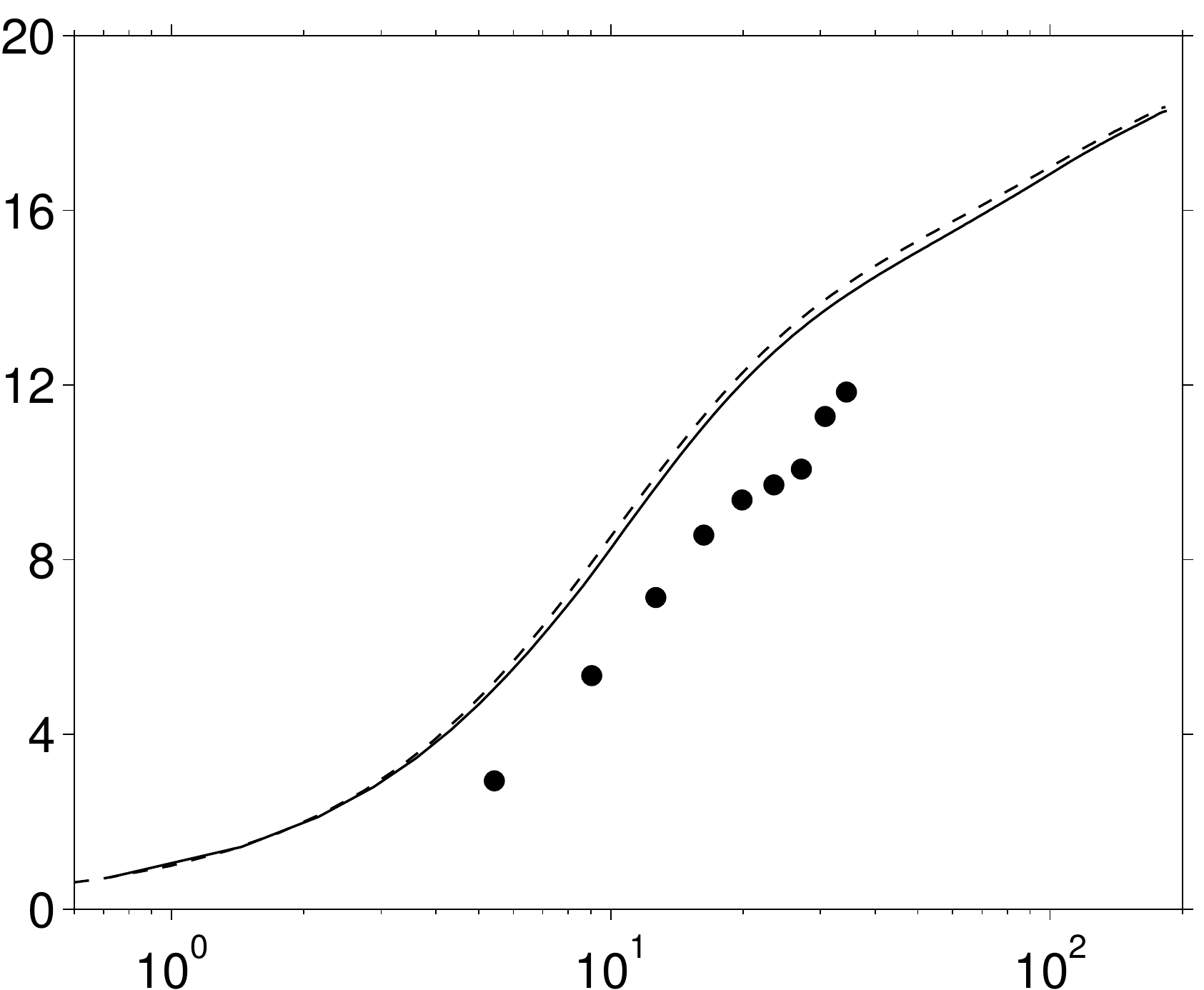}
        \hspace{-0.5\linewidth}\raisebox{.85\linewidth}
        {\small $(b)$}
        \\
        \centerline{\raisebox{1ex}{\small $ y^+ $}}
        \end{minipage}
         \vspace{2ex}
        \caption{Wall-normal profile of the mean streamwise fluid velocity 
                (solid line) and particle velocity (filled symbols) both in 
                (\textit{a}) outer units and (\textit{b}) wall units. 
                The dashed line represents the corresponding profile of a
                single phase flow at a bulk Reynolds number 
                identical to  the current simulation.
                }
        \label{fig:velocity_profile}
\end{figure}
A free-slip condition is 
imposed at the top boundary while the no-slip condition is imposed at 
the bottom wall. 
The channel is driven by a horizontal mean pressure 
gradient imposing a constant flow rate.
The Reynolds number of the flow based on the channel height 
$h$ and the bulk velocity 
$u_\mathrm{b}~\equiv
~\frac{1}{h}\int_{0}^{h}\langle u_\mathrm{f} \rangle \mathrm{d}y$ is
set to a value $Re_\mathrm{b} \equiv u_\mathrm{b}h/\nu= 2870$ 
generating a turbulent flow at
a friction Reynolds number $Re_\tau\equiv u_\tau h/\nu = 184$, where
$u_\tau \equiv \sqrt{\langle \tau_\mathrm{w} \rangle/\rho}$ 
is the friction velocity,
and $\langle\tau_\mathrm{w} \rangle$ 
is the mean shear stress averaged in time
and 
over the solid wall at $y=0$.
The value of the friction velocity matches the value obtained
in single-phase flow at the same bulk Reynolds number to within $1\%$.
The spherical particles considered are relatively small in size when 
compared to the channel depth (particle diameter 
$D/h=10/256$) 
but they are larger than the 
viscous length scale ($D^+=7.21$). 
Note that the ``$^+$'' superscript indicates 
scaling in wall units.
The particle-to-fluid density ratio is set to a value of
$\rho_\mathrm{p}/\rho_\mathrm{f} = 1.7$. 
The Stokes number is defined as the ratio between the
particle relaxation time scale
$\tau_\mathrm{p} \equiv \rho_\mathrm{p}/\rho_\mathrm{f}D^2/(18\nu)$, 
assuming Stokes drag law,
and a fluid time scale. 
The Stokes number based on the near wall fluid time scale 
has a value $St^+\equiv\tau_\mathrm{p}u_\tau^2/\nu = 4.9$ and the 
Stokes number based on bulk fluid time scale $T_\mathrm{b} \equiv
h/u_{\mathrm{b}}$ has a value of 
$St_\mathrm{b}\equiv\tau_{\mathrm{p}}/T_\mathrm{b} = 0.41$.
The global solid volume fraction was set to  
$\Phi_\mathrm{s} = 5 \times 10^{-4}$.
This means that there are $N_\mathrm{p} = 518$ particles in the
computational domain $\Omega$ of size $12h\times h\times 3h$.
The non-dimensional gravity constant $gh/u_\mathrm{b}^2$ 
was set to a value $0.8$ such that the Galileo number, defined
as $Ga^2 = (\rho_\mathrm{p}/\rho_\mathrm{f}-1)gD^3/\nu^2$ has a value
\revision{%
  $Ga = 16.58$
}{%
  $Ga = 16.49$
} 
and the Shields number, defined as
$\theta^{-1} = (\rho_\mathrm{p}/\rho_\mathrm{f}-1)gD/u_\tau^2$
has a value $\theta = 0.19$. 
The computational domain is discretized by a uniform
isotropic grid with grid spacing 
$\Delta x = \Delta y = \Delta z = h/256$.
This grid spacing yields a particle resolution of $D/\Delta x = 10$
and in terms of wall units $\Delta x^+ \approx 0.7$.
Table \ref{tab:physical_parameters} and 
table \ref{tab:numerical_parameters} summarize the important 
physical and numerical parameter values adopted.

\revision{%
  The simulation was initiated from 
  single-phase flow, inserting the 
  particles at randomly-chosen positions.
  After starting-up the 
  current 
  simulation, the particles rapidly 
  settle to the bottom. For the 
  flow to develop into a new statistically steady state regime,
  a transient period of $40T_\mathrm{b}$  was required and the data
  which corresponds to this time interval is not considered in the
  statistical analysis which follows. 
  After a 
  statistically 
  steady state has been attained,
  the accumulation of statistics has been carried out 
  for a period of 
  $162T_\mathrm{b}$.}
{%
  The simulation was first initiated on a coarser grid (by a factor of
  two in each coordinate direction), using a fully turbulent flow
  field with particles distributed throughout the computational domain
  as initial condition. 
  A transient of roughly $40\,T_\mathrm{b}$ was observed, after which
  the particles had formed patterns very similar to those observed and
  described below.
  Then the final data of the coarser-grid simulation was interpolated
  upon the current grid, and the simulation was continued for another
  $252\,T_\mathrm{b}$.  
  The entire initial interval of $560\,T_\mathrm{b}$ was
  unintentionally computed with an incorrect value of the particles'
  moment of inertia  
  ($\rho_pD^5\pi5/48$ instead of $\rho_pD^5\pi/60$). 
  After correcting the value of the moment of inertia, the simulation
  was run for additional $260\,T_\mathrm{b}$, the first
  $5\,T_\mathrm{b}$ of which have been discarded. 
  Therefore, the temporal observation interval over which the present
  data was generated measures $T_\mathrm{obs}=255\,T_b$. 
}

Figure \ref{fig:energy_trend} shows the time evolution 
\revision{}{over the observation interval} 
of the 
box-averaged turbulent kinetic energy,
defined as 
$\langle k \rangle_{\mathrm{xyz}} = 
 \frac{1}{2}
(\langle u_{\mathrm{f},i}u_{\mathrm{f},i} \rangle_{\mathrm{xyz}} - 
 \langle u_{\mathrm{f},i}\rangle_{\mathrm{xyz}}
 \langle u_{\mathrm{f},i} \rangle_{\mathrm{xyz}})$, 
where $\langle \cdot \rangle_{\mathrm{xyz}}$ denotes an instantaneous
average over the domain occupied by the fluid, as defined in
(\ref{equ-def-avg-operator-box-fluid-only}).  
The time evolution of $Re_\tau$ 
based upon the instantaneous value of the friction velocity 
is also presented in the figure. 
Several cycles of large-scale fluctuations can be observed. 
The reader is referred to 
\ref{sec-statistical-convergence-fluid-phase} for a
statistical convergence check based on the
balance of the mean streamwise momentum.

Figure \ref{fig:3d_flow_field} gives an impression of the 
size of the domain, the coherent flow structures and the 
particle distribution.

The simulation was typically performed on $48$ to $256$ processor cores
and required roughly 
\revision{%
  one million core hours of CPU time.
}{%
  half a million core hours of CPU time.
}

In addition, we have generated single-phase reference data at the 
same parameter point by means of a pseudo-spectral method
\citep{kim:1987}. This reference simulation has been run
for over $4400$ bulk time units, 
i.e.\ a considerably longer interval than
that of the main simulation.
The single-phase data will be used for the purpose of 
comparison in the following.

\subsection{Notation}\label{subsec-notation}
%
%
Before turning to the results, let us fix the basic notation
followed throughout the present text. Velocity vectors and their
components corresponding to the fluid and the particle phases
are distinguished by subscript ``f'' and ``p'', respectively,
as in $\mathbf{u}_{\mathrm{f}} = (u_{\mathrm{f}},v_{\mathrm{f}},w_{\mathrm{f}})^T$ 
and $\mathbf{u}_{\mathrm{p}} = (u_{\mathrm{p}},v_{\mathrm{p}},w_{\mathrm{p}})^T$. 
Similarly, the particle position vector is denoted as 
$\mathbf{x}_{\mathrm{p}} = (x_{\mathrm{p}},y_{\mathrm{p}},z_{\mathrm{p}})^T$ 
and the vector of angular particle velocity as 
$\boldsymbol{\omega}_{\mathrm{p}} = 
(\omega_{\mathrm{p,x}},\omega_{\mathrm{p,y}},\omega_{\mathrm{p,z}})^T$.

Fluctuations of the fluid velocity field with respect to the average
over wall-parallel planes and time are henceforth denoted by a single
prime, i.e.\
$\mathbf{u}_{\mathrm{f}}^\prime(\mathbf{x},t)=
\mathbf{u}_{\mathrm{f}}(\mathbf{x},t)
-\langle\mathbf{u}_{\mathrm{f}}\rangle(y)$.   
Likewise, the fluctuations of the particle velocity are defined as the
difference between the instantaneous value and the average (over time
and all particles located in predefined wall-normal intervals) at
the corresponding location, viz.\ 
$\mathbf{u}_{\mathrm{p}}^\prime(\mathbf{x}_{\mathrm{p}}(t),t)=
\mathbf{u}_{\mathrm{p}}(\mathbf{x}_{\mathrm{p}}(t),t)
-\langle\mathbf{u}_{\mathrm{p}}\rangle(y_\mathrm{p}(t))$.   
The reader is referred to  \ref{sec-ensemble-averaging} for 
definitions of the various averaging operators used in the present
study.   

The particle radius is henceforth denoted as $R=D/2$. 

\begin{figure}
   \centering
        \begin{minipage}{2ex}
        \rotatebox{90}{\small $\quad y/h$}
        \end{minipage}
        \begin{minipage}{.5\linewidth}
        \includegraphics[width=\linewidth]
        {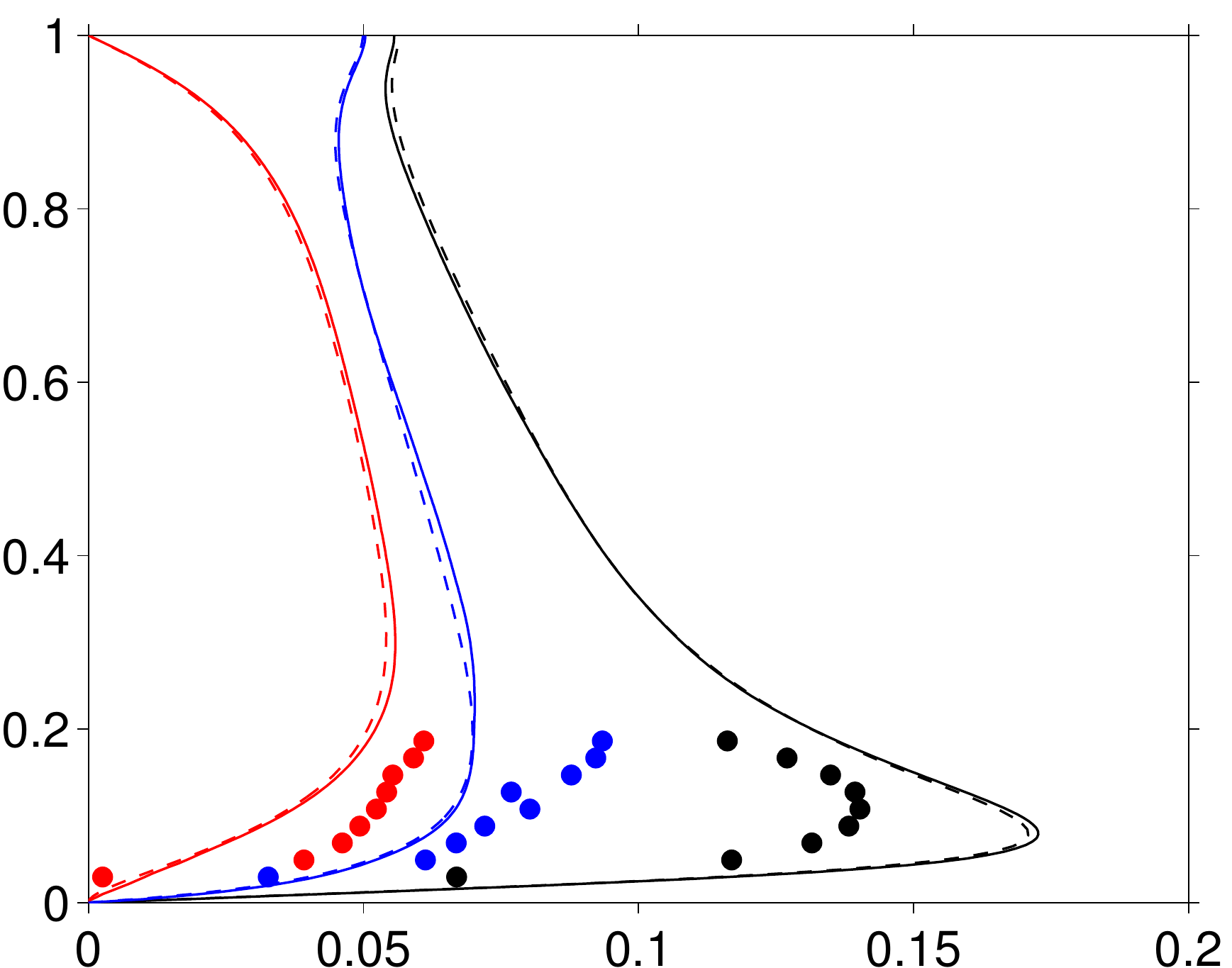}
        \hspace{-0.5\linewidth}\raisebox{.85\linewidth}
        {\small $(a)$}
        \\
        \centerline
        {\small $\langle u'_{\mathrm{f},\alpha} u'_{\mathrm{f},\alpha} 
                  \rangle^{\frac{1}{2}}
                 /u_\mathrm{b}$,
                $\langle u'_{\mathrm{p},\alpha} u'_{\mathrm{p},\alpha} 
                  \rangle^{\frac{1}{2}}
                /u_\mathrm{b}$}
        \end{minipage}\\[20pt]
        %
        \begin{minipage}{3ex}
        \rotatebox{90}
        {\small $\langle u'_{\mathrm{f},\alpha} u'_{\mathrm{f},\alpha} 
                 \rangle^{\frac{1}{2}}
                 /u_\tau$,
                $\langle u'_{\mathrm{p},\alpha} u'_{\mathrm{p},\alpha} 
                \rangle^{\frac{1}{2}}
                /u_\tau$}
        \end{minipage}
        \begin{minipage}{.5\linewidth}
        \includegraphics[width=\linewidth]
        {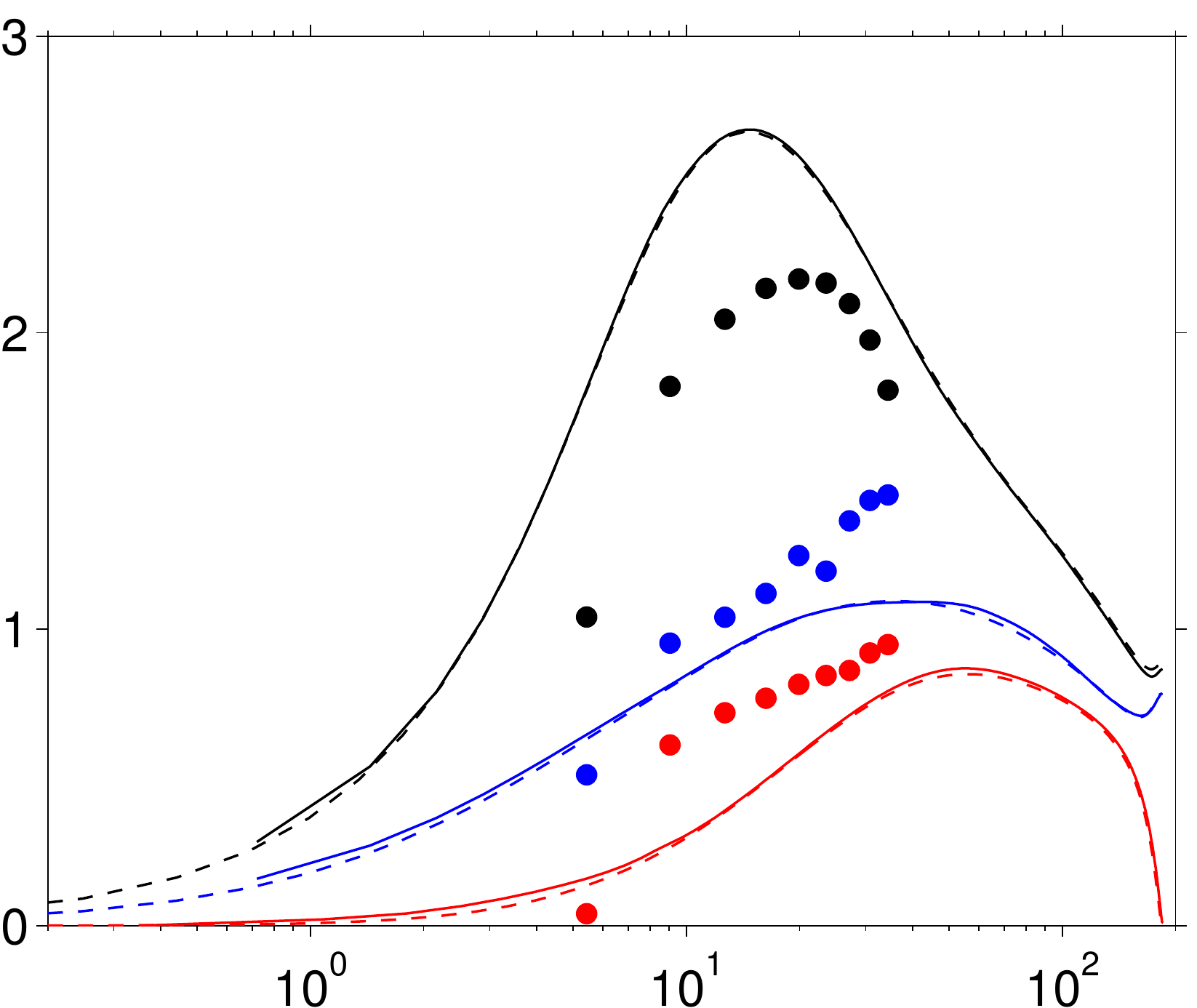}
        \hspace{-0.5\linewidth}\raisebox{.85\linewidth}
        {\small $(b)$}
        \\
        \centerline{\raisebox{1ex}{\small $ y^+ $}}
        \end{minipage}\\[-191pt]
         \hspace{-.21\linewidth}
         \begin{minipage}{.19\linewidth}
           \includegraphics[width=\linewidth]
           {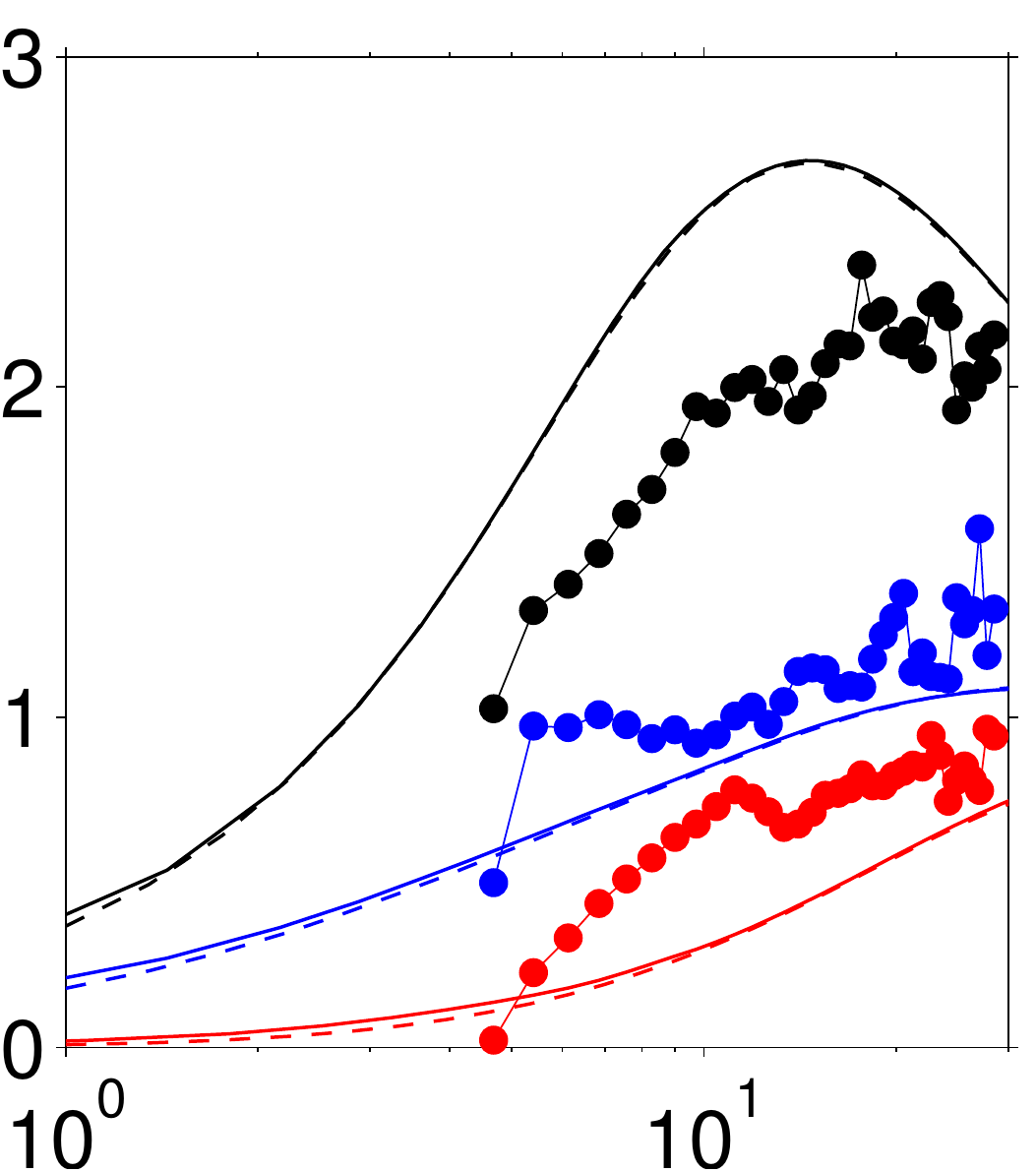}
         \end{minipage}
         \vspace{.21\linewidth}
        \caption{Wall-normal profile of the rms velocity fluctuations of
                the fluid phase (lines) and particles (filled symbols) 
                in (\textit{a}) outer units and (\textit{b}) wall units. 
                The dashed line 
                indicates the single phase data. 
                The black, blue and red colors 
                represent the streamwise, spanwise and wall-normal 
                components of the fluctuations, respectively.
                The 
                insert 
                shows the same quantities but 
                the particle data is 
                evaluated at finer bins of width $\Delta h = \Delta x$.}
        \label{fig:rms_velocity_profile}
\end{figure}
\section{Results}
\label{sec-results}
\subsection{Eulerian statistics}
\label{subsec-eulerian-statistics}
\subsubsection{Mean solid volume fraction}
\label{subsubsec-mean-solid-volume-fraction}
  The global solid volume fraction has such a low value that
  even if all particles are in contact with the solid wall, their
  projected area covers less than two percent of the wall-plane, i.e.\ 
  $N_p\pi R^2/(L_xL_z)=0.0172$.  
  Figure~\ref{fig:particle_solid_volume_fraction} shows the
  wall-normal profile of the average solid volume fraction
  $\langle\phi_s\rangle$ (cf.\ averaging operator defined in 
  \revision{}{%
    equation~\ref{equ-def-avg-solid-volume-frac} in 
  }
  \ref{subsec-binned-averages-over-particle-related-quantities}).  
  Due to the gravitational settling effect, a strong 
  wall-normal particle concentration gradient forms near the 
  wall. 
  The wall-normal location of the particle centers 
  of the overwhelming majority of all particles 
  \revision{%
    ($98.99\%$)
  }{%
    ($99.15\%$)
  } 
  is in the 
  interval $D/2\le y_\mathrm{p}<D$ (i.e.\ corresponding to the second
  averaging bin).  
  At larger wall-distances, the number of particle samples becomes
  scarce for the given temporal observation interval. In order to
  ensure an adequate quality of the statistics, we have chosen to 
  consider only those bins with $y_\mathrm{p}^+\leq36$ (up to the
  dashed line in figure~\ref{fig:particle_solid_volume_fraction}), for
  which the number of samples per bin ranges from a maximum of
  \revision{%
    $6.6\cdot10^6$ down to $993$. 
  }{%
    $10^7$ down to $1386$. 
  }
  Therefore, in the following presentation, all Eulerian particle
  statistics are only provided for wall distances $y^+\leq36$. 
\subsubsection{Mean streamwise velocity}
\label{subsubsec-mean-streamwise-velocity}
%
The profiles of the mean streamwise velocity of fluid and
    solid phase are 
shown in figure \ref{fig:velocity_profile} in 
outer and inner scaling.  
%
Additionally, the 
data for single-phase flow is included for comparison.
In both scalings, the
velocity profile of the fluid phase,  
$\langle u_{\mathrm{f}} \rangle$, almost coincides with the 
results of the single phase flow.
The 
difference
between the curves 
is 
smaller
than 
\revision{%
  $8\times 10^{-3}$
}{%
  $9.3\times 10^{-3}$
} 
times the maximum velocity, 
and, therefore, within the range of statistical uncertainty.
Thus, for the given 
parameters,
the presence of particles has a negligible effect on the mean fluid 
velocity profile. 

The mean streamwise 
particle velocity, $\langle u_{\mathrm{p}} \rangle$,
is observed to be 
systematically
smaller than that of the fluid phase
$\langle u_{\mathrm{f}} \rangle$. 
The difference in the mean velocities of the two phases, 
i.e.\ the apparent velocity lag, is denoted as
\begin{equation}   
  u_{\mathrm{lag}} = \langle u_{\mathrm{f}} \rangle - \langle u_{\mathrm{p}} \rangle
  \,,
  \label{eq:apparent-velocity-lag}
\end{equation}
%
which is found to vary in the range of 
\revision{%
  $2.1 u_\tau$ to $3.4 u_\tau$, 
  corresponding to $0.14 u_{\mathrm{b}}$ to $0.22 u_{\mathrm{b}}$  
}{%
  $2.1 u_\tau$ to $3.2 u_\tau$, 
  corresponding to $0.14 u_{\mathrm{b}}$ to $0.21 u_{\mathrm{b}}$  
}
(cf.\ figure~\ref{fig:relative_velocity}$b$ which will be discussed
below). 
%
\begin{figure}
   \centering
        \begin{minipage}{2ex}
        \rotatebox{90}{\small $\quad y/h$}
        \end{minipage}
        \begin{minipage}{.5\linewidth}
        \includegraphics[width=\linewidth]
        {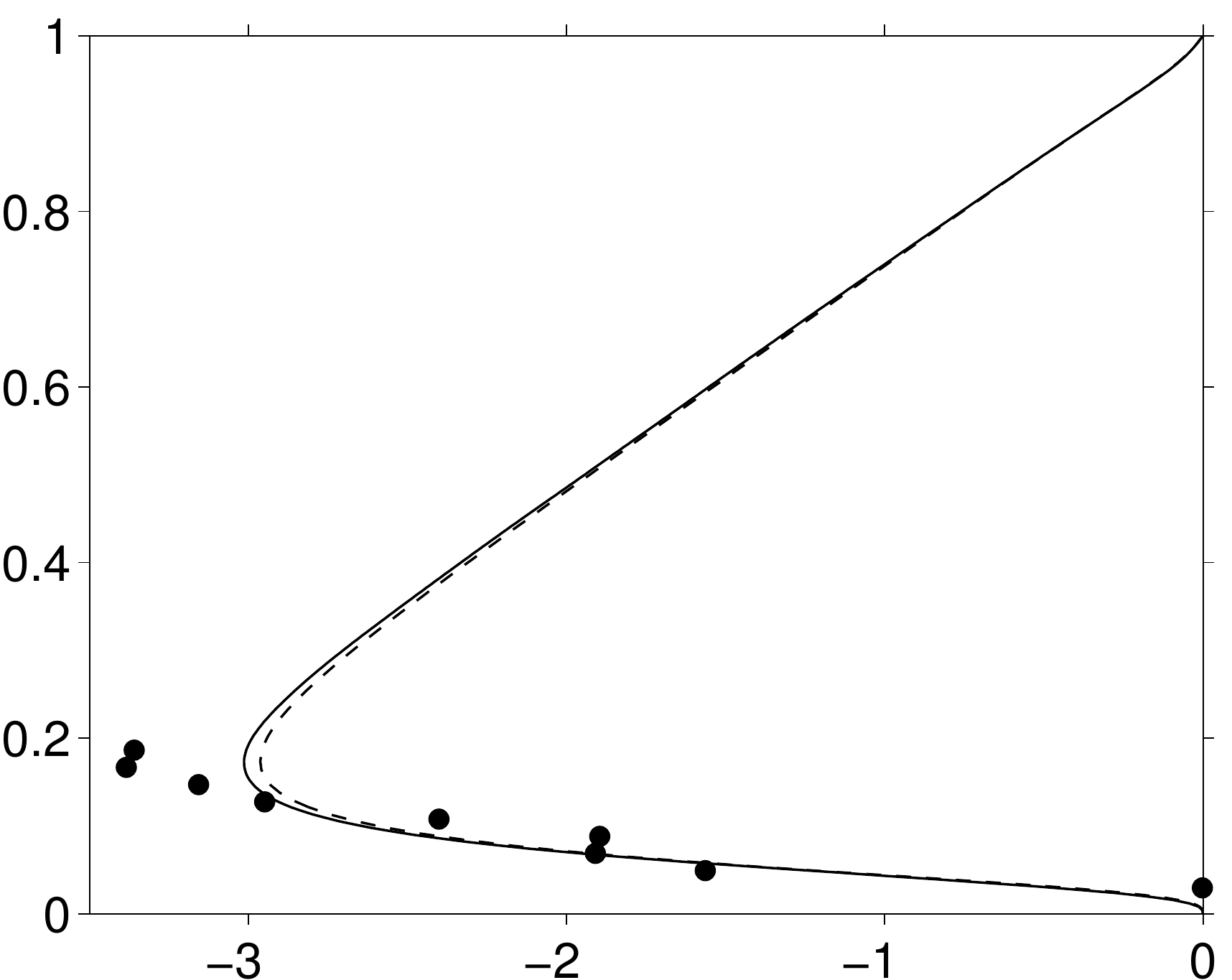}
        \hspace{-0.5\linewidth}\raisebox{.85\linewidth}
        {\small $(a)$}
        \\
        \centerline
        {\small $[ \langle u'_\mathrm{f} v'_\mathrm{f} \rangle/u^2_\mathrm{b}$,
                $\langle u'_\mathrm{p} v'_\mathrm{p} \rangle/u^2_\mathrm{b} ]
                \times 10^3$ }
        \end{minipage}\\[20pt]
        %
        \begin{minipage}{3ex}
        \rotatebox{90}
        {\small $\langle u'_\mathrm{f} v'_\mathrm{f} \rangle/u^2_\tau$,
                $\langle u'_\mathrm{p} v'_\mathrm{p} \rangle/u^2_\tau$ }
        \end{minipage}
        \begin{minipage}{.5\linewidth}
        \includegraphics[width=\linewidth]
        {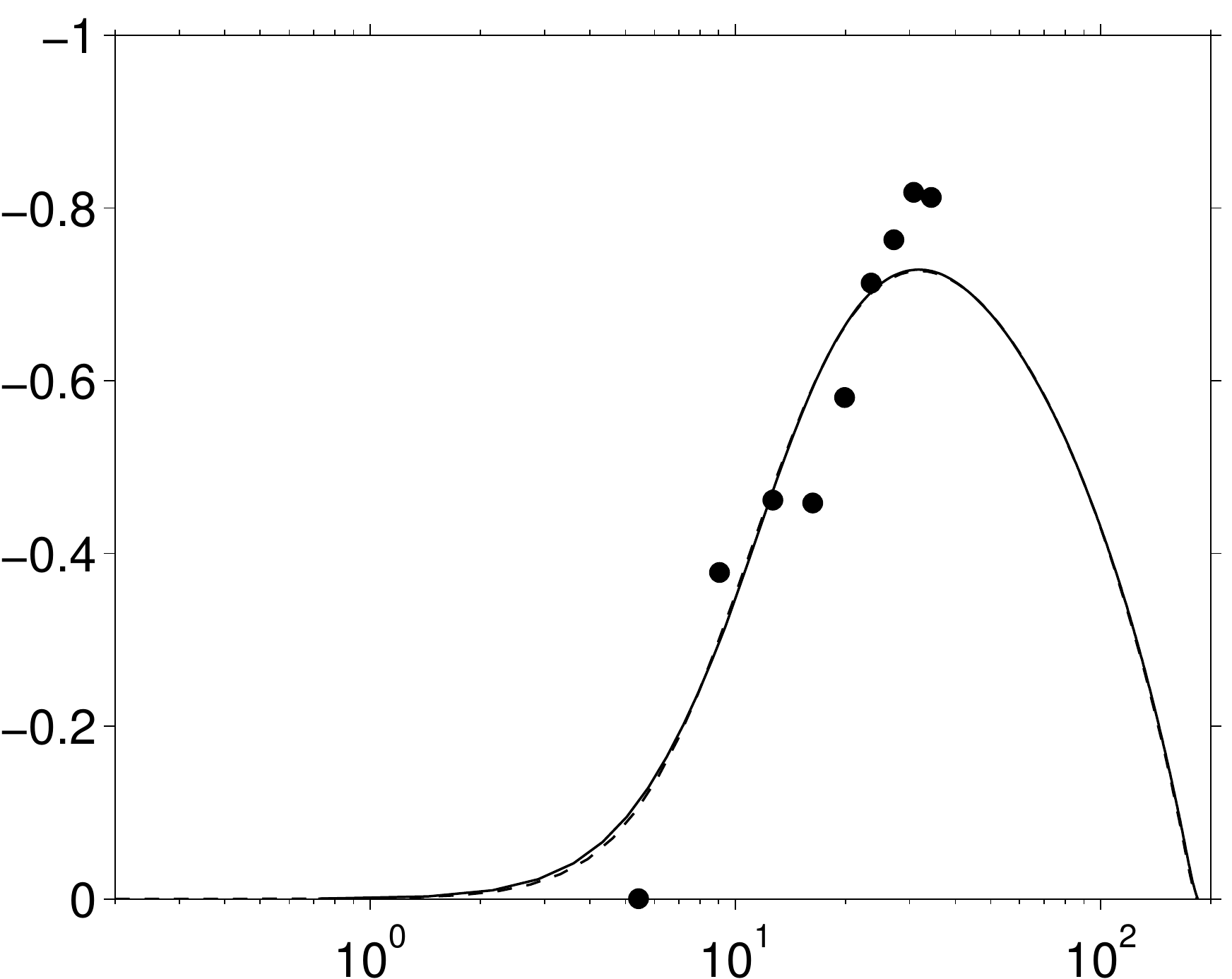}
        \hspace{-0.5\linewidth}\raisebox{.85\linewidth}
        {\small $(b)$}
        \\
        \centerline{\raisebox{1ex}{\small $ y^+ $}}
        \end{minipage}\\[-180pt]
         \hspace{-.12\linewidth}
         \begin{minipage}{.21\linewidth}
           \includegraphics[width=\linewidth]
           {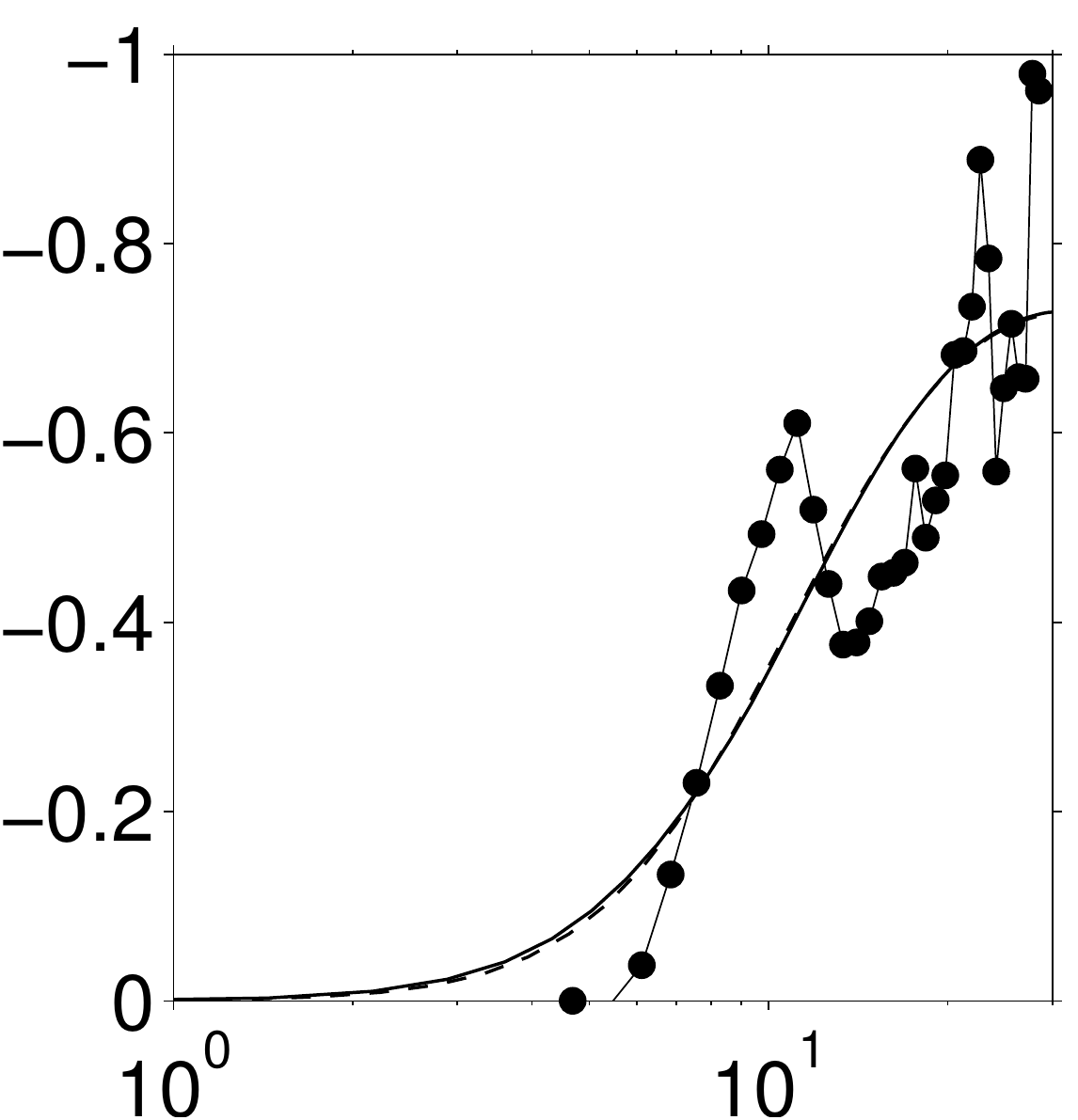}
         \end{minipage}
         \vspace{.21\linewidth}
        \caption{Wall-normal profile of the Reynolds shear stress of
                the fluid phase (lines) and correlation between 
                streamwise and wall-normal velocity fluctuations of 
                particles (filled symbols) 
                in (\textit{a}) outer units and (\textit{b}) wall units. 
                The dashed line 
                indicates the 
                single phase Reynolds stress 
                profiles. 
                The 
                insert 
                shows the same quantities but 
                the particle data is 
                evaluated at finer bins of width $\Delta h = \Delta x$.}
        \label{fig:reynold_stress_profile}
\end{figure}
  The general feature of a positive apparent velocity lag $u_{\mathrm{lag}}>0$
  has been reported in a number of experimental studies on horizontal
  wall-bounded shear flow 
  \citep{rashidi:1990,kaftori:1995b,taniere_oesterle_monnier_EIF_97,kiger:2002,righetti:2004,muste:09,noguchi:09} 
  as well as in the interface-resolved DNS of horizontal plane channel
  flow by \cite{shao:2012}.  
  The aforementioned references cover a relatively broad range of 
  parameter values in terms of Stokes number, Galileo number and solid
  volume fraction. 
  Therefore, the phenomenon, which is commonly linked to particles
  residing preferentially in low-speed regions of the flow,  
  appears to be a relatively `robust' feature of particulate flow in
  horizontal, wall-bounded, shear turbulence. 
  
  However, it should be mentioned that in most of the aforementioned
  experimental studies a change of sign of the apparent velocity lag
  is observed -- contrary to the present simulations and contrary to
  those by \cite{shao:2012}. 
  \cite{kaftori:1995b} measured particle
  velocities exceeding the average fluid velocity in a small interval
  ($y\lesssim D$) very near the wall. The authors, however, attribute this
  result to measurement errors (influence of the angular particle
  velocity) rather than to a physical effect. 
  \cite{kiger:2002} observed average particle velocities
  leading the average fluid velocity at wall-distances
  $y^+\lesssim35$; however, they did not further comment on the
  possible cause for this result.  
  \cite{righetti:2004} also detected a change of sign of the
  apparent slip velocity, i.e.\ $\langle u_{\mathrm{p}}\rangle$ being larger than
  $\langle u_{\mathrm{f}}\rangle$ for $y^+\lesssim20$ (slightly depending on particle
  diameter), and vice versa further away from the wall. 
  The authors explained the leading particle velocity near the wall as a
  consequence of sweep events (i.e.\ in the fourth quadrant of the
  $u_{\mathrm{p}}^\prime,v_{\mathrm{p}}^\prime$-plane) dominating the particle transport
  from the outer flow towards the wall. 
  %
  In the present case, on the contrary, the vast majority of the
  particles are residing in the direct vicinity of the wall (cf.\
  concentration profile in
  figure~\ref{fig:particle_solid_volume_fraction}), 
  and excursions of individual particles to larger wall-distances
  occur only very infrequently. Therefore, the importance of 
  particles carrying high streamwise momentum while being swept
  towards the wall is of little importance to the mean streamwise
  velocity budget in the region near the wall. 
  %
  %
  As a consequence, negative values of $u_{\mathrm{lag}}$ are not observed in
  the present case. 

  In \S~\ref{subsec-fluid-seen-by-particles} 
  we will return
  to the  
  apparent velocity lag in order to determine the contribution from 
  the instantaneous lag with respect to a characteristic fluid
  velocity in the vicinity of the particles. 
  In \S~\ref{subsec-particle-position-and-coherent-structures} it will
  be shown that indeed the decisive contribution stems from particles
  preferentially sampling the low-speed regions of the flow.   
  %
\begin{figure}
   \centering
        \begin{minipage}{3ex}
        \rotatebox{90}{\small $n_\mathrm{coll}/n_\mathrm{p}^{(j)}$}
        \end{minipage}
        \begin{minipage}{.5\linewidth}
        \includegraphics[width=\linewidth]
        {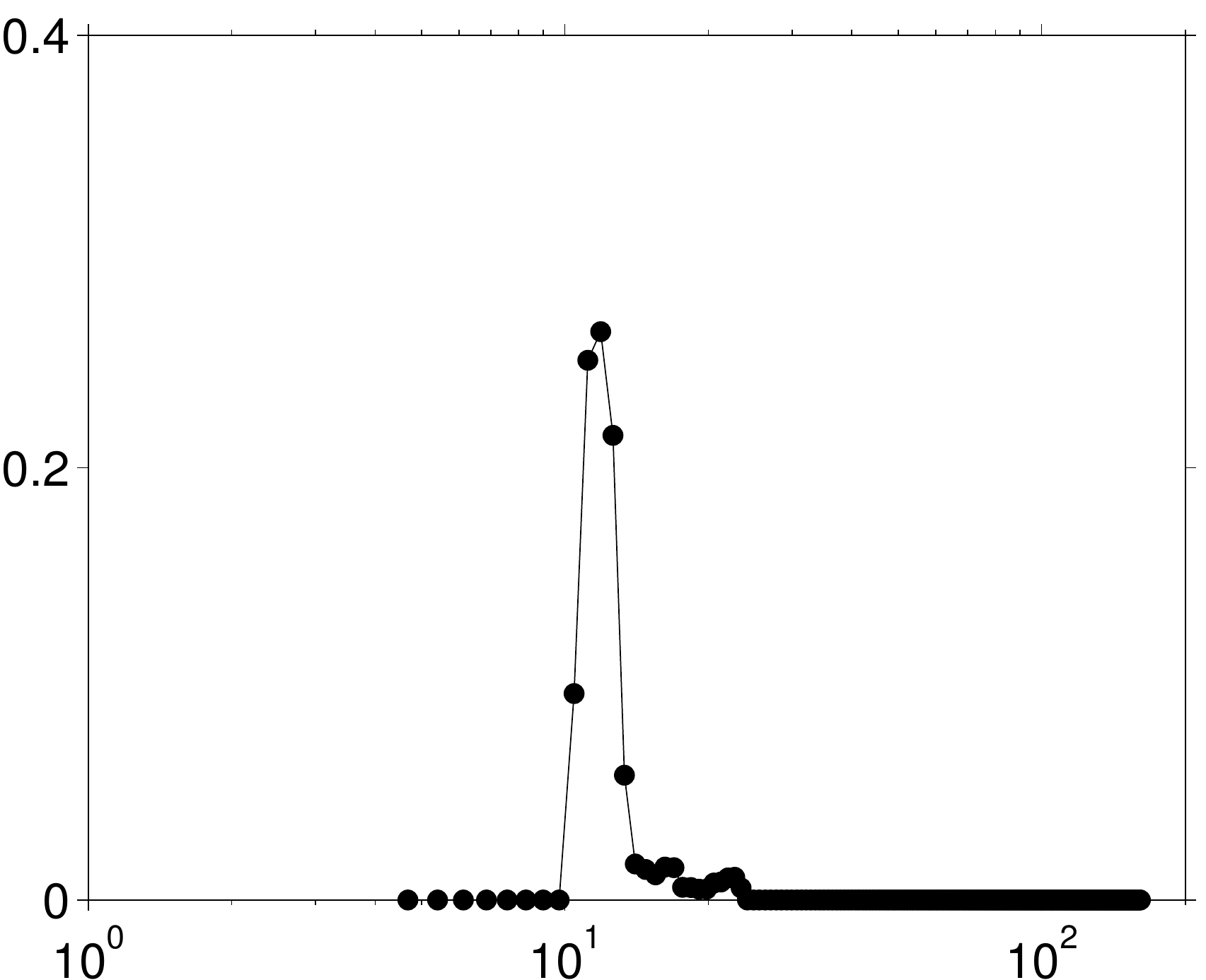}
        \\
        \centerline
        {\small $y^+$ }
        \end{minipage}
         \vspace{2ex}
        \caption{Wall-normal profile of the number of 
                 collision events happening mainly in the
                 vertical direction, i.e.\
                 the orientation of the line joining the centers 
                 of the two colliding particles is closer to a 
                 vertical line than it is to a horizontal line.
                 The collision event is attributed to the particle
                 which is at a larger wall-normal distance than 
                 its partner. Data points correspond to the same
                 fine bins adopted in the 
                 insert 
                 of figure
                 \ref{fig:reynold_stress_profile}.}
        \label{fig:collision_count_profile}
\end{figure}
\subsubsection{Covariances of fluid and particle velocity fluctuations} 
\label{subsubsec-covariances-fluid-particle-velocity-fluctuations}
Figure \ref{fig:rms_velocity_profile} and figure 
\ref{fig:reynold_stress_profile} show the wall-normal profiles of 
the rms velocity fluctuation components of the fluid phase 
$\langle u'_{\mathrm{f},\alpha} u'_{\mathrm{f},\alpha} \rangle^{\frac{1}{2}}$ 
and of the particle phase
$\langle u'_{\mathrm{p},\alpha} u'_{\mathrm{p},\alpha} \rangle^{\frac{1}{2}}$, 
the Reynolds shear stress profile of the fluid phase 
$\langle u'_{\mathrm{f}} v'_{\mathrm{f}} \rangle$ as well as the correlation 
between the streamwise and wall-normal particle velocity fluctuations 
$\langle u'_{\mathrm{p}} v'_{\mathrm{p}} \rangle$. 
All components of the fluid velocity fluctuation covariances deviate 
only marginally from the 
single phase counterparts when 
normalized both in outer and wall scales. This confirms that the presence 
of the particles, at the presently adopted parameter values,
has a negligible influence on the one-point statistics of the 
fluid
velocity field. 

Concerning the dispersed phase, in the near wall region within
$y^+ < 36$, the particle 
velocity
fluctuation intensity is observed to be
less than that of the fluid 
fluctuation in the streamwise direction. The reverse is true for 
the cross-stream components, i.e.\ in the wall-normal and spanwise 
directions particle velocity components are observed to 
fluctuate more strongly than fluid velocities.
An exception are 
those particles which are in the near proximity of the wall
(with their centers located at wall normal distance 
within one particle diameter from the
wall) where $\langle v_{\mathrm{p}}'^2 \rangle$ 
($\langle w_{\mathrm{p}}'^2 \rangle$),
is also 
smaller
than $\langle v_{\mathrm{f}}'^2 \rangle$ 
($\langle w_{\mathrm{f}}'^2 \rangle$). 
%
  Note that the apparent `jump' from the first to the second data
  point in figure~\ref{fig:rms_velocity_profile}$(b)$ is due to the
  width of the bins, as can be seen from the insert where much finer
  sampling bins where used and 
  smoother 
  curves are obtained adjacent to
  the wall. The insert also demonstrates the limited amount of
  particle statistics for locations $y^+>20$. 
  The overall trend of more isotropic particle velocity fluctuations
  as compared to the fluid phase can be explained by the action of
  particle collisions which tend to re-distribute the kinetic energy
  among the components. 
  Furthermore, as already noted by \cite{kaftori:1995b}, 
  the action of gravity in the negative wall-normal direction leads to
  an additional enhancement of the fluctuation intensity of the
  wall-normal particle velocity component. 
  On the other hand, the profiles in
  figure~\ref{fig:rms_velocity_profile}$(b)$ (particularly the insert)
  demonstrate the effect of the finite size of the particles: 
  the damping of the particle velocity fluctuations very close to the
  wall is felt at larger distances than what is experienced by the
  fluid, and the wall-normal component has a nearly vanishing
  fluctuation intensity at center locations around $y^+=5$. 

  Let us turn to the covariance between streamwise and wall-normal
  velocity fluctuations. Figure~\ref{fig:reynold_stress_profile}$(a)$
  shows that the fluid and the particle phase exhibit similar values
  in the near-wall region $y/h\leq0.2$. 
  Upon closer inspection (figure~\ref{fig:reynold_stress_profile}$b$
  and insert) it can be seen that the particle `Reynolds stress' tends
  towards zero at the wall faster than the fluid counterpart, and a
  cross-over is observed at $y^+\approx7$, which again manifests the 
  finite-size of the particles. 
  However, we detect a clear local maximum of 
  $\langle u'_{\mathrm{p}} v'_{\mathrm{p}} \rangle$ 
  at a wall-distance of $y^+\approx10$. The insert shows that the
  particle `Reynolds stress' first decreases significantly for
  $y^+>10$, then rises again for $y^+>15$. 
  Although the fluctuation intensity of the two components 
  ($\langle u_{\mathrm{p}}^{\prime}u_{\mathrm{p}}^{\prime}\rangle$ and 
  $\langle v_{\mathrm{p}}^{\prime}v_{\mathrm{p}}^{\prime}\rangle$) does not change much 
  in the interval $10\leq y^+\leq 20$, their covariance does exhibit the
  observed `kink'. This means that the correlation between the two
  components is somehow disturbed in that region. 
  Therefore, we have more carefully investigated the particle paths in
  the corresponding region. 
  It turns out that particle-particle contact is responsible for the 
  observed behavior. 
  Figure~\ref{fig:collision_count_profile} shows the number of
  detected collisions as a function of wall-distance. In computing the
  counter, only those collisions were considered which take place in
  a predominantly vertical direction (i.e.\ where the line connecting
  the centers of the two colliding particles is closer to the
  vertical than to any horizontal axis). 
  By this choice we eliminate those collisions which occur
  between particles in motion predominantly in a wall-parallel plane. 
  As can be seen, these collisions are concentrated in the same
  interval of wall-normal distances as the reduction of particle
  `Reynolds stress' observed in
  figure~\ref{fig:reynold_stress_profile}$(b)$. 
  We have analyzed the trajectories of all particles colliding in the
  mentioned region (figure omitted), and found that these collisions
  mostly correspond to either:
  (i) particles being picked up from the wall by high-speed fluid,
  then colliding at oblique angles with those particles being located
  on their downstream side; 
  (ii) particles approaching the near-wall region from the outer flow,
  then colliding with near-wall particles. 
  In the former case, an ejection (i.e.\ in the second quadrant of the
  $u_{\mathrm{p}}^\prime,v_{\mathrm{p}}^\prime$-plane) will be reduced in
  intensity or converted into a third quadrant event. 
  In the latter case, the collision will tend to convert what is most
  probably a `sweep' event (fourth quadrant) into a first quadrant event. 
  In both cases, the consequence will be a reduction in amplitude of
  the (overall negative) correlation between $u_{\mathrm{p}}^\prime$
  and $v_{\mathrm{p}}^\prime$.  
%
\begin{figure}
   \centering
        \begin{minipage}{2ex}
        \rotatebox{90}{\small $\quad y/h$}
        \end{minipage}
        \begin{minipage}{.5\linewidth}
        \includegraphics[width=\linewidth]
        {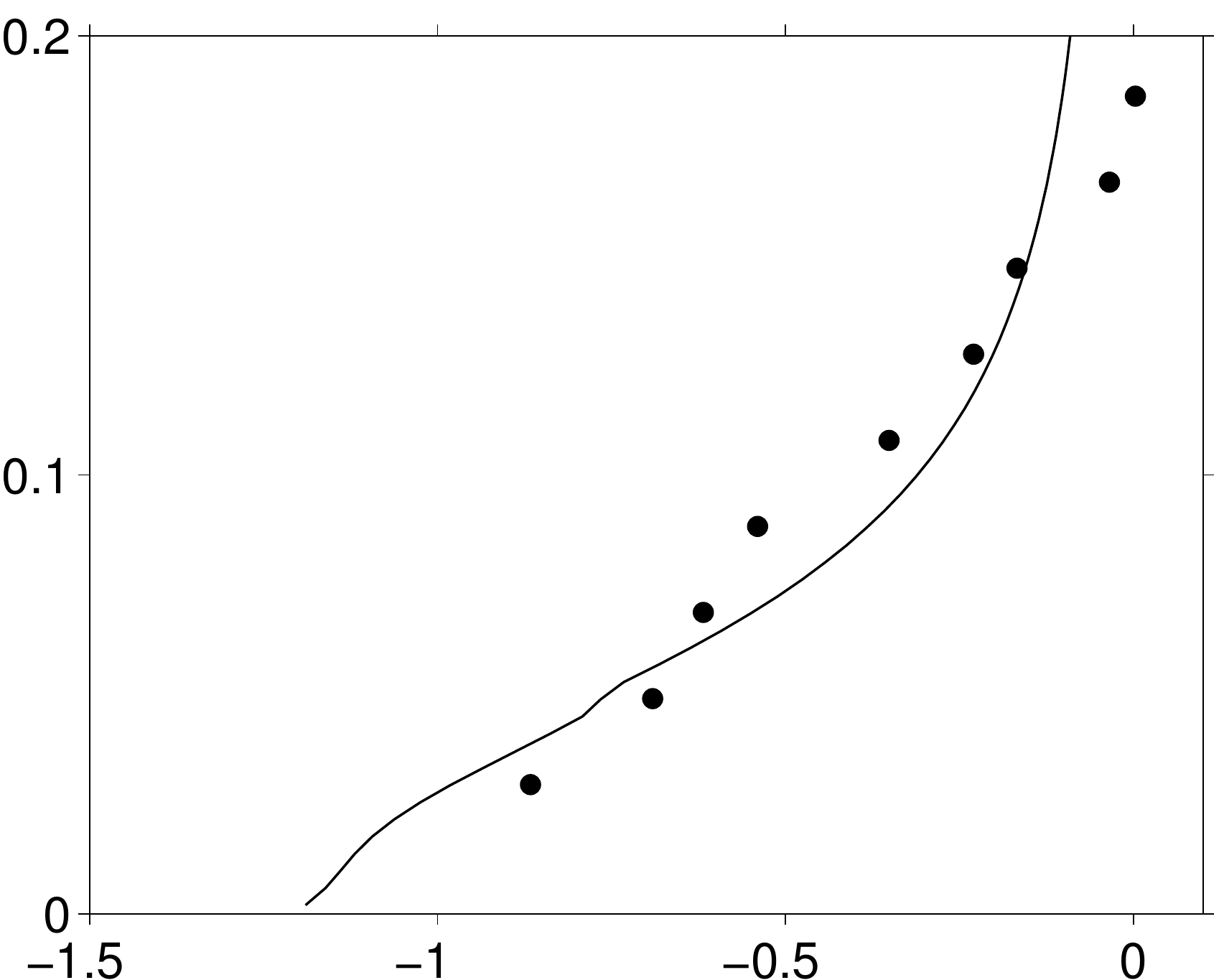}
        \\
        \centerline{\small 
                   [$-A\,\rmd\langle u_\mathrm{f} \rangle/\rmd y$;
                   $\langle\omega_\mathrm{p,z}\rangle$]$R/u_\tau$}
        \end{minipage}      
         \vspace{2ex}
        \caption{Wall-normal profiles of the mean fluid shear  
                 (solid line) and mean
                 value of the spanwise component of
                 particle  angular velocity 
                 (symbols).
                 %
                 The proportionality factor 
                 is given as 
                 $A = 0.33.$}    
        \label{fig:mean-shear-mean-rotation}
\end{figure}
\subsubsection{Mean shear of the fluid phase and mean angular velocity
               of the particle phase} 
\label{subsubsec-mean-shear-fluid-mean-shear-particle}
Figure \ref{fig:mean-shear-mean-rotation} shows the profiles of the
mean fluid shear, $\rmd\langle u_{\mathrm{f}}\rangle/\rmd y$, and the mean angular 
particle velocity around the spanwise axis, 
$\langle \omega_{\mathrm{p,z}} \rangle$. 
%
The figure suggests
that the
average particle rotation is driven by the mean shear, as the data points
are approximately proportional to the mean fluid velocity gradient. 
We observe that 
$\langle \omega_{\mathrm{p,z}} \rangle \approx -A \rmd \langle u_{\mathrm{f}}\rangle/\rmd y$
with a proportionality factor of $A\approx 0.33$ (visual fit).
An analogous result has already been observed in particulate
flow through a vertical channel
\citep{uhlmann:2008,garcia-villalba:2012}
where 
a proportionality factor of $A\approx 0.15$ was
    obtained.
%
Note that the sign of the angular particle velocity corresponds to
forward ``rolling'' motion. Incidentally, the 
angular velocity 
of particles adjacent to the wall is significantly 
smaller 
than
would be required by a condition of traction on the wall
for which 
$-\langle \omega_{\mathrm{p,z}} \rangle R/\langle u_{\mathrm{p}} \rangle = 1$. 
In the present simulation, the observed value
of $-\langle \omega_{\mathrm{p,z}} \rangle R/\langle u_{\mathrm{p}}\rangle $ 
in the averaging bin adjacent to the wall 
is equal to $0.3$.
Please note that there is no need for particles to 
actually
obey a traction
condition in the present setup 
even if they remain in contact with the wall, 
since tangential forces were 
neglected in our 
contact model.
%
\begin{figure}
   \centering
   \begin{minipage}{2ex}
     $(a)$\\[4ex]
          \rotatebox{90}
          {\small \hspace{3ex} $z/h$}
        \end{minipage}
        \begin{minipage}{.8\linewidth}
          \includegraphics[width=\linewidth,clip=true,
          viewport=250 650 2180 1180]
          {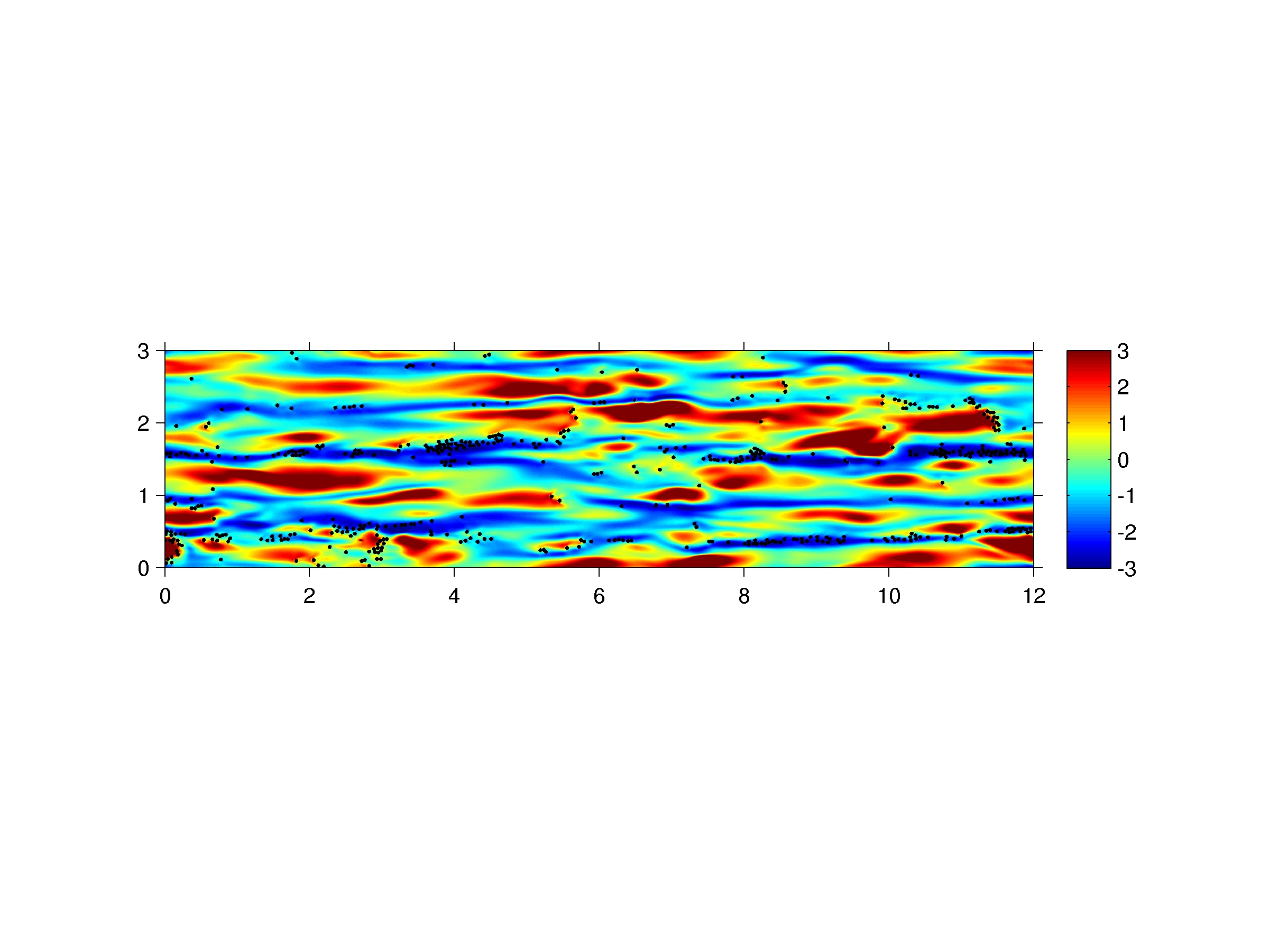}
        \end{minipage}
         \begin{minipage}{2ex}
          \rotatebox{90}{\raisebox{2ex}
            {\small $(u_\mathrm{f}  - \langle u_\mathrm{f} \rangle)/u_\tau$}}
         \end{minipage}\\
        \begin{minipage}{2ex}
          $(b)$\\[4ex]
          \rotatebox{90}
          {\small \hspace{3ex} $z/h$}
        \end{minipage}
        \begin{minipage}{.8\linewidth}
          \includegraphics[width=\linewidth,clip=true,
          viewport=250 650 2180 1180]
          {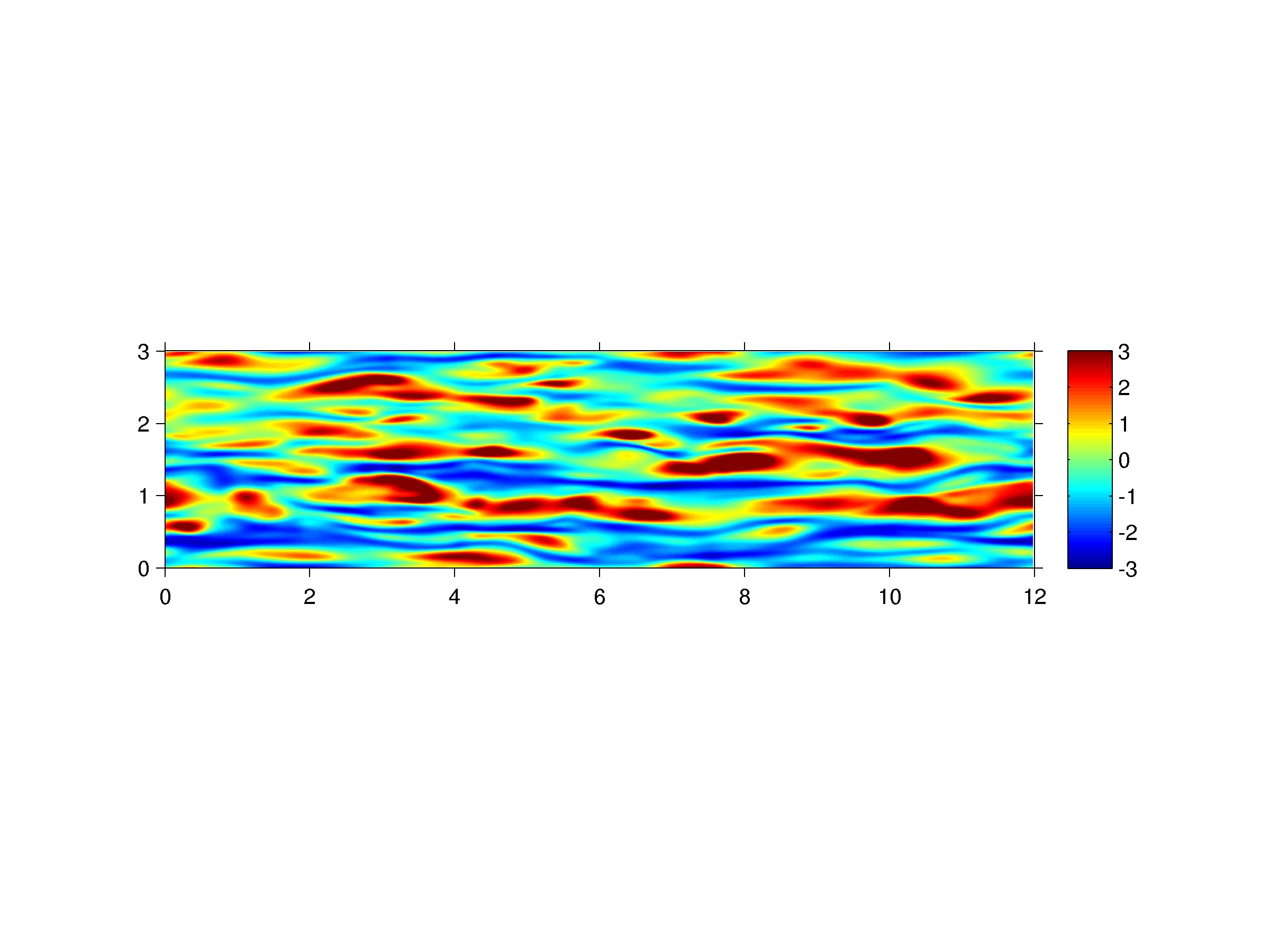}
        \end{minipage}
         \begin{minipage}{2ex}
          \rotatebox{90}{\raisebox{2ex}
            {\small$(u_\mathrm{f}  - \langle u_\mathrm{f} \rangle)/u_\tau$}}
          \end{minipage}\\
        \begin{minipage}{2ex}
          $(c)$\\[4ex]
          \rotatebox{90}
          {\small \hspace{3ex} $z/h$}
        \end{minipage}
        \begin{minipage}{.8\linewidth}
          \includegraphics[width=\linewidth,clip=true,
          viewport=250 650 2180 1180]
          {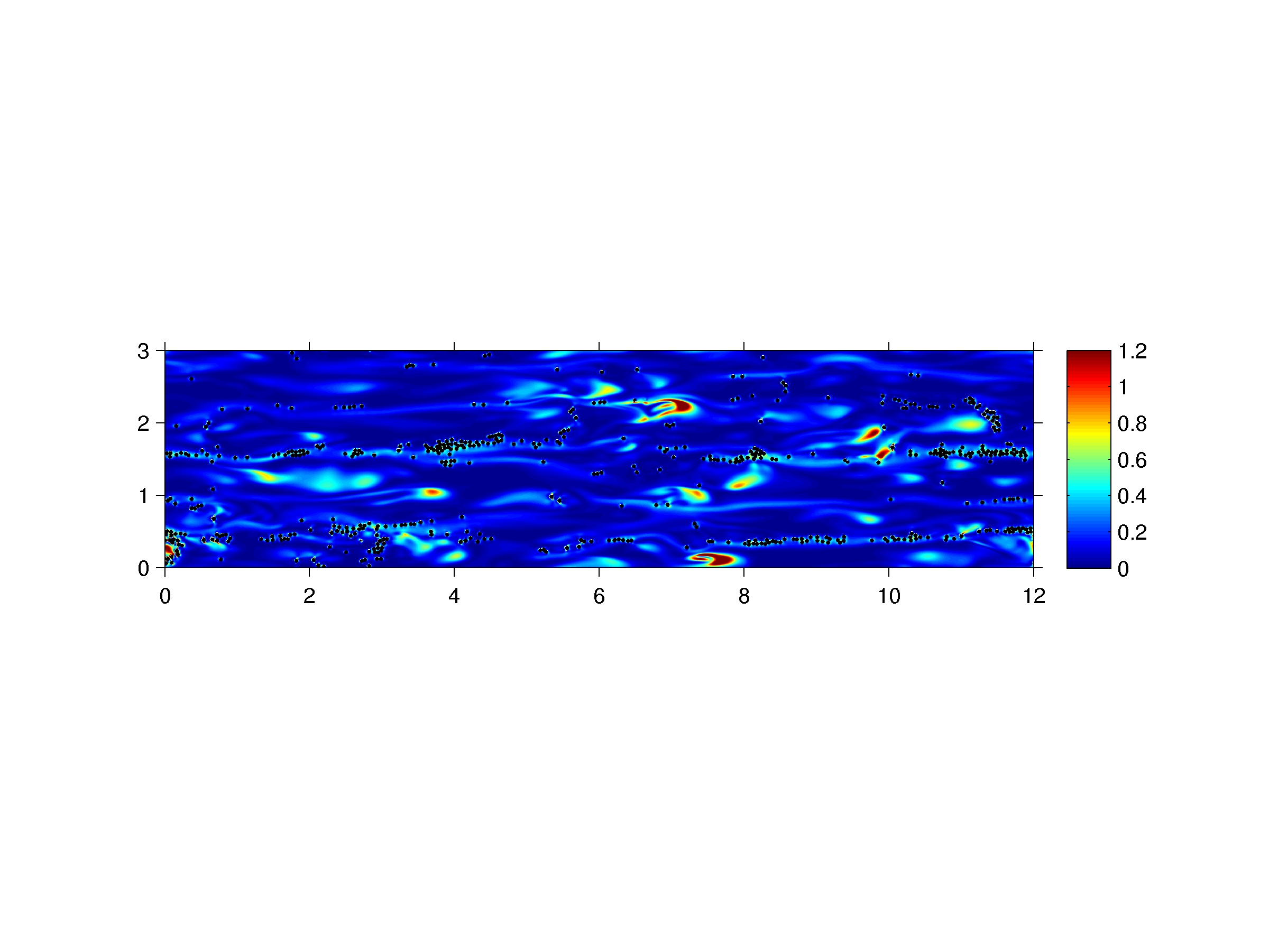}
        \end{minipage}
         \begin{minipage}{2ex}
          \rotatebox{90}{\raisebox{2ex}
            {\small$\varepsilon^+$}}
         \end{minipage}\\
        \begin{minipage}{2ex}
          $(d)$\\[4ex]
          \rotatebox{90}
          {\small \hspace{3ex} $z/h$}
        \end{minipage}
        \begin{minipage}{.8\linewidth}
          \includegraphics[width=\linewidth,clip=true,
          viewport=250 650 2180 1180]
          {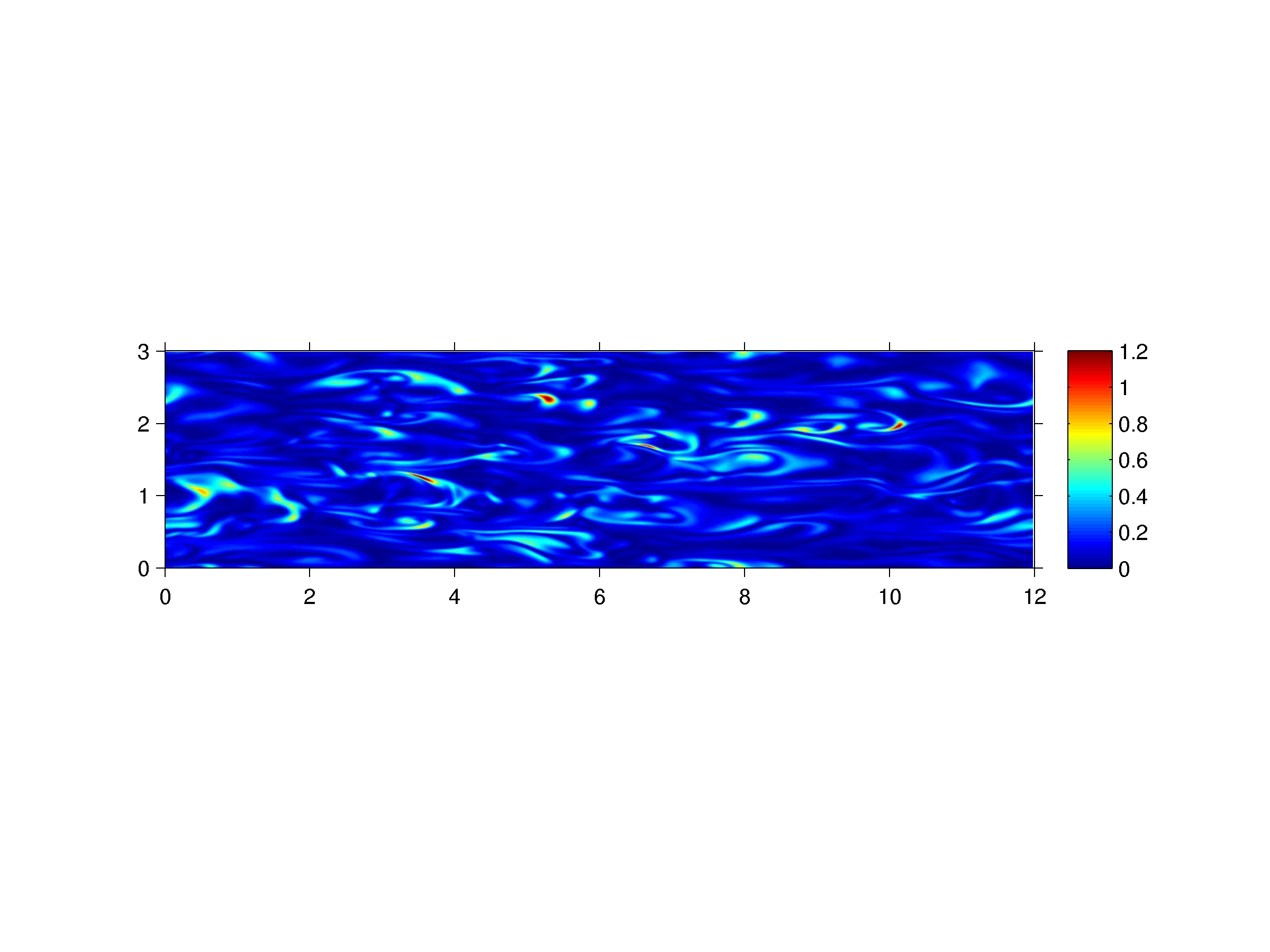}
          \\
         \centerline{\small $x/h$}
        \end{minipage}
         \begin{minipage}{2ex}
          \rotatebox{90}{\raisebox{2ex}
            {\small$\varepsilon^+$}}
         \end{minipage}
        \caption{
          \revision{%
            Instantaneous streamwise velocity field in a wall-parallel
            plane located at $y^+=5$:
            ($a$) in the present case and 
            ($b$) in single phase flow. 
            Intersections with particles in graph ($a$) are shown in
            black.}{%
            Visualization of instantaneous flow fields in a wall-parallel
            plane located at $y^+=5$. 
            Graphs $(a)$ and $(b)$ show fluctuations of the streamwise
            fluid velocity component, 
            $(c)$ and $(d)$ show the instantaneous dissipation rate of
            fluctuating kinetic energy, $\varepsilon=
            2\nu s_{ij}^\prime s_{ij}^\prime$, where 
            $s_{ij}^\prime=(u_{i,j}^\prime+u_{j,i}^\prime)/2$. 
            Graphs $(a)$, $(c)$ refer to the present particulate flow
            case (intersections with particles are shown in black),  
            and  $(b)$, $(d)$ to single phase flow. 
            Note that one matching instant in time is shown in $(a)$,
            $(c)$ and similarly in $(b)$, $(d)$. 
            Moreover, the instant when $(a)$,
            $(c)$ are taken corresponds to the three-dimensional
            visualization shown in figure~\ref{fig:3d_flow_field}. 
          }
        }
        \label{fig:uprime_xz_yplus5}
\end{figure}
\subsubsection{Instantaneous flow field and two-point correlations} 
\label{subsubsec-two-point-correlations-fluid-velocity-field}
%
Figure \ref{fig:uprime_xz_yplus5} shows the instantaneous value
of the streamwise velocity component in a wall-parallel plane
at a wall-distance of $y^+=5$. 
The particulate flow result (figure~\ref{fig:uprime_xz_yplus5}$a$)
can be compared to the result of single-phase flow at otherwise
identical parameters provided in figure~\ref{fig:uprime_xz_yplus5}$b$.
In both cases the flow field exhibits the well-known 
streaks of high- and low- speed fluid regions. 
From visual observation the result of the two-phase flow differs little 
from that of the single-phase flow and the particles do not seem to
alter the turbulence structure much at the given wall distance.  
%
\begin{figure}
  \centering
  \begin{minipage}{2ex}
    \rotatebox{90}{\small 
      $\quad R_\mathrm{u'u'}(r_x,y^+\!=\!10,0)$
    }
  \end{minipage}
  \begin{minipage}{.45\linewidth}
    \centerline{\small $(a)$}
    \includegraphics[width=\linewidth]
    {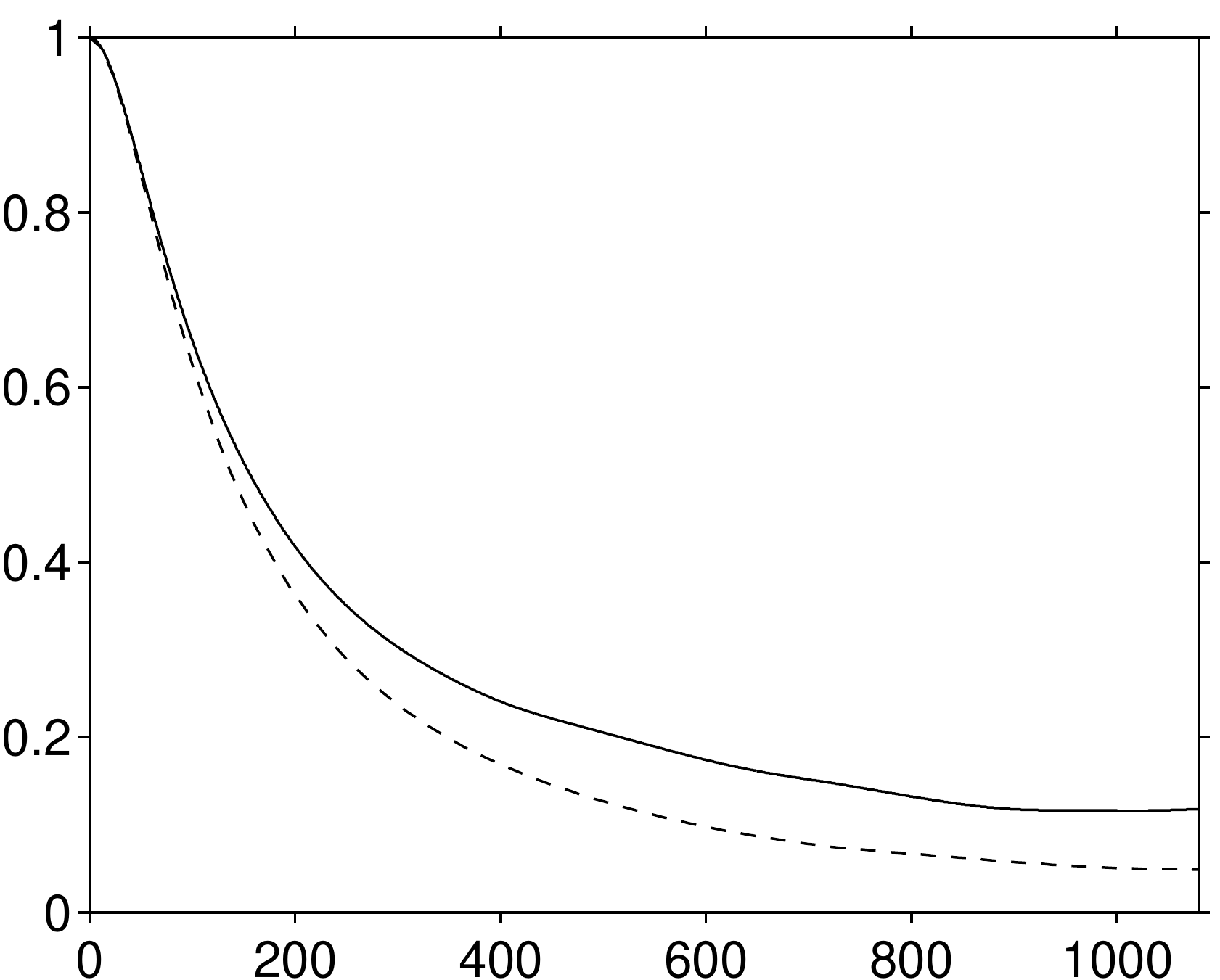}
    \\
    \centerline{\small $r_\mathrm{x}^+$}
  \end{minipage}
  \hfill
  \begin{minipage}{2ex}
    \rotatebox{90}
    {\small 
      $\quad R_\mathrm{u' u'}(0,y^+\!=\!10,r_z)$ 
    }
  \end{minipage}
  \begin{minipage}{.45\linewidth}
    \centerline{\small $(b)$}
    \includegraphics[width=\linewidth]
    {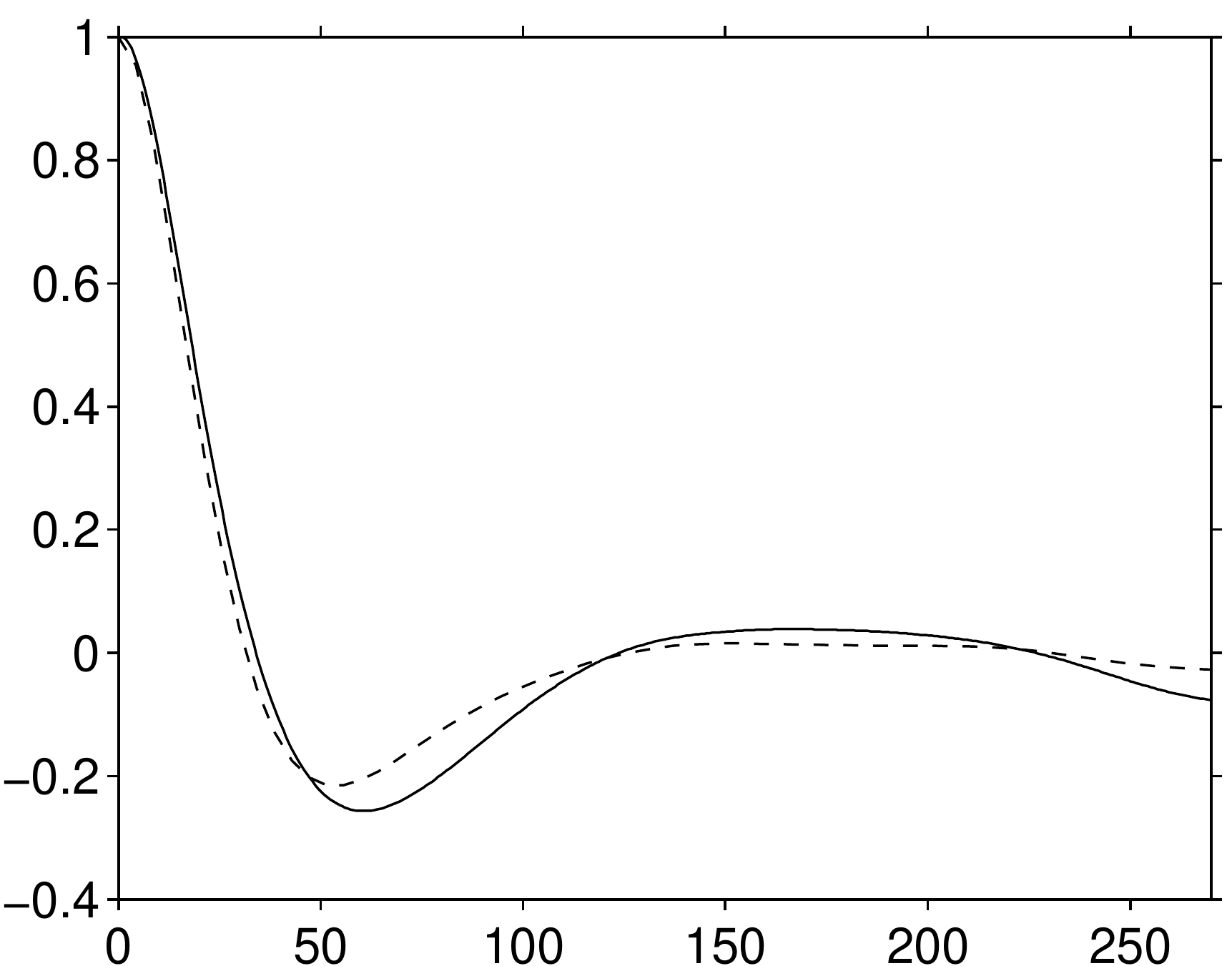}
    \\
    \centerline{\small $r_\mathrm{z}^+$}
  \end{minipage}
  \\[3ex]
  \begin{minipage}{2ex}
    \rotatebox{90}{\small 
      $E_\mathrm{u' u'}(k_x,y^+\!=\!10,0)/\langle u^\prime u^\prime\rangle$
    }
  \end{minipage}
  \begin{minipage}{.45\linewidth}
    \centerline{\small $(c)$}
    \centerline{\small $k_\mathrm{x}\,D$}
    \includegraphics[width=\linewidth]
    {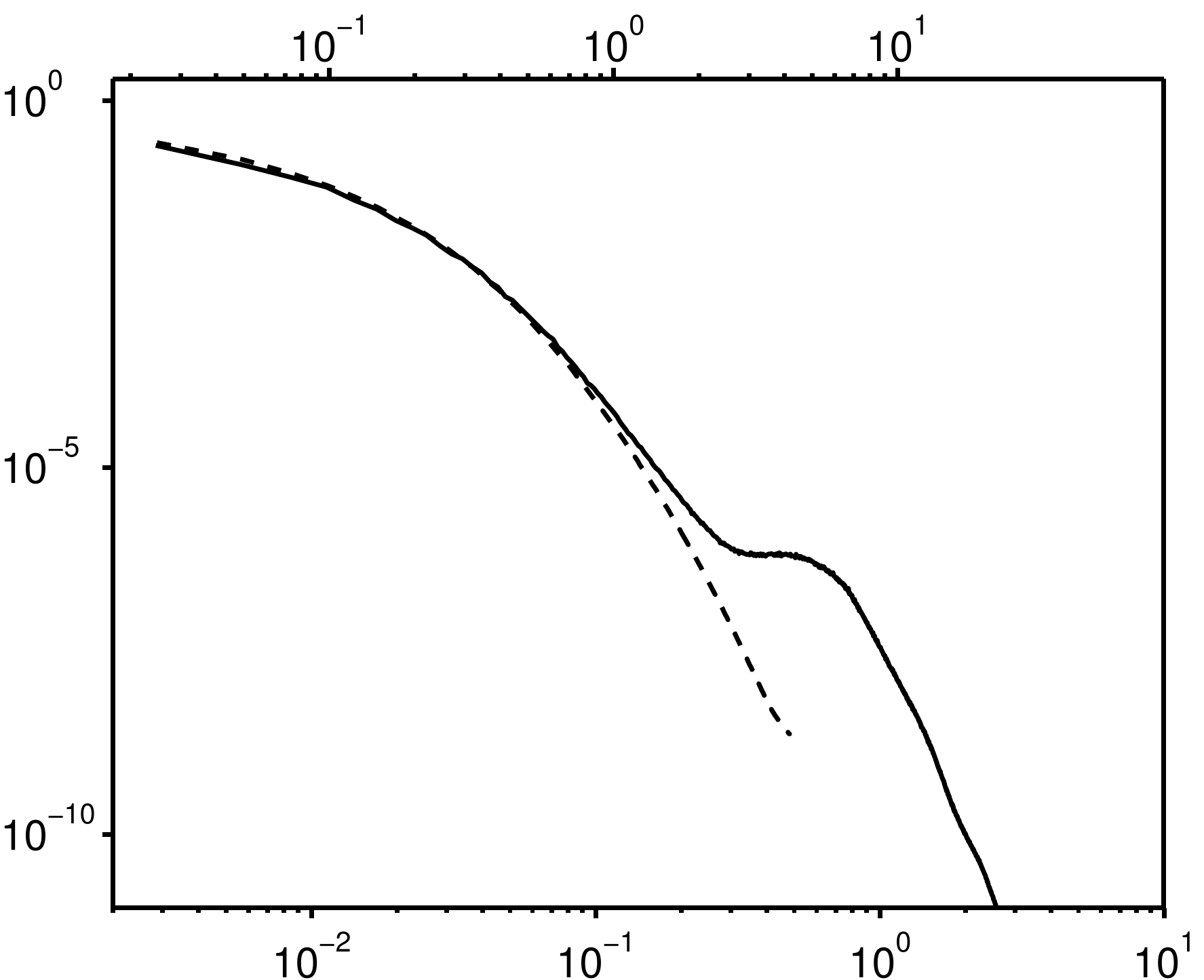}
    \\
    \centerline{\small $k_\mathrm{x}^+$}
  \end{minipage}
  \hfill
  \begin{minipage}{2ex}
    \rotatebox{90}
    {\small 
      $E_\mathrm{u' u'}(0,y^+\!=\!10,k_z)/\langle u^\prime u^\prime\rangle$
    }
  \end{minipage}
  \begin{minipage}{.45\linewidth}
    \centerline{\small $(d)$}
    \centerline{\small $k_\mathrm{z}\,D$}
    \includegraphics[width=\linewidth]
    {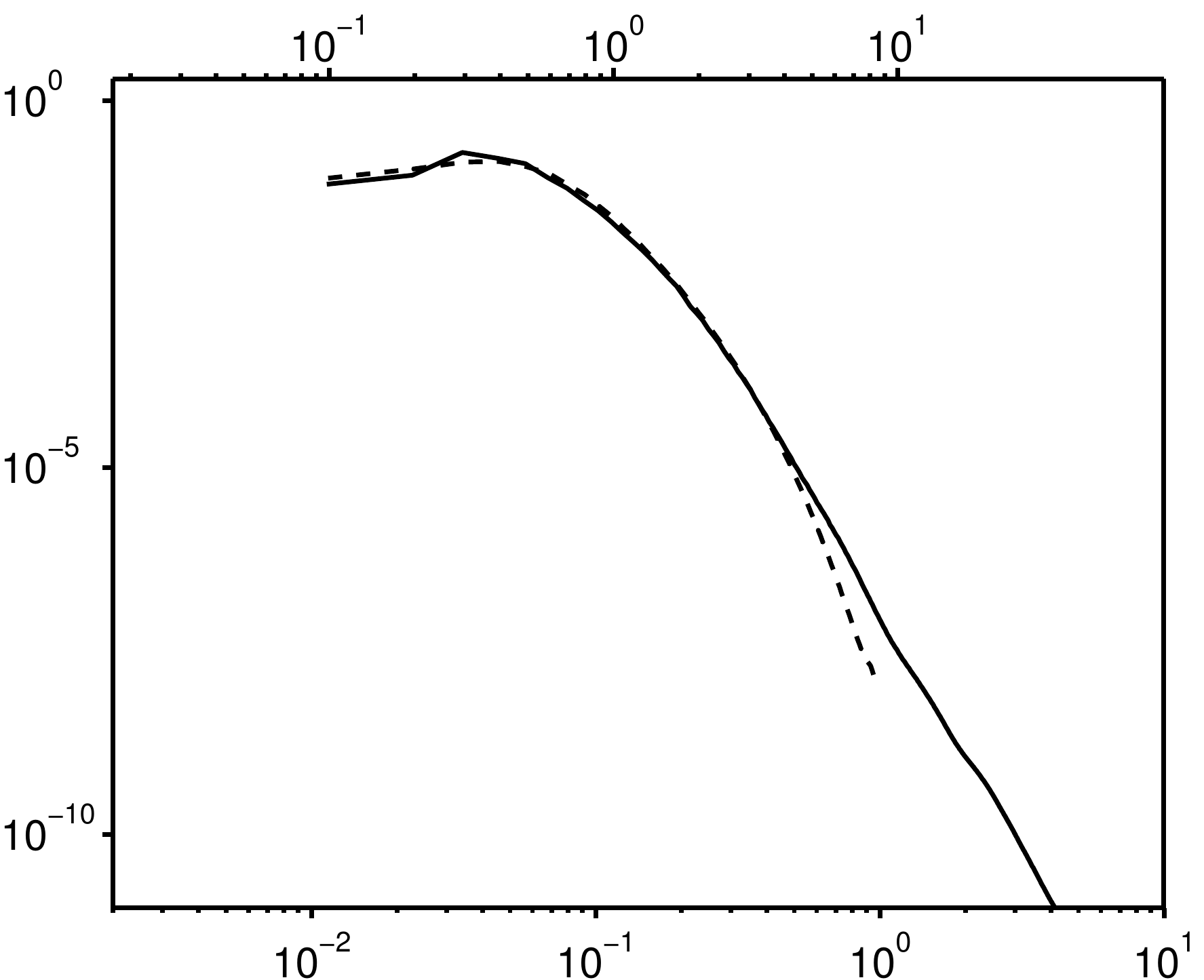}
    \\
    \centerline{\small $k_\mathrm{z}^+$}
  \end{minipage}
  %
  \caption{%
    \revision{%
      Auto-correlation coefficient of the streamwise velocity 
      component as defined in (\ref{eqn:twopoint_corr}): 
      (\textit{a}) as a function of the streamwise separation and
      (\textit{b}) as a function of the spanwise separation at a
      wall-normal distance $y^+=10$.
      The dashed line indicates the data of single-phase flow.}
    {%
     Graphs $(a)$, $(b)$ show the auto-correlation coefficient of the
     streamwise velocity component as defined in
     (\ref{eqn:twopoint_corr}), for separations in the streamwise and 
     spanwise direction, respectively. 
     The corresponding one-dimensional energy spectra are shown in
     graphs $(c)$ and $(d)$ as a function of the wavenumber $k_\alpha$
     (with directional index $\alpha=1$ and $3$). 
     Solid lines are for the present particulate flow case, dashed
     lines for corresponding single-phase flow. 
    }
  }
  \label{fig:auto_corr_x_z}
\end{figure}
\revision{}{%
  A more sensitive measure of local flow field modification is
  provided by the dissipation rate of fluctuating kinetic energy,
  $\varepsilon=2\nu s_{ij}^\prime s_{ij}^\prime$, where 
  $s_{ij}^\prime=(u_{i,j}^\prime+u_{j,i}^\prime)/2$ is the fluctuating
  rate-of-strain tensor. A map of $\varepsilon$ is provided in
  figure~\ref{fig:uprime_xz_yplus5}$c$ for the present case and in
  figure~\ref{fig:uprime_xz_yplus5}$d$ for the corresponding
  single-phase flow.  
  Note that in the particulate case, where data is generated by means
  of a finite-difference method on a staggered grid, consistent
  computation of the dissipation rate requires interpolation, thereby
  introducing a certain amount of smoothing. 
  Contrarily, the single-phase data is obtained via a pseudo-spectral
  method in a collocated grid arrangement, where dissipation can be
  directly computed point-wise. 
  Indeed, the figure shows that localized contributions to the
  dissipation rate can be observed in the vicinity of the 
  phase-interfaces. However, the largest values of dissipation are in 
  general found in regions outside the low-speed streaks, which appear
  not to be much affected by the presence of particles. 
  Also, note that such instantaneous dissipation maps exhibit a
  considerable temporal variation due to the intermittent nature of
  this quantity. Therefore, a one-on-one comparison is at best
  qualitative. 
  When plane-averaged and averaged in time, the
  contribution to dissipation stemming from the particles should
  become negligible, since we have seen above that the turbulent
  kinetic energy is practically unaffected by their presence. 
}

The possible influence of the particles can be further quantified
by analyzing 
the two-point auto-correlations of the
streamwise velocity fluctuations defined as
\begin{equation}
\!\!\!\!\!\!\!\!\!\!\!\!\!\!\!\!\!\!\!\!\!\!\!
R_{\mathrm{u}^\prime\,\mathrm{u}^\prime}(r_\mathrm{x},y,r_\mathrm{z})=
       \frac{1}
            {L_\mathrm{x} L_\mathrm{z}\langle u' u'\rangle(y)}
       \int_0^{L_\mathrm{x}}\int_0^{L_\mathrm{z}} u^\prime(x,y,z)
          u^\prime(x+r_\mathrm{x},y,z+r_\mathrm{z})\,\mathrm{d}x\mathrm{d}z \,
\label{eqn:twopoint_corr}
\end{equation}
where $r_\mathrm{x}$ and $r_\mathrm{z}$ are streamwise and spanwise
separations, 
respectively, 
and $u^\prime$ refers to a velocity fluctuation with respect to the
average over wall-parallel planes and time. 
Note, that in (\ref{eqn:twopoint_corr}) it is not
distinguished between regions of fluid or solid phase. 
%
Figure \ref{fig:auto_corr_x_z}$a$ shows the auto-correlation
coefficient of the streamwise velocity component as a function of
streamwise separations at a wall distance of $y^+= 10$ at zero 
spanwise separation.
In addition the result from the single-phase reference case is 
included in the figure.
At increasing separations $r_\mathrm{x}$ the correlation
coefficient in the present case is larger than in the single-phase
flow and exhibits values of  $0.12$ 
at the largest streamwise distance compared to a value of
$0.04$ in the single-phase case.
Figure~\ref{fig:auto_corr_x_z}$b$ shows the correlation coefficient
as a function of spanwise separation, $r_\mathrm{z}$ for zero 
streamwise separation.
The magnitude and location of the minima differs,
i.e.\ the minimum is located at 
$r_\mathrm{z}^+= 60$ and has a value of $-0.26$ in the present case
while in the single-phase flow the minimum is located 
at $r_\mathrm{z}^+=55$ and has a value of $-0.19$. 
In contrast to the single-phase flow the auto-correlation in the
present case reaches
positive values for $120<r_\mathrm{z}^+< 230$, leading to
a local maximum at $r_\mathrm{z}^+= 166$ ($R_{\mathrm{u}^\prime\,\mathrm{u}^\prime}=0.04$).

\revision{}{%
  Since small-scale contributions are not easily observed from the
  spatial correlation functions, we present the corresponding
  one-dimensional energy spectra in
  figure~\ref{fig:auto_corr_x_z}$(c,d)$.  
  It can be seen that particles affect the distribution of kinetic
  energy among the scales only at scales around and below their
  diameter $D$. The effect is more pronounced in the streamwise
  direction, where a visible `bump' is observed. 
} 

In conclusion, a small but distinct effect of the particles on the 
turbulence structure is found in the auto-correlation of the flow
field close to the wall. 
The increase in the magnitude of the correlation coefficient 
of the streamwise velocity 
fluctuations for separation in the streamwise as well as in the spanwise
direction 
%
can be explained by a stabilising effect of the particles
on the near wall structures leading to somewhat more elongated flow
structures.

\begin{figure}
    \centering 
    \includegraphics[width=.5\textwidth]
    {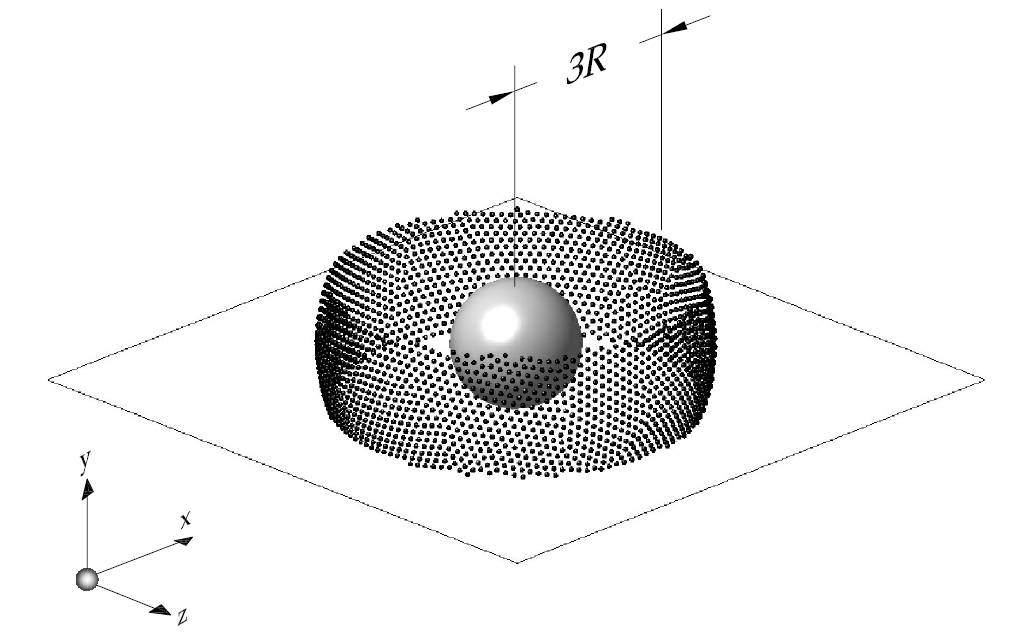}
    \caption{
      The spherical segment surface ${\cal S}$ over which 
             fluid velocity 
             is averaged 
             in order 
             to define a representative instantaneous velocity 
             at the particle location. The spherical segment has a radius 
             $R_\mathrm{s} = 3R$. 
             The parallel planes which define the segment are
             intersecting the sphere with radius $R_s$ at a distance
             equal to $R$ on either side of the center, both being
             parallel to the wall-plane.
             }    
    \label{fig:particle_shell_sketch}
\end{figure}
\subsection{Definition of the fluid velocity seen by finite-size
  particles} 
\label{subsec-fluid-seen-by-particles}
  The concept of a fluid velocity ``seen'' by suspended particles is
  extensively used in the context of particulate flow systems. 
  It represents a simplification in the sense that one supposes
  particles to undergo a certain forcing due to an ``incoming'' flow,
  similar to a fixed immersed object in a cross-stream, while ignoring
  the modification to the flow due to the particle itself. 
  When considering point particles, particularly in the case of
  one-way coupling, the characteristic fluid
  velocity is simply taken as the fluid velocity evaluated at the particle's
  location.\footnote{%
    In the case of two-way coupling, strictly speaking one would need
    to consider the velocity at the particle location in a companion
    simulation including all particles except the one under
    consideration.  
  }
  Finite-size particles however, 
  modify the carrier flow around them by constraining the fluid
  velocity at the fluid/solid interface. 
  Consequently, there is no unique definition of a representative
  fluid velocity seen by finite-size particles. 

\begin{figure}
   \centering
        \begin{minipage}{2ex}
        \rotatebox{90}{\small$\tilde{x}/R$}
        \end{minipage}
        \begin{minipage}{.25\linewidth}
        \vspace{10ex}
        \includegraphics[width=\linewidth]
        {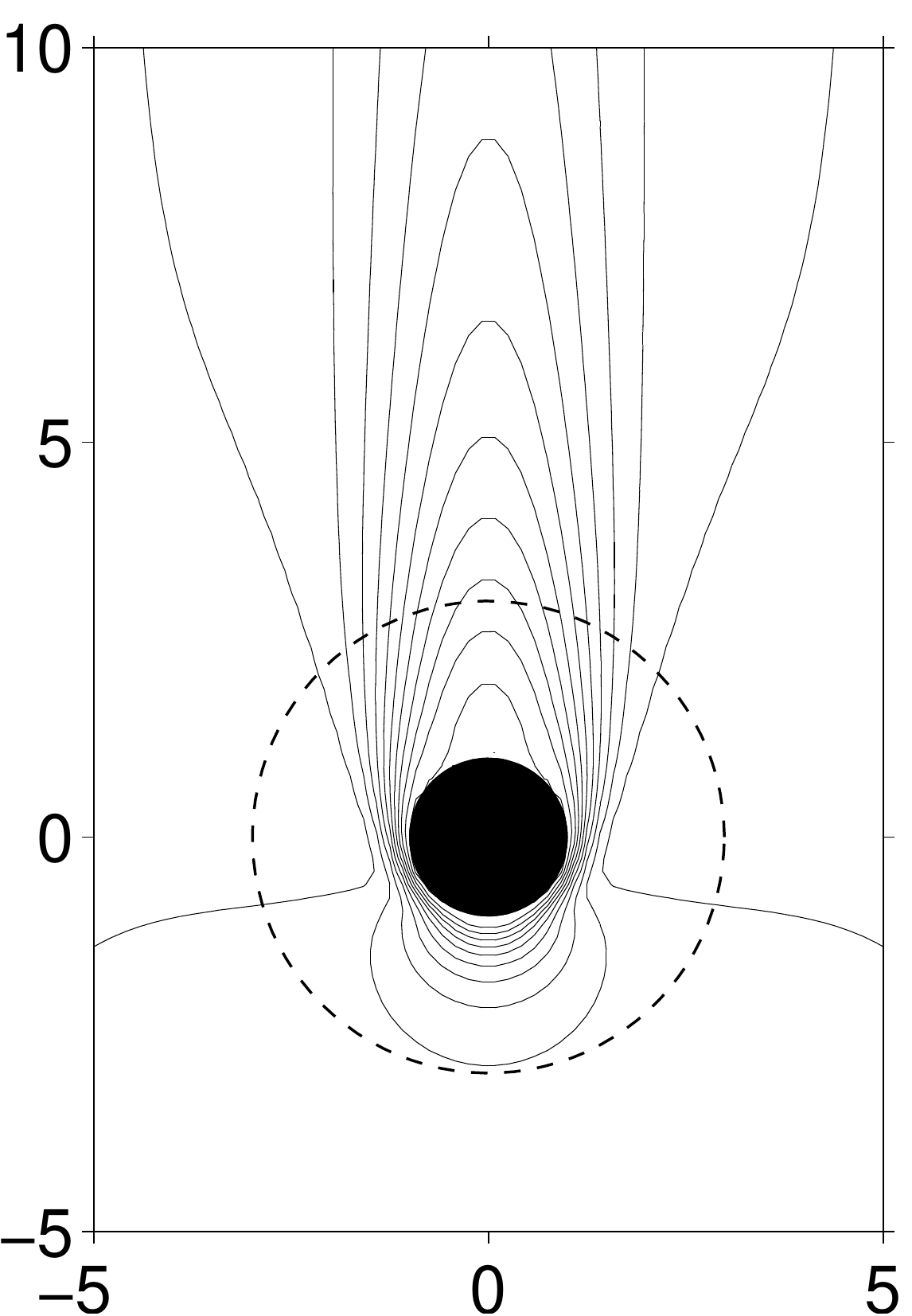}
        \hspace{-0.6\linewidth}\raisebox{1.5\linewidth}
        {\small $(a)$}
        \\
        \centerline{\hspace{-2ex}\small $\tilde{z}/R$}
        \end{minipage}
        %
        \begin{minipage}{2ex}
        \rotatebox{90}{\small$\tilde{x}/R$}
        \end{minipage}
        \begin{minipage}{.25\linewidth}
        \vspace{10ex}
        \includegraphics[width=\linewidth]
        {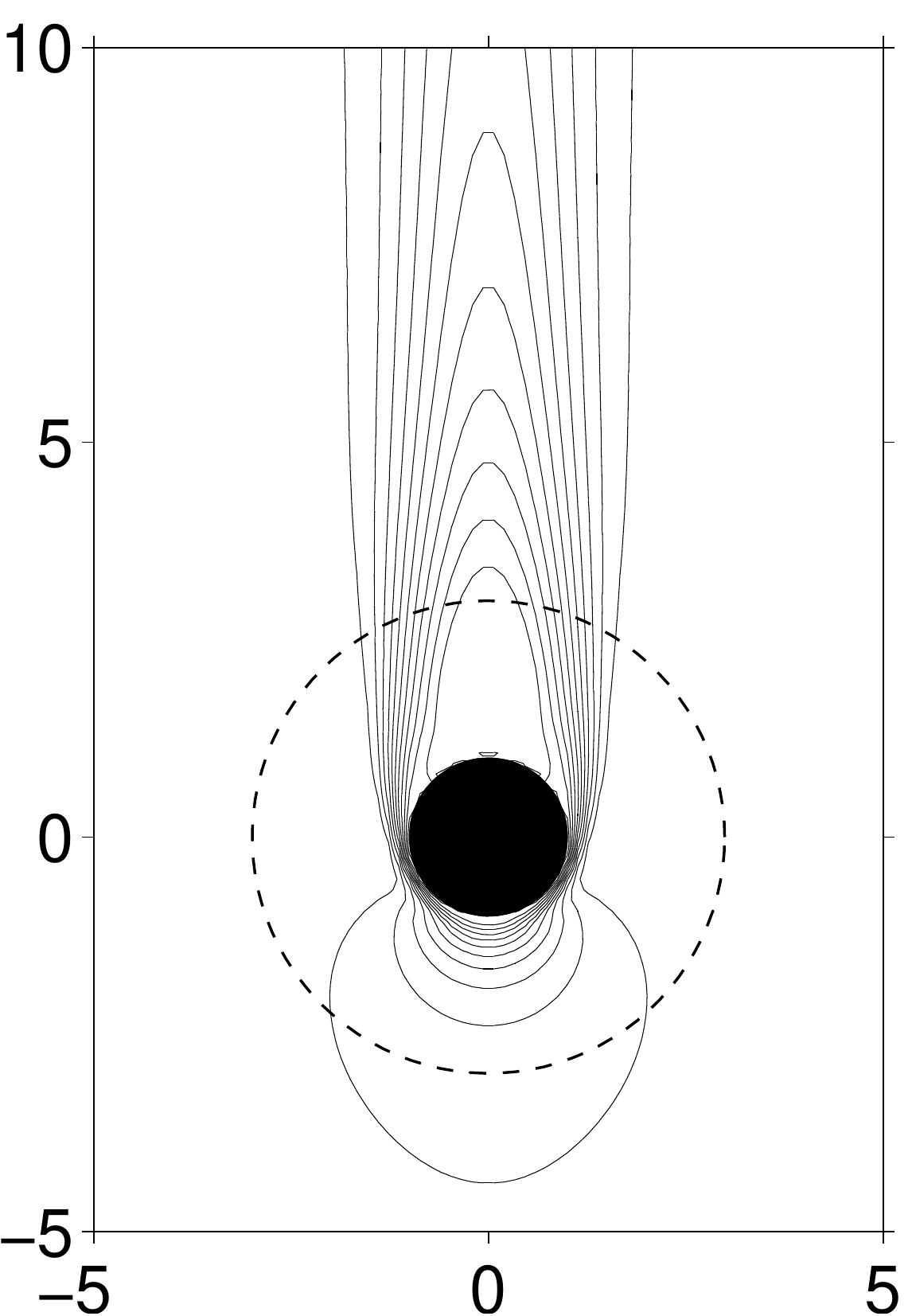}
        \hspace{-0.6\linewidth}\raisebox{1.5\linewidth}
        {\small $(b)$}
        \\
        \centerline{\hspace{-2ex}\small $\tilde{z}/R$}
        \end{minipage}
        %
        \begin{minipage}{2ex}
        \rotatebox{90}{\small$\tilde{x}/R$}
        \end{minipage}
       \begin{minipage}{.25\linewidth}
        \vspace{10ex}
        \includegraphics[width=\linewidth]
        {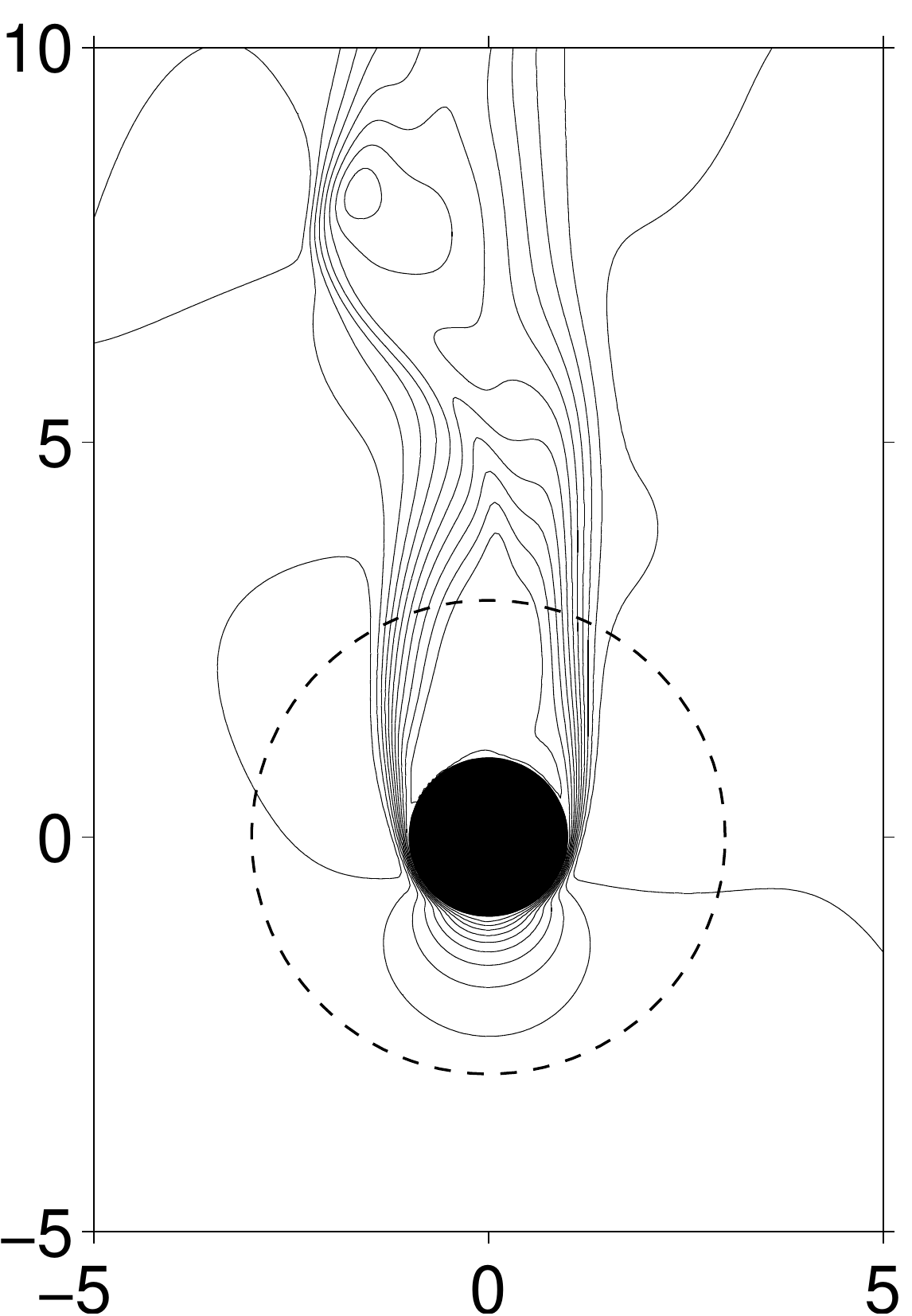}
        \hspace{-0.6\linewidth}\raisebox{1.5\linewidth}
        {\small $(c)$}
        \\
        \centerline{\hspace{-2ex}\small $\tilde{z}/R$}
        \end{minipage}
        \\
        \begin{minipage}{2ex}
        \rotatebox{90}
        {\small $ {u}^{\cal S}_\mathrm{f}/
                 {u}_{\mathrm{f},\infty}$ }
        \end{minipage}
        \begin{minipage}{.4\linewidth}
        \includegraphics[width=\linewidth]
        {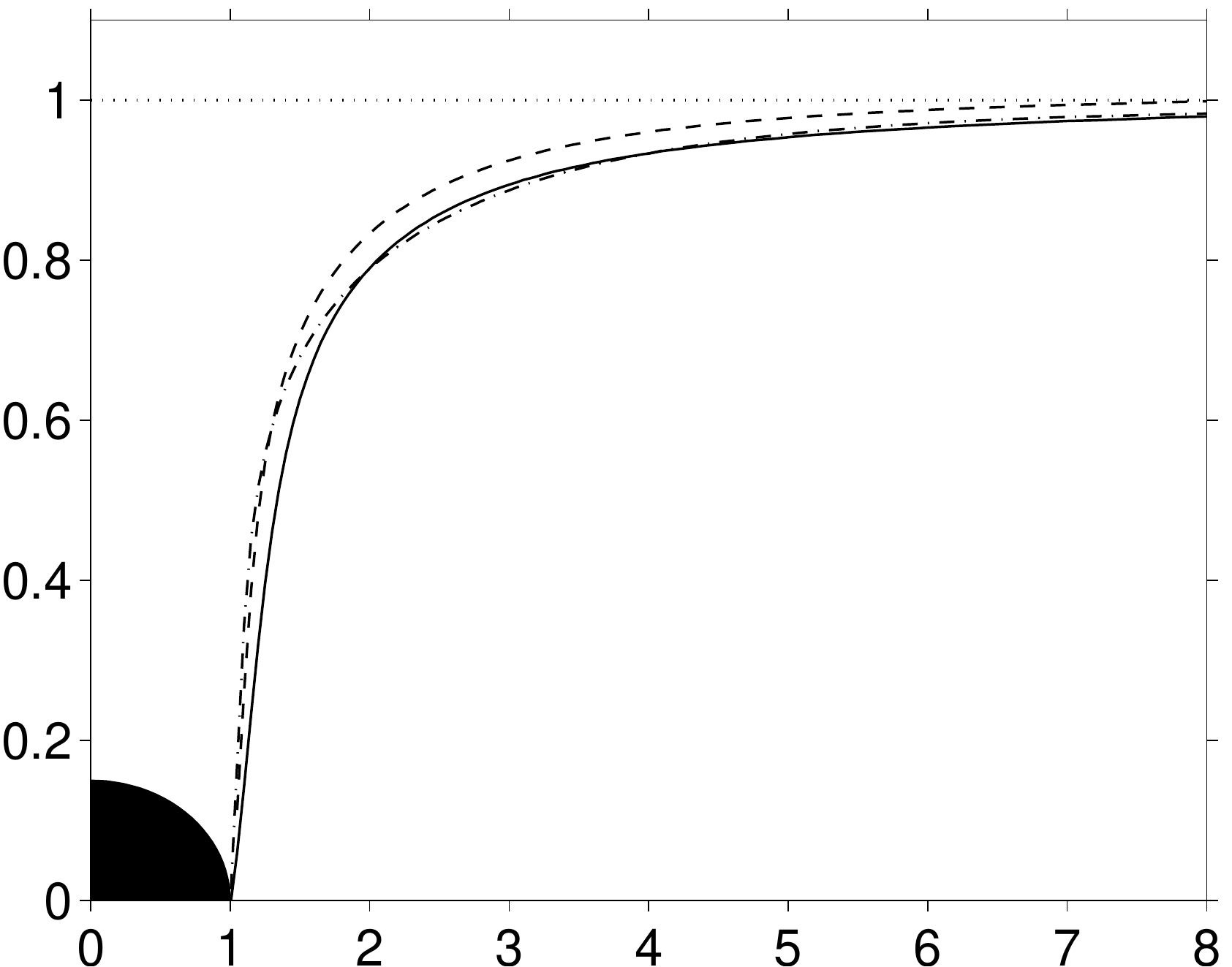}
        \hspace{-0.5\linewidth}\raisebox{.8\linewidth}
        {\small $(d)$}
        \\[1ex]
        \centerline{\small $  R_\mathrm{s}/R $}
        \end{minipage}
        \caption{(\textit{a})-(\textit{c})
                 Contour plots
                 of instantaneous streamwise velocity 
                 fields of an unbounded flow past
                 a fixed sphere in the steady axisymmetric
                 and the unsteady wake regimes.
                 (\textit{a}) $Re_\infty = 50$, (\textit{b}) $Re_\infty = 140$, 
                 and (\textit{c}) $Re_\infty = 350$.
                 Contourlines correspond to 
                 (0:0.1:1) 
                 times the magnitude of ${u}_{\mathrm{f},\infty}$.
                 The 
                 circle indicated by a dashed line
                 %
                 has a radius of $3R$.
                (\textit{d}) 
                The computed value ${u}^{\cal S}_\mathrm{f}$ of the 
                axial 
                component of the instantaneous fluid velocity
                ``seen by the particle'', as given by the definition 
                (\ref{eq:mean_local_velocity}), plotted as a function
                of $R_\mathrm{s}$. 
                (\solid) $Re_\infty = 50$;
                (\dashed) $Re_\infty = 140$;
                (\chndot) $Re_\infty = 350$.
                }
        \label{fig:slip_pure_sedimentation}
\end{figure}
  Previously, there have been several attempts 
  to define the undisturbed fluid velocity seen by 
  finite-size particles. \citet{bagchi:2003} 
  simulated isotropic turbulence swept past a fixed sphere and 
  \citet{zeng:2008} considered non-periodic 
  channel flow with a fixed sphere.
  In both of these studies, the undisturbed fluid velocity at
  a particle location was estimated from the fluid velocity at the
  same location of a separately computed
  turbulent flow without the sphere. 
  \citet{merle:2005}, using Taylor hypothesis, 
  approximated the undisturbed 
  fluid velocity at a certain downstream location 
  ($2.5D$ from the center of the particle) 
  and at an earlier time. Note that all of these three studies
  considered fixed spheres. The authors remark that these approaches
  are only adequate for estimating the velocity seen by particles
  with sizes smaller than the flow scales and they question
  the applicability in the case of finite-size particles.
  \citet{lucci:2010} considered mobile finite-size particles in
  decaying isotropic turbulence in the absence of gravity. 
  They defined the characteristic fluid velocity as an average
  over a small spherical cap with center located in the direction
  given by the particle velocity (measured in an inertial reference
  frame). 
  In turbulent flows, however, 
  this choice of directional bias appears questionable since it is the
  relative (not the absolute) velocity which distinguishes the
  direction of the ``incoming'' fluid flow. 

  Here, we propose a definition for the characteristic fluid velocity in the
  vicinity of the particles which avoids the directional
  assumption. 
  The instantaneous fluid velocity $\mathbf{u}^{\cal S}_\mathrm{f}$ in the
  vicinity of the $i$th particle is approximated by the
  average of the velocity of the fluid located on a  
  spherical surface ${\cal S}$ of radius $R_s$ centered at the particle's
  center location $\mathbf{x}^{(i)}_\mathrm{p}$. 
  In order to avoid sampling bias due to the inhomogeneity in
  wall-normal direction  
  in the considered channel flow, the averaging spherical surface is 
  trimmed by two wall-parallel planes at one particle radius $R$ below and 
  above the center leading to the definition of a spherical surface
  segment as shown in figure~\ref{fig:particle_shell_sketch}. 
  Any point $l$ on this surface has a position vector $\mathbf{x}_l^{(i)}(t) =
  \mathbf{x}_\mathrm{p}^{(i)}(t) + R_\mathrm{s}\mathbf{n}$ where $\mathbf{n}$ is an
  outward pointing unit vector normal to ${\cal S}$ at $\mathbf{x}_l^{(i)}(t)$. 
  With the surface ${\cal S}$ discretized by 
  $N_\mathrm{l}$ 
  Lagrangian marker 
  points, 
  the average velocity of the fluid 
  instantaneously located on the surface ${\cal S}$ 
  is defined as 
  (using 
  the indicator 
  function $\phi_{\mathrm{f}}(\mathbf{x},t)$ defined in 
  \ref{equ-def-fluid-indicator-fct}):
  \begin{equation}
    \mathbf{u}_\mathrm{f}^{\cal S}(\mathbf{x}_\mathrm{p}^{(i)},R_\mathrm{s},t) =
    \frac{1}{n_{\cal S}^{(i)}(t)}\sum_{l=1}^{N_\mathrm{l}}
    \phi_{\mathrm{f}}(\mathbf{x}^{(i)}_l(t),t)
    \mathbf{u}_\mathrm{f}(\mathbf{x}^{(i)}_l(t),t)\quad
    \forall\,i=1\ldots N_\mathrm{p}\,,
    \label{eq:mean_local_velocity}
  \end{equation}
  where the counter $n_{\cal S}^{(i)}(t)$ is defined as
  \begin{equation}\label{eq:shell-sample-counter}
    n_{\cal S}^{(i)}(t) = \sum_{l=1}^{N_\mathrm{l}}\phi_{\mathrm{f}}(\mathbf{x}^{(i)}_l(t),t)\,.
  \end{equation}
  Note that the presence of $\phi_{\mathrm{f}}$ in (\ref{eq:mean_local_velocity})
  avoids 
  undesired 
  sampling of velocity data inside a neighboring
  solid particle.
  Generally $\mathbf{x}_l^{(i)}(t)$ does 
  not coincide with the fixed Cartesian grid
  where the velocity field $\mathbf{u}_\mathrm{f}(\mathbf{x},t)$ is
  available. Therefore we determine 
  $\mathbf{u}_\mathrm{f}(\mathbf{x}_l^{(i)}(t),t)$ by a trilinear interpolation 
  from $\mathbf{u}_\mathrm{f}$ values at 
  the grid 
  nodes in the fluid domain $\Omega_\mathrm{f}(t)$. 
  
  The choice of the radius $R_\mathrm{s}$ needs to meet two requirements. 
  First, it should be chosen sufficiently large in order to avoid
  a possible influence of the particle's own near field upon the computed
  value of the surrounding fluid velocity. Secondly, the value of 
  $R_\mathrm{s}$ should not be chosen too large such that the resulting
  $\mathbf{u}_\mathrm{f}^{\cal S}$ is still of direct relevance to the
  motion of the $i$th particle. 
%
\begin{figure}
   \centering
        \begin{minipage}{3ex}
        \rotatebox{90}{\small $y/h$}
        \end{minipage}
        \begin{minipage}{.5\linewidth}
        \centerline{\small $(a)$ }\vspace{2ex}
        \centerline{\small 
                      $[\langle u_\mathrm{f}\rangle;\langle u_\mathrm{p}\rangle;
                        \langle u^{\cal S}_\mathrm{f}\rangle ]^+$ }
        \includegraphics[width=\linewidth]
        {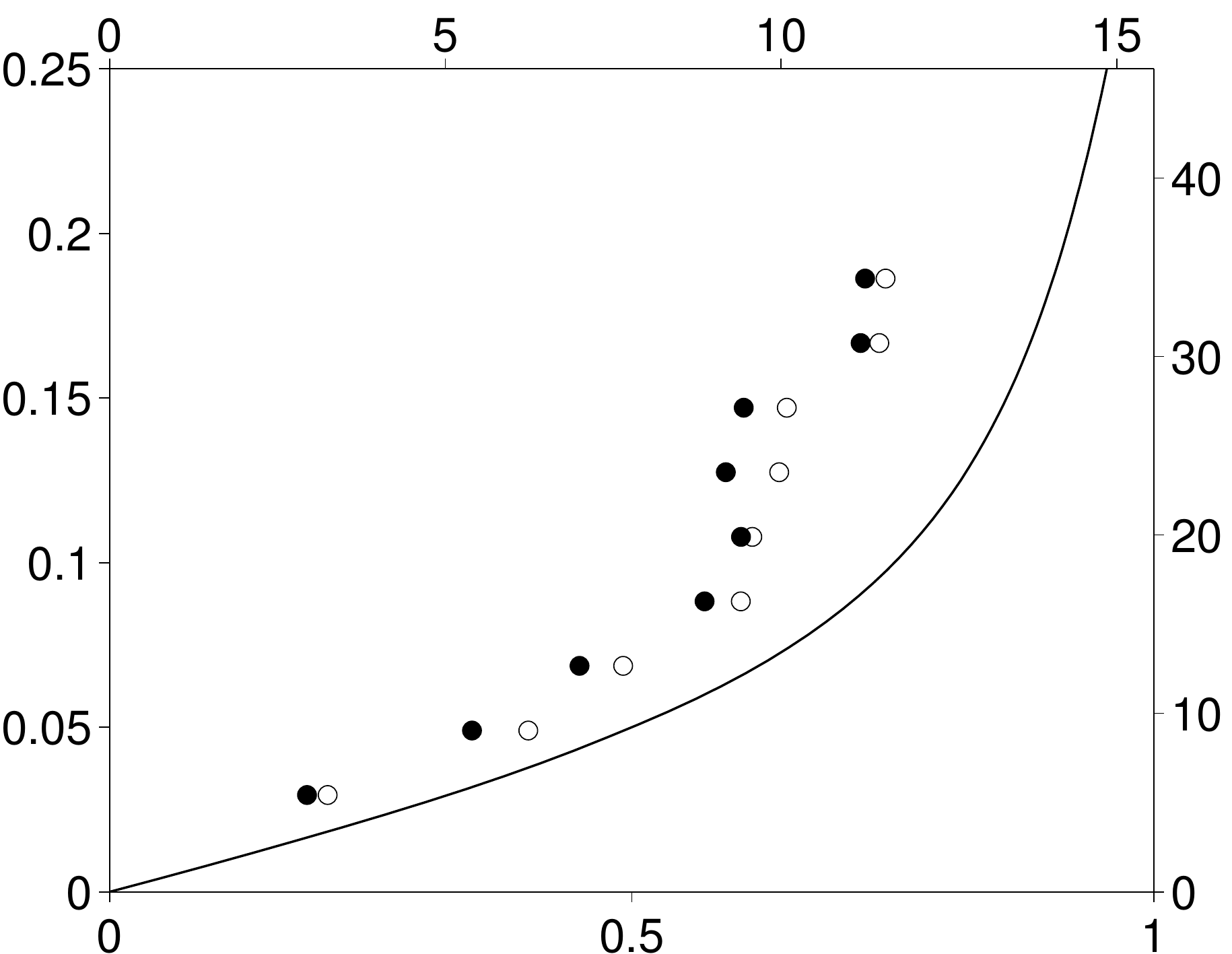}
        \centerline{\small 
              $[\langle u_\mathrm{f}\rangle;\langle u_\mathrm{p}\rangle;
                \langle u^{\cal S}_\mathrm{f}\rangle]/u_\mathrm{b}$
             }
        \end{minipage}  
        \begin{minipage}{4ex}
          \rotatebox{90}{\small $ y^+$}
        \end{minipage}   
        \\
        \begin{minipage}{3ex}
        \rotatebox{90}{\small $y/h$}
        \end{minipage}
        \begin{minipage}{.5\linewidth}
        \centerline{\small $(b)$ }\vspace{2ex}
        \centerline{\small 
                    $[\langle u_\mathrm{rel}^{\cal S}\rangle; u_\mathrm{lag}]^+$ }
        \includegraphics[width=\linewidth]
        {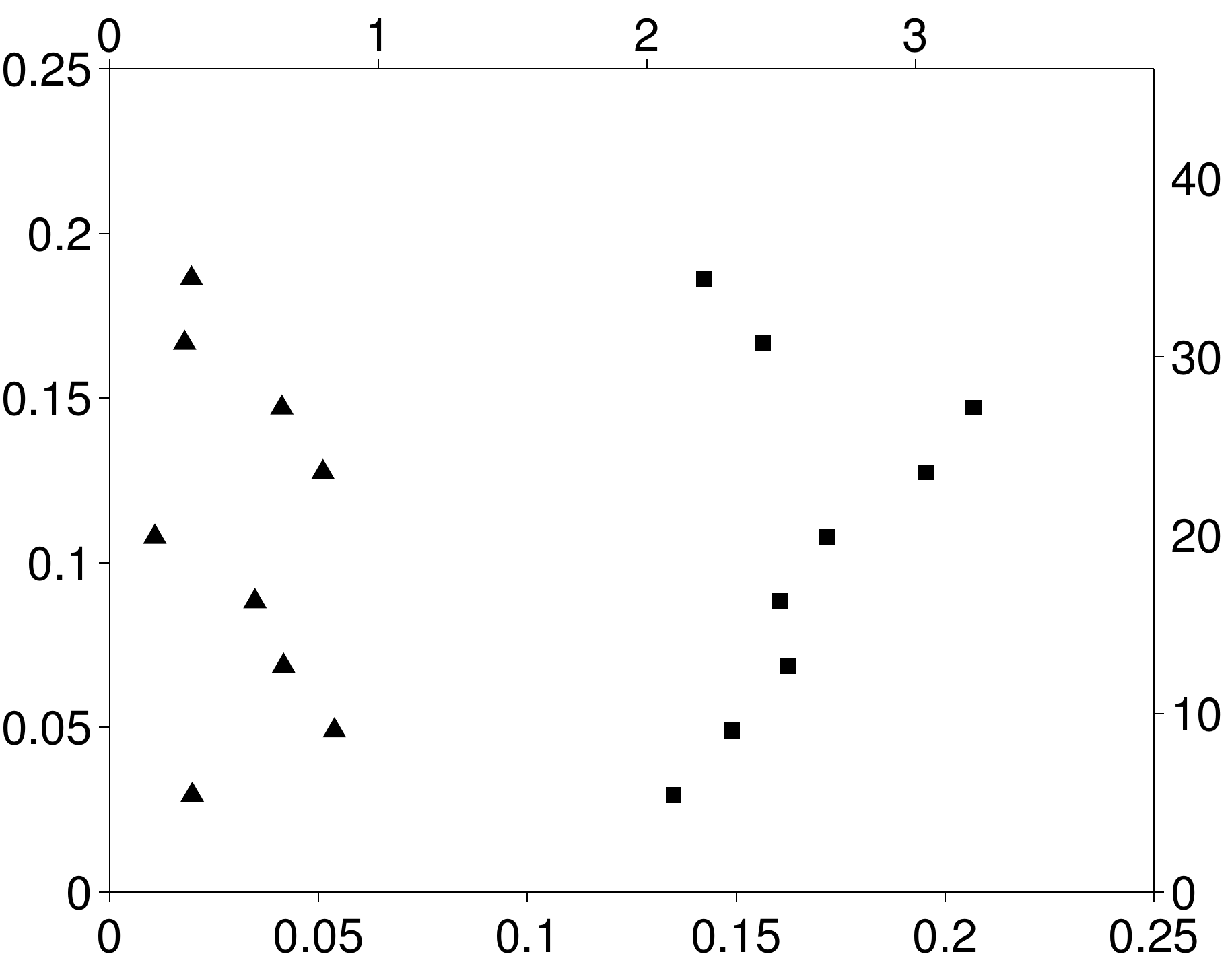}
        \centerline{\small 
              $[\langle u_\mathrm{rel}^{\cal S}\rangle; u_\mathrm{lag}]/u_b$
             }
        \end{minipage}  
        \begin{minipage}{4ex}
          \rotatebox{90}{\small $ y^+$}
        \end{minipage}   
        \caption{(\textit{a})
                 Wall-normal profile of 
                 the mean streamwise fluid velocity 
                 $\langle u_\mathrm{f}\rangle$ (\solid);
                 the mean streamwise particle velocity 
                 $\langle u_\mathrm{p}\rangle$ (\solidcircle);  
                 and the mean streamwise fluid velocity in the vicinity
                 of the particles 
                 $\langle u^{\cal S}_\mathrm{f}\rangle$ as defined in
                 (\ref{eq:mean_local_velocity})(\circle).
                 %
                 (\textit{b})
                 Wall-normal profile of 
                 the  average streamwise 
                relative velocity between the phases 
                $\langle u_\mathrm{rel}^{\cal S}\rangle$ as defined in 
                (\ref{eq:instantaneous-relative-velocity}) 
                (\solidtriangle), as well as  
                \revision{the the}{the} apparent velocity lag
                $u_\mathrm{lag} = \langle u_\mathrm{f}\rangle - 
                 \langle u_\mathrm{p}\rangle$ 
                (\solidsquare). 
                In both figures, outer units are shown in the 
                left and bottom axes and wall units in the 
                right and top axes.
                }    
        \label{fig:relative_velocity}
\end{figure}
  \subsubsection{Testing the definition in uniform unbounded flow past a
    fixed sphere}
\label{subsubsec-validation-uniform-unbounded-flow-past-sphere}
  In order to define a reasonable value for  $R_\mathrm{s}$, we have applied 
  definition (\ref{eq:mean_local_velocity}) to resolved numerical
  data of a uniform unbounded flow past a fixed sphere. 
  In this simple flow the relative velocity is equal to the
  incoming fluid velocity ${u}_{\mathrm{f},\infty}$. 
  Therefore, this test case allows to estimate
  the smallest value of $R_\mathrm{s}$ which yields a sufficiently close
  approximation of the relative velocity, i.e.\ for which
  ${u}_\mathrm{f}^{\cal S} \approx {u}_{\mathrm{f},\infty}$. 
  Here the only parameter is the Reynolds number based on the particle
  diameter and the incoming flow velocity, $Re_\infty =
  {u}_{\mathrm{f},\infty} D/\nu$, for which we have chosen three
  values, namely $Re_\infty = 50$,  $Re_\infty = 140$ 
  (both in the steady axisymmetric wake regime), and $Re_\infty = 350$
  in the unsteady vortex shedding regime \citep[see e.g.][for details
  on the structure of wakes behind spheres]{johnson:1999}. 
  Please note that the value of the Reynolds number based on particle 
  diameter and the apparent velocity lag ($Re_{\mathrm{lag}} = u_{\mathrm{lag}}D/\nu$)
  in the horizontal channel flow considered in the main part of this
  manuscript is in the range of $15$ to $20$.  
  Returning to the fixed sphere in unbounded uniform flow, snapshots
  of the velocity field in a plane through the center of the sphere
  are depicted in figure~\ref{fig:slip_pure_sedimentation}.
  It should be mentioned that in this unbounded flow configuration
  trimming the spherical averaging surface is not necessary. 
  However, for reasons of consistency with the application of
  definition (\ref{eq:mean_local_velocity}) to wall-bounded flow,
  we have performed the averaging over the previously defined
  spherical surface segment (cf.\
  figure~\ref{fig:particle_shell_sketch}).  
  
  Figure \ref{fig:slip_pure_sedimentation}\textit{d} shows the
  average value of the instantaneous fluid velocity 
  as defined by (\ref{eq:mean_local_velocity}) while
  varying the value of $R_\mathrm{s}$.  
  Note that in the unsteady case ($Re_\infty = 350$) an average in time over
  several flow fields was performed. 
  As expected, 
  the value of the fluid velocity defined in
  (\ref{eq:mean_local_velocity}) tends towards the value of the
  incoming  
  fluid velocity ${u}_{\mathrm{f},\infty}$ for large values of $R_\mathrm{s}$. 
  As $R_\mathrm{s}$
  tends towards $R$, the surrounding fluid appreciates the presence of
  the sphere and ${u}_\mathrm{f}^{\cal S}$ tends to zero due to the
  no-slip condition.  
  Depending on the Reynolds number, the degree to which the 
  computed value of the surrounding fluid velocity is affected
  by the presence of the sphere varies somewhat. 
  However, it is seen that 
  for all the three cases considered, approximately 90\% of the value of 
  the incoming flow velocity is obtained at a distance of $3R$ 
  from the particle center. 
  Therefore, based on the above results, 
  we define the velocity ``seen'' by the finite-size particles through 
  (\ref{eq:mean_local_velocity}) with a choice of $R_\mathrm{s}=3R$. 
\revision{}
{%
  Note that a similar analysis of the fluid velocity field in the
  vicinity of a single fixed particle in forced turbulence has been
  performed by \cite{naso:10}.  
}
%
\subsubsection{%
  The contribution of the mean relative velocity  
  to the apparent velocity lag
}
\label{subsubsec-mean-streamwise-relative-velocity}
  After having found a method to determine the fluid velocity
  ``seen'' by a particle as given in (\ref{eq:mean_local_velocity}),
  we can now define an instantaneous relative velocity for the $i$th
  particle, viz.  
  \begin{equation}\label{eq:instantaneous-relative-velocity}
    \mathbf{u}^{\cal S}_\mathrm{rel}(i,R_s,t) = 
    \mathbf{u}_\mathrm{f}^{\cal S}(\mathbf{x}^{(i)}_\mathrm{p},R_\mathrm{s},t) 
    - \mathbf{u}^{(i)}_\mathrm{p}(t) \,,
  \end{equation}
  with $R_\mathrm{s}=3R$. 
  The average  over all particles and time of $u_\mathrm{f}^{\cal S}$ and of 
  $u_\mathrm{rel}^{\cal S}$ can be computed by the averaging operator
  (\ref{equ-def-avg-operator-binned-particles}) defined in
  \ref{subsec-binned-averages-over-particle-related-quantities}.
  %
  Note that the values of $\langle u_\mathrm{f}^{\cal S} \rangle$ and $\langle
  u_\mathrm{rel}^{\cal S} \rangle$ discussed in the following stem
  from 
  \revision{%
    45
  }{%
    70
  } 
  instantaneous flow fields. Therefore the statistics
  are based on a smaller number of samples 
  than those computed from data collected during run-time, e.g.\ the
  values of $\langle u_{\mathrm{f}} \rangle$ and $\langle u_\mathrm{p} \rangle$.
  \revision{}{%
    Furthermore, we have verified that the following results are not
    very sensitive to a variation of the chosen value of $R_s$. 
  }

  Figure \ref{fig:relative_velocity}$a$ shows
  $\langle u_\mathrm{f}^{\cal S} \rangle$ in comparison to 
  $\langle u_\mathrm{f} \rangle$ and $\langle u_\mathrm{p} \rangle$
  in the wall region $y^+ < 36$.
  The figure clearly demonstrates that the average velocity of
  the fluid in the vicinity of the particles $\langle u_\mathrm{f}^{\cal S}
  \rangle$ is significantly smaller than the unconditioned fluid
  velocity $\langle u_\mathrm{f} \rangle$ and that the former is comparable in
  magnitude to the mean particle velocity $\langle u_\mathrm{p} \rangle$ for
  most of the wall-normal interval. 
  %
  %
  Figure \ref{fig:relative_velocity}$b$ which shows
  $\langle u_\mathrm{rel}^{\cal S} \rangle$ in comparison to $u_{\mathrm{lag}}$ 
  highlights this finding.
  At $y^+ \approx 5$, which holds most of the particles,
\revision{
    $\langle u_\mathrm{rel}^{\cal S} \rangle \approx 0.31u_\tau $ which
    is considerably smaller than $u_{\mathrm{lag}} \approx
    2.13u_\tau$.
    }{%
    $\langle u_\mathrm{rel}^{\cal S} \rangle \approx 0.31u_\tau $ which
    is considerably smaller than $u_{\mathrm{lag}} \approx
    2.10u_\tau$.
    }
  At larger wall-normal distances the value of $\langle u_\mathrm{rel}^{\cal S}\rangle$ 
  increases slightly, attaining a
  \revision{
    maximum value of approximately $0.85u_\tau$ 
  }{%
    maximum value of approximately $0.91u_\tau$ 
  }
  although a clear trend with wall-normal distance is difficult to infer
  due to the 
  decrease of the number of data samples away from the wall (cf.\
  figure~\ref{fig:particle_solid_volume_fraction}).  
  However, it is beyond doubt that for the considered wall distances 
  $\langle u_\mathrm{rel}^{\cal S} \rangle$ is considerably smaller than $u_{\mathrm{lag}}$.
  The relative particle Reynolds number
  $Re_{\mathrm{rel}} = \langle u_\mathrm{rel}^{\cal S}\rangle D/\nu$ for the particles 
  closest to the wall has a value $Re_{\mathrm{rel}} \approx 2.2$ which is
  much smaller than $Re_{\mathrm{lag}} \approx 15$. 
  The very small value of $Re_{\mathrm{rel}}$ indicates that there are no
  substantial particle-induced wakes, 
  which is confirmed by the conditionally averaged flow field
  around the particles discussed in more detail in
  \S~\ref{subsec-particle-position-and-coherent-structures}.
  It can be concluded from the present results that 
  the dominant contribution to $u_{\mathrm{lag}}$ stems indeed from the
  preferential location of particles in the low-speed regions of the
  flow, in agreement with previous experimental findings
  \citep[][]{kaftori:1995b,kiger:2002}.  
%

\begin{figure}
  \centering
  \begin{minipage}{1ex}
    \rotatebox{90}{\small }
  \end{minipage}
  \hspace{1ex}
  \begin{minipage}{.9\linewidth}
    \includegraphics[width=\linewidth,clip=true,
    viewport=350 650 2100 1200]
    {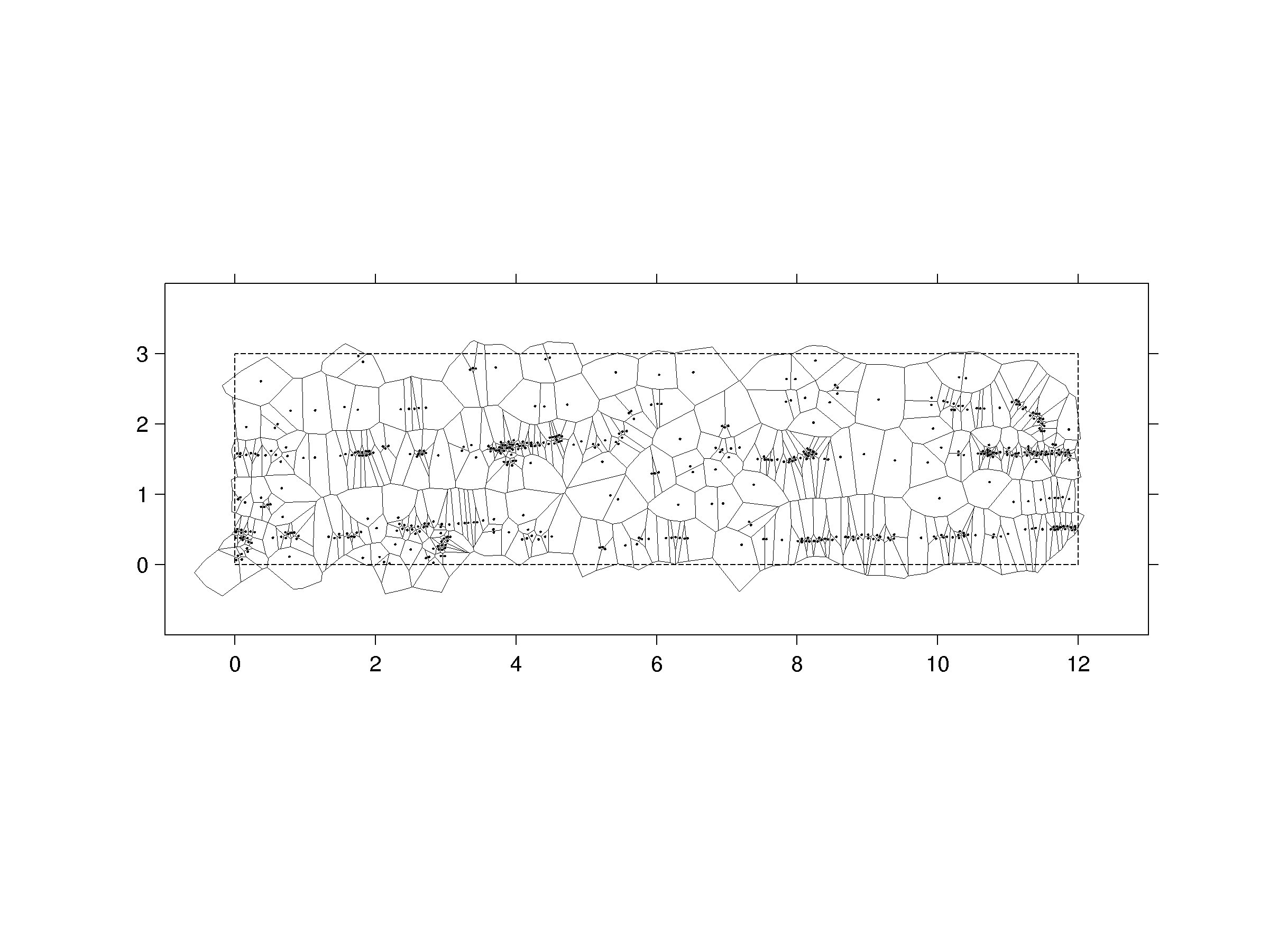}
    \\
    \centerline{\small}
  \end{minipage}
  \caption{%
    A Voronoi tessellation of a periodic horizontal
    plane corresponding to the particle positions
    of the snapshot shown in figure 
    \ref{fig:uprime_xz_yplus5}$a$.}
  \label{fig:voronoi_tesselation}
\end{figure}
\begin{figure}
   \centering
        \begin{minipage}{2ex}
        \rotatebox{90}{\small $\quad pdf$}
        \end{minipage}
        \begin{minipage}{.5\linewidth}
        \includegraphics[width=\linewidth]
        {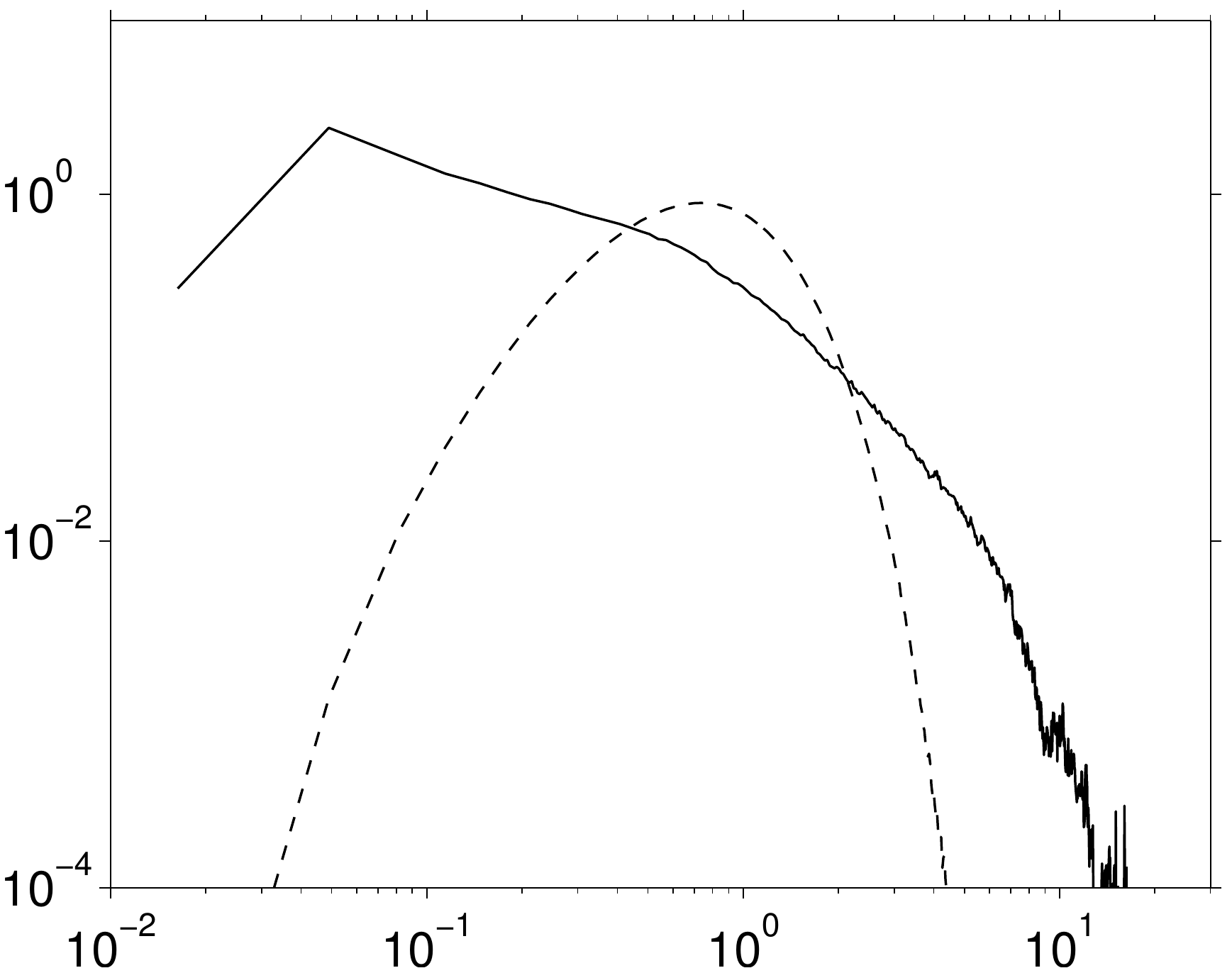}
        \hspace{-0.5\linewidth}\raisebox{.85\linewidth}
        {\small $(a)$}
        \\
        \centerline
        {\small $ A_\mathrm{V}/\langle A_\mathrm{V} \rangle $ }
        \end{minipage}\\[20pt]
        %
        \begin{minipage}{3ex}
        \rotatebox{90}
        {\small $\quad pdf$ }
        \end{minipage}
        \begin{minipage}{.5\linewidth}
        \includegraphics[width=\linewidth]
        {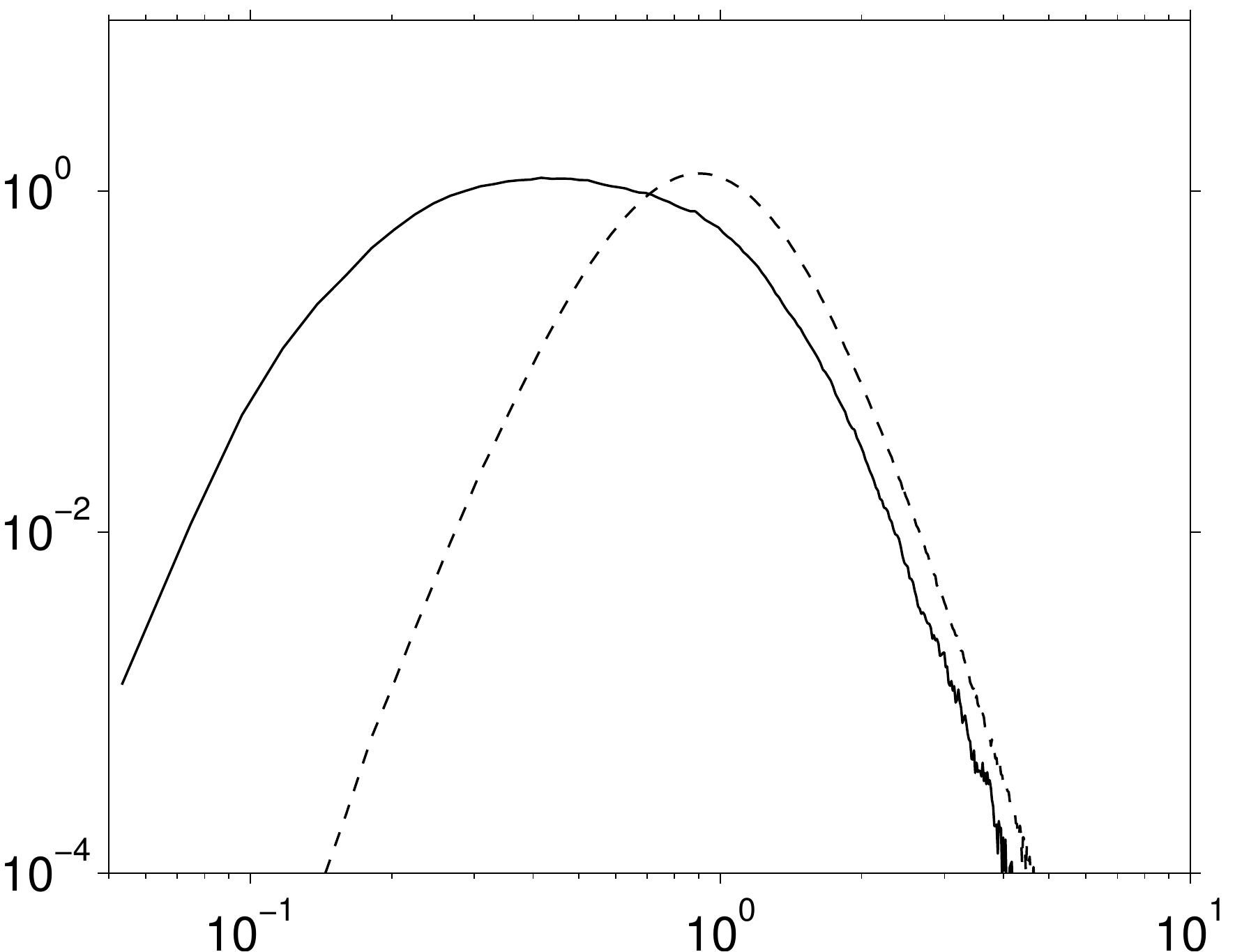}
        \hspace{-0.5\linewidth}\raisebox{.85\linewidth}
        {\small $(b)$}
        \\
        \centerline{\raisebox{1ex}{\small $ l_\mathrm{x,V}/l_\mathrm{z,V} $}}
        \end{minipage}
        \caption{Probability density function of
                 (\textit{a}) the normalized Voronoi 
                 cell 
                 areas and 
                 (\textit{b}) the ratio of the distances between 
                 the extreme streamwise and
                 spanwise extents of the Voronoi 
                 cells. 
                 The dashed
                 lines represent corresponding distributions of  
                 particle positions generated randomly with uniform
                 probability.  
               }
        \label{fig:pdf_voronoi}
\end{figure}
\subsection{Distribution and motion of particles at the bottom wall}
\label{subsec-distribution-motion-of-particles}
%
As a combined effect of gravity and turbulent dynamics, 
particles in the present study are observed to spend most of their
time residing near the bottom wall. Occasionally they are 
entrained and lifted up by the carrier flow and 
either 
return to the bottom or are ejected to the outer region of the flow.
During their residence time at the wall, 
the behaviour of particles (their streamwise 
and spanwise motion as well as their spatial distribution) 
is closely related to the dynamics of the near wall coherent 
flow structures. In this section we study the 
near-wall particle behaviour by analyzing particle positions and
velocities. 

It is clearly observable from the instantaneous 
flow field and from the corresponding particle positions shown
in figure \ref{fig:uprime_xz_yplus5}$a$
that particles  
form elongated structures, resembling streamwise aligned chains.
Visualization of sequences of such images shows that 
these particle clusters remain in the form of 
quasi-streamwise aligned streaky structures which maintain
a coherence over substantial time scales. In line with previous
experimental findings \citep{kaftori:1995a,nino:1996}, 
these particle streaks are observed to reside preferentially 
in low-speed fluid regions.
Animations show that, during their travel downstream, 
particles 
remain 
in these relatively 
quiescent low-speed streaks for relatively long time 
intervals, exhibiting only slight spanwise ondulations. 
In those animations it can also be seen that incoming sweeps of high
momentum fluid sometimes act to dissolve the particle accumulations,
after which those particles relatively quickly migrate back to a
neighboring low-speed region.
%

\begin{figure}
   \centering
        \begin{minipage}{2ex}
        \rotatebox{90}{\small $\quad R_{\mathrm{L},Q_\mathrm{V}}$}
        \end{minipage}
        \begin{minipage}{.5\linewidth}
        \includegraphics[width=\linewidth]
        {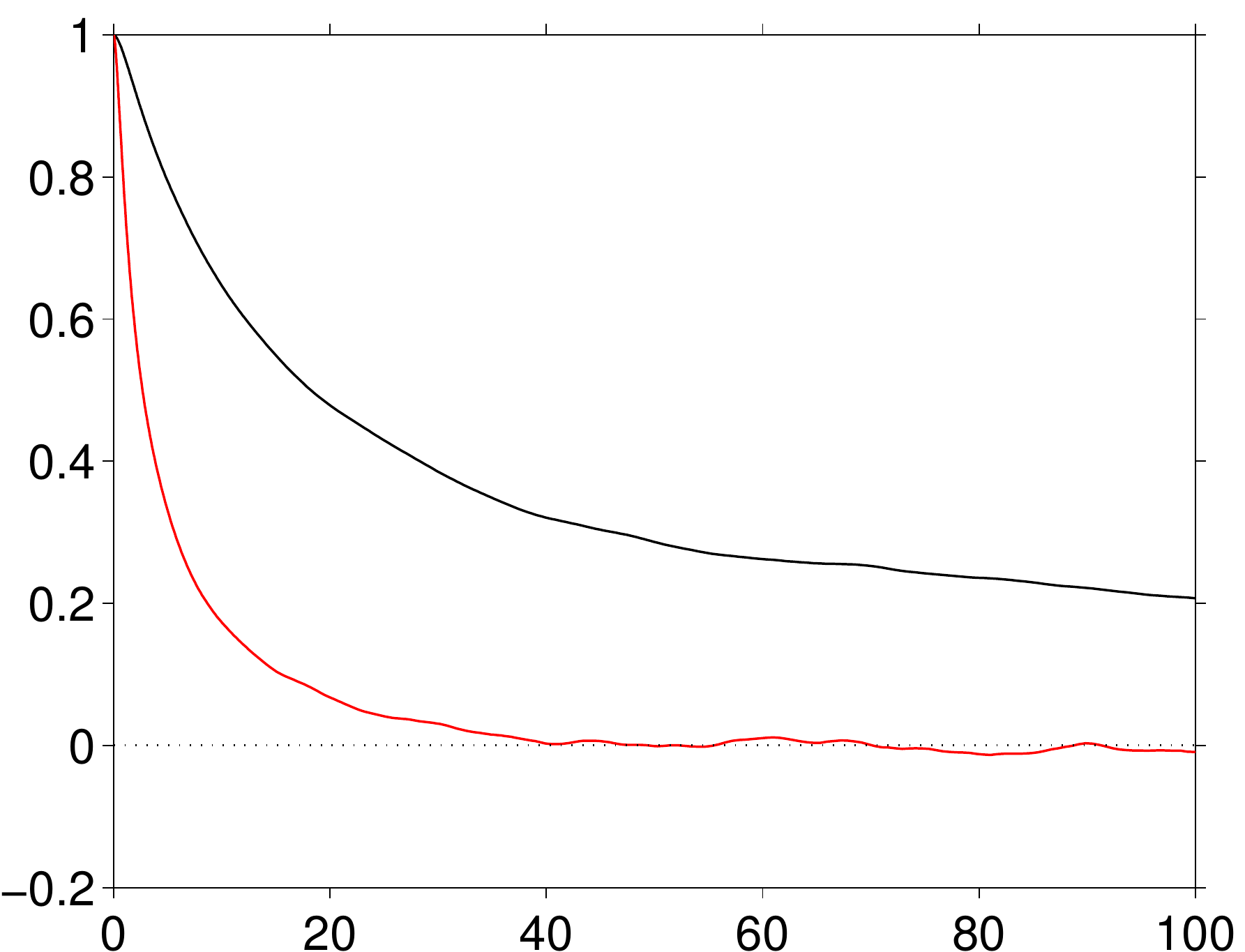}
        \\
        \centerline
        {\small $ \tau_\mathrm{sep}/T_\mathrm{b} $ }
        \end{minipage}
         \caption{
                 Lagrangian autocorrelation as a function of 
                 time lag $\tau_\mathrm{sep}$ for the Voronoi
                 area (\solid) and the Voronoi 
                 cell 
                 aspect ratio ({\color{red} \solid}).}
        \label{fig:lagrangian_autocorrelation_voronoi}
\end{figure}
As a quantitative measure of the spatial distribution of particles,
we 
have 
carried out a cluster analysis based on Voronoi diagrams. 
Voronoi diagrams are tessellations which partition a space
based on a set of given center positions (the 'sites')
into regions such that each point inside a given region is 
closer to the region's site than to any other site 
\citep{okabe:1992}.
Voronoi diagram analysis is widely utilized in many 
scientific areas 
and 
has been introduced as a tool for the analysis of preferential
particle 
concentration 
in turbulent flows 
\citep{monchaux:2010b,monchaux:2012}.

In order to investigate the spatial distribution of particles in the
immediate vicinity of the wall, 
a horizontal plane of the bi-periodic
domain is 
tessellated based on instantaneous 
%
streamwise and spanwise
positions of all particles 
whose center is located at a wall-normal distance within one particle diameter from the 
bottom wall.
Each of the resulting Voronoi regions has an area 
$A_\mathrm{V}^{(i)}$ 
associated to the $i$th particle. 
The example given in figure \ref{fig:voronoi_tesselation} demonstrates
that the tesselation is area-filling, taking into account the
bi-periodicity. 
The inverse of a 
Voronoi cell
area is directly related to the local particle 
concentration: 
small areas correspond to high local particle 
concentration and large Voronoi cell areas to voids.
We have analyzed the distribution of 
these areas compared to the distribution of the Voronoi 
areas associated with
randomly-positioned particle sets. The p.d.f.\ of the 
Voronoi areas (normalized by the mean value) of 
randomly positioned points is 
independent of the particle number density \citep{Ferenc:2007}.
However, finite-size particles have the constraint that they cannot 
overlap, and at close-packing (the maximum possible particle 
number density), 
all 
Voronoi areas are identical and have a Dirac delta distribution. 
Here we generate reference data from a 
sufficiently large number of randomly-positioned particle sets 
(with uniform probability in space) 
applying the constraint that they do not overlap and adopting a 
particle number density equal to the one of the present flow case. 

\begin{figure}
   \centering
        \begin{minipage}{2ex}
        \rotatebox{90}{\small $\quad pdf$}
        \end{minipage}
        \begin{minipage}{.5\linewidth}
        \includegraphics[width=\linewidth]
        {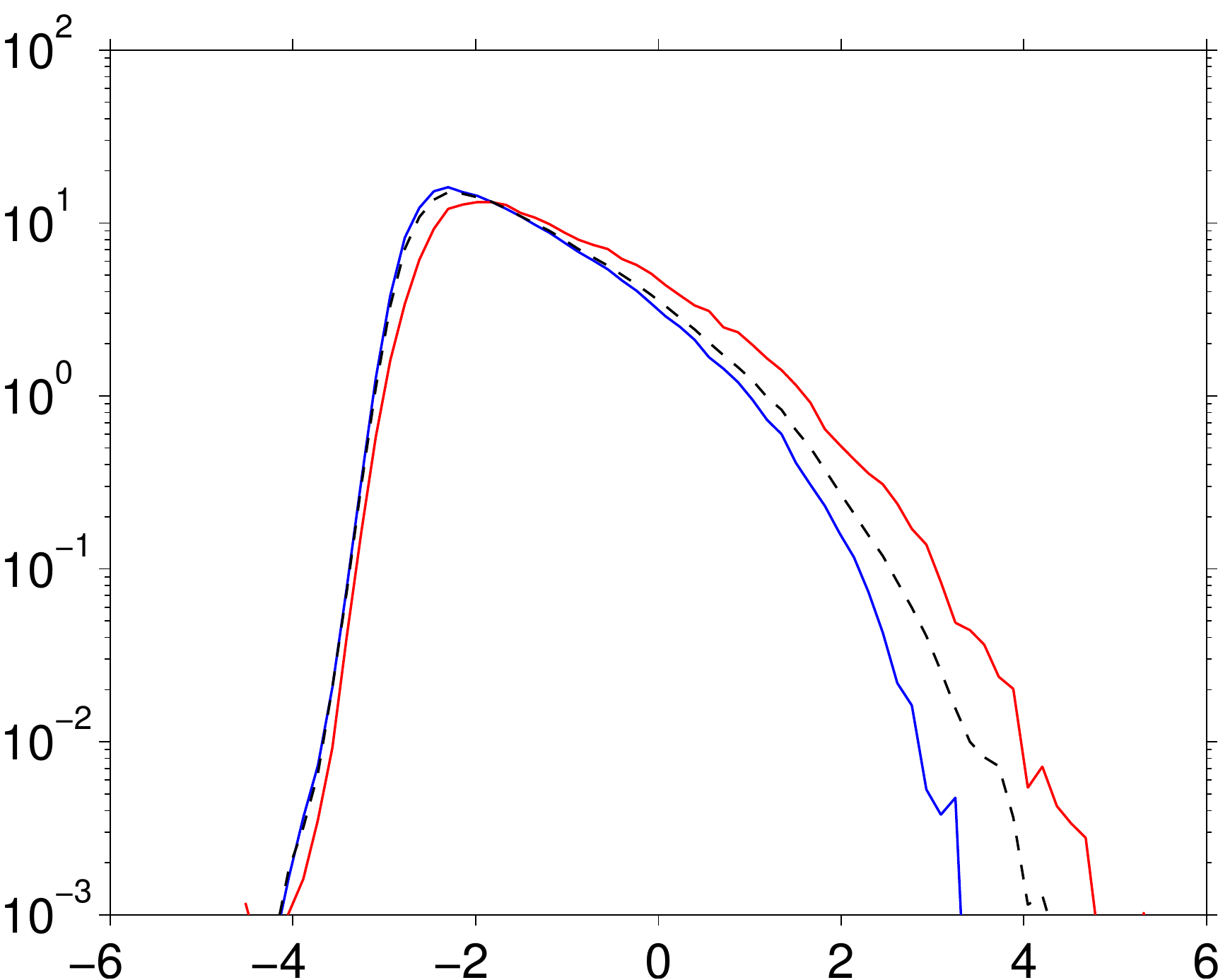}
        \\
        \centerline
        {\small $ [u_\mathrm{p} - \langle u_\mathrm{f} \rangle]^+ $ }
        \end{minipage}
         \caption{P.d.f.s 
                  of the streamwise particle velocity
                  relative to the mean flow 
                  $u^{(i)}_\mathrm{p}(t) - \langle u_\mathrm{f} 
                   \rangle(y^{(i)}_\mathrm{p}(t))$
                  conditioned upon 
                  \revision{the the}{the}
                  Voronoi cell area $A^{(i)}_\mathrm{V}(t)$ 
                  of all particles whose centers are located at
                  a wall-normal distance from the  bottom 
                  $y^{(i)}_\mathrm{p}(t) < D$. The color coding represents:
                  conditional p.d.f. of those particles
                  such that their associated Voronoi cell area
                  $A^{(i)}_\mathrm{V}(t) < A^c_\mathrm{V}$, ({\color{blue} \solid});
                  conditional p.d.f. of those particles
                  such that their associated Voronoi cell area
                  $A^{(i)}_\mathrm{V}(t) > A^v_\mathrm{V}$, ({\color{red} \solid}).
                  The dashed line corresponds to the
                  unconditioned p.d.f. of the same quantity
                  in this wall-normal interval. Note that
                  the p.d.f.s are not centered about their
                  respective mean.
                  }
        \label{fig:u_pdf_conditioned_to_voronoi_area}
\end{figure}
Figure \ref{fig:pdf_voronoi}(\textit{a}) shows the p.d.f.\ of the normalized 
Voronoi areas computed from 
\revision{%
  12900 
}{%
  20314
}
snapshots 
of the particle `field'.
A 
significant 
deviation of the p.d.f.\ 
from the corresponding randomly-positioned particles 
indicates that the particle distribution
at the bottom wall is far from random 
-- as already observed visually, e.g.\ in
figure~\ref{fig:voronoi_tesselation}. 
%
The DNS data exhibits a much 
higher probability
of finding very small and very large Voronoi areas 
than in a 
random case
with uniform probability.
%
The fact that
there is a high probability for particles in the present case to be 
associated with very small 
Voronoi cell 
areas
indicates that, consistent with the visual evidence, particles 
have a clear tendency to accumulate. 

In order to provide a measure of the anisotropy of particle
accumulation regions, 
we have further analyzed the 
shape of the Voronoi 
cells.
%
More 
specifically, we have 
examined the 
slenderness (i.e.\ the aspect ratio) 
of these regions by computing the ratio between the largest
extent  
of each Voronoi cell 
in the streamwise direction ($l_\mathrm{x,V}$) and 
in the spanwise direction ($l_\mathrm{z,V}$). 
Figure \ref{fig:pdf_voronoi}(\textit{b}) shows the
p.d.f.\ of the aspect ratio of the Voronoi 
cells 
along with 
the p.d.f.\ of the same quantity for the randomly-positioned particles.
Again, there is
an appreciable difference between the two p.d.f.s. 
On average, the Voronoi regions associated with 
the particles from the present case have a smaller aspect ratio 
\revision{%
  ($\langle l_\mathrm{x,V}/l_\mathrm{z,V}\rangle = 0.711$) 
}{%
  ($\langle l_\mathrm{x,V}/l_\mathrm{z,V}\rangle = 0.722$) 
}
when compared to that of the random 
case ($\langle l_\mathrm{x,V}/l_\mathrm{z,V}\rangle = 1.062$ -- which
should tend to unity when increasing the sampling size).  
Moreover, there is a higher probability of finding a small aspect 
ratio than in the random case and vice versa. This indicates
that the regions are 
significantly 
squeezed (stretched) in the streamwise
(spanwise) directions which means that 
particles are more likely to be streamwise aligned than aligned along
the spanwise axis.
%
Therefore, the Voronoi cell aspect ratio analysis confirms the visual
observation of alignment of particles into streamwise elongated `streaks'.

\begin{figure}
   \centering
        \begin{minipage}{2ex}
        \rotatebox{90}{\small $\quad R_{\mathrm{L},u_{\mathrm{p},\alpha}}$}
        \end{minipage}
        \begin{minipage}{.5\linewidth}
        \includegraphics[width=\linewidth]
        {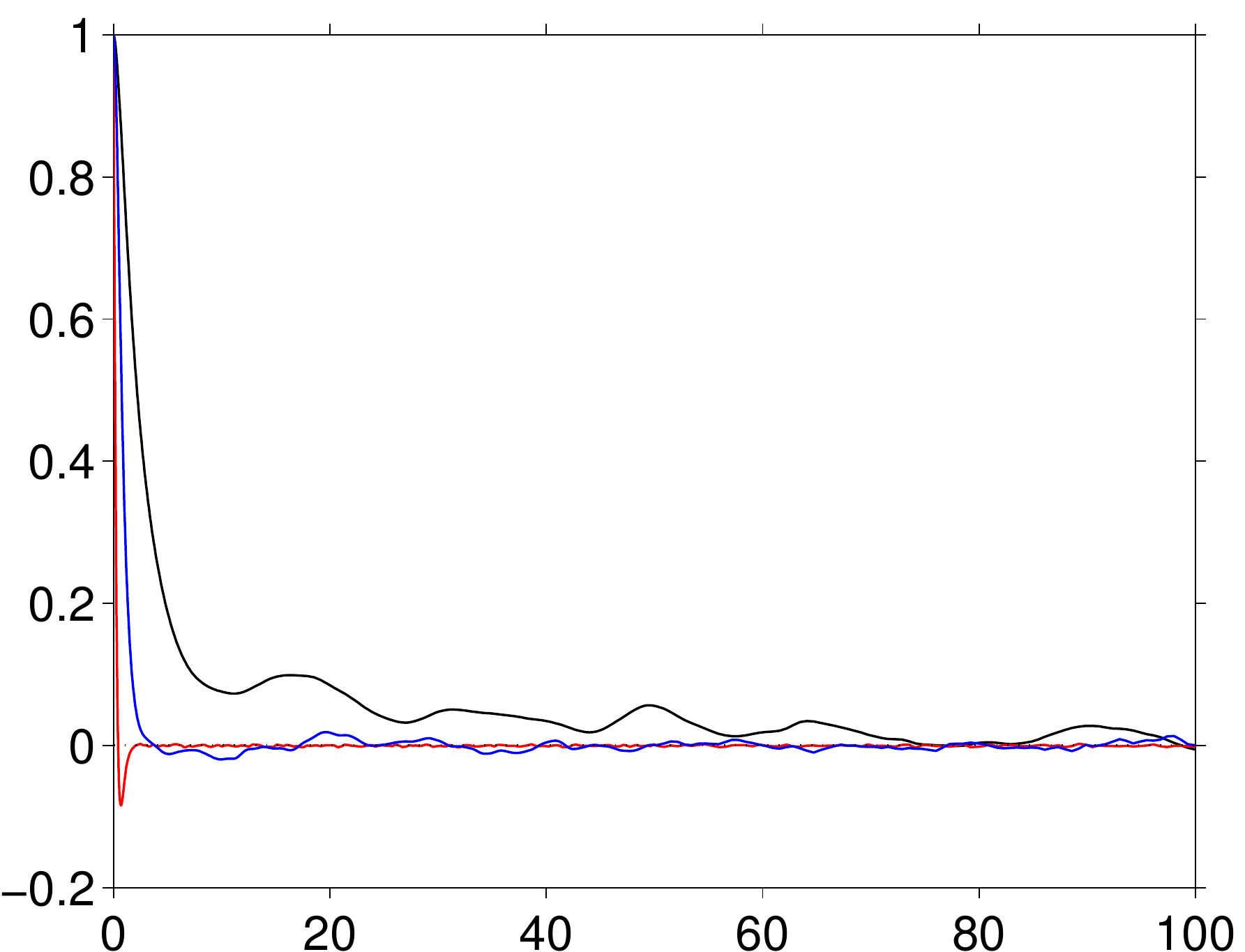}
        \\
        \centerline
        {\small $ \tau_\mathrm{sep}/T_\mathrm{b} $ }
        \end{minipage}\\[-175pt]
         \hspace{70pt}
         \begin{minipage}{.3\linewidth}
           \includegraphics[width=\linewidth]
           {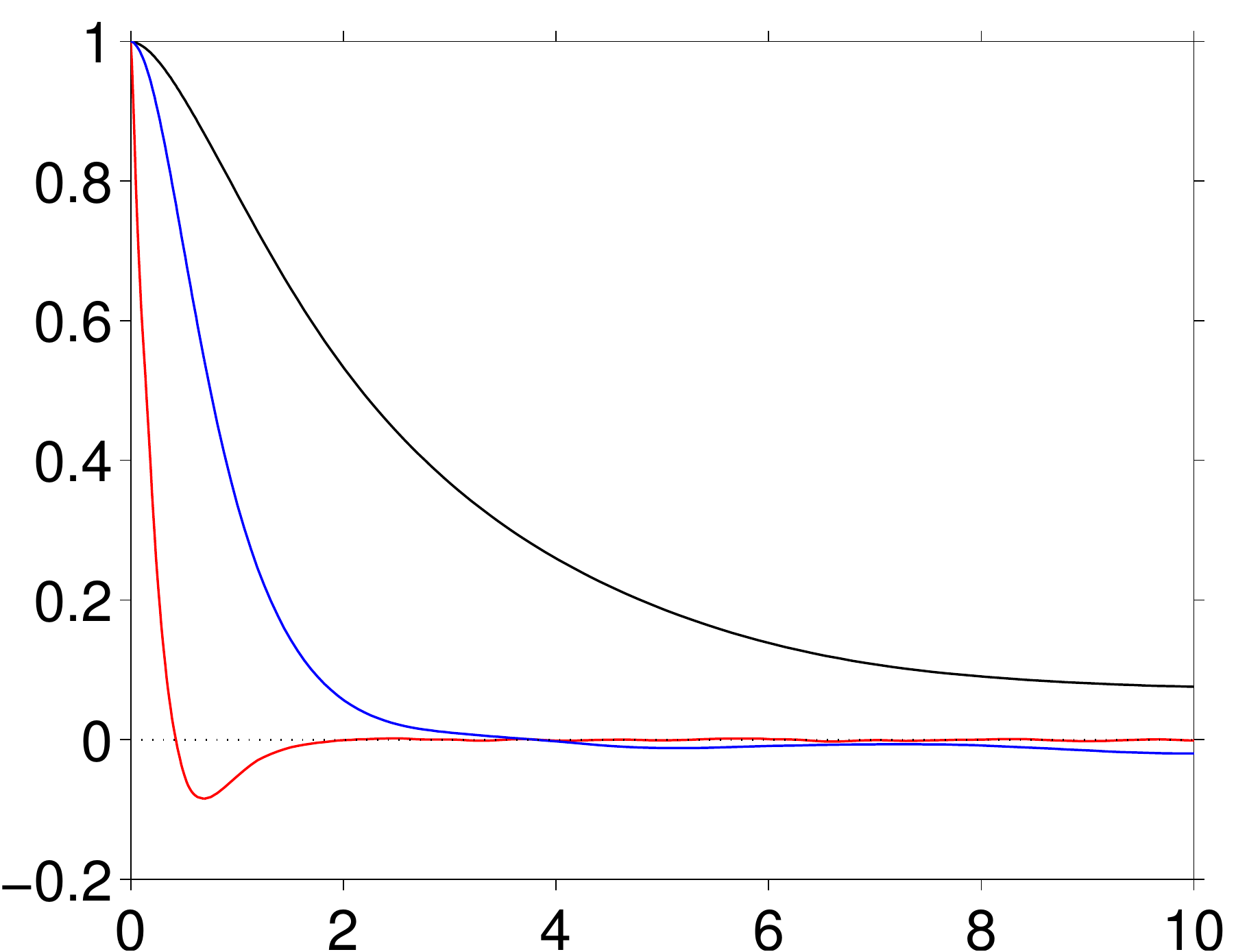}
         \end{minipage}
         \vspace{.21\linewidth}
         \caption{Lagrangian autocorrelation as a function of 
                 time lag $\tau_\mathrm{sep}$  for the 
                 velocity of all particles 
                 whose centers are located at a wall-normal 
                 wall distance 
                 $y^{(i)}_\mathrm{p}(t) < D$.
                 The color code indicates: 
                 (\solid) streamwise component, $\alpha = 1$; 
                 ({\color{red}\solid}) wall-normal component, $\alpha =
                 2$; 
                 ({\color{blue}\solid}) spanwise component, $\alpha =
                 3$. 
                 The 
                 insert 
                 shows a closeup of the same data
                 for small separation times.
                 }
        \label{fig:lagrangian_autocorrelation_velocities}
\end{figure}
Furthermore, we have tracked the Voronoi area and 
Voronoi aspect ratio associated to 
each particle in time 
in order 
to study the behaviour of these 
quantities in a Lagrangian sense. 
The Lagrangian autocorrelation function of a Voronoi
cell quantity 
as a function of separation time
$\tau_{\mathrm{sep}}$ 
can be defined as
\begin{equation}
  \label{equ:def-lagrangian-auto-corr-voronoi-quant}
  R_{\mathrm{L},Q_\mathrm{V}}(\tau_{\mathrm{sep}}) = 
  \frac{\langle Q'_\mathrm{V}(t)
    \,
    Q'_\mathrm{V}(t+\tau_{\mathrm{sep}})\rangle
  }
  {\langle Q'_\mathrm{V}(t) 
    \,
    Q'_\mathrm{V}(t) \rangle
  }
  \,,
\end{equation}
where  the quantity $Q_\mathrm{V}(t)$ 
is either the Voronoi area or the Voronoi aspect 
ratio at time $t$ and 
$Q'_\mathrm{V}(t) = Q_\mathrm{V}(t)-\langle Q_\mathrm{V}
\rangle
$.
%
The 
averaging in
(\ref{equ:def-lagrangian-auto-corr-voronoi-quant}) 
is performed over all particles and over time, where 
segments of time signals are used for which the considered particle
resides close to the wall, i.e. $y_p^{(i)} (t^\prime)  < D \ \forall\,t^\prime \in
[t,t+\tau_{sep}]$ . 

Figure \ref{fig:lagrangian_autocorrelation_voronoi} shows the 
quantity $R_{\mathrm{L},Q_\mathrm{V}}$ as a function of $\tau_{\mathrm{sep}}$ for both
the Voronoi area and Voronoi aspect ratio. 
  The data shown in
  figure~\ref{fig:lagrangian_autocorrelation_voronoi} 
  demonstrates that the spatial particle distribution is indeed  
  highly persistent in time. The Voronoi cell area is still
  significantly correlated 
  \revision{%
    (with a coefficient value above $0.2$)
    after $80$ bulk time units. 
  }{%
    (with a coefficient value of approximately $0.2$)
    even after $100$ bulk time units. 
  }
  The Voronoi cell aspect ratio, on the other hand, changes much
  faster, with the correlation coefficient practically vanishing after
  approximately $40$ bulk time units. 
  The Taylor micro scale 
  \revision{%
    (i.e.\ the osculating parabola at zero
    separation) measures $5.7\,T_\mathrm{b}$ ($1.3\,T_\mathrm{b}$) for the cell area (cell
    aspect ratio).
  }{%
    (given by the osculating parabola at zero separation) 
    measures $5.2\,T_\mathrm{b}$ ($1.3\,T_\mathrm{b}$) for the cell area (cell
    aspect ratio). 
  }
  The difference in time scales measured for these two quantities
  becomes obvious from a closer inspection of time sequences of
  Voronoi diagrams. Therein it can be observed that while particles
  typically remain located for a long time inside one specific
  accumulation region, 
  the shape of their associated Voronoi cell fluctuates significantly
  without necessarily changing the cell's area. 

The intersection points
of the two curves in figure~\ref{fig:pdf_voronoi}(\textit{a})
(DNS data versus random particle distribution) 
can be used to define an objective criterion for determining whether a
particle is located inside a cluster or inside a void
\citep{monchaux:2010b}. 
For this purpose, the cell area corresponding to the lower ($A^c_\mathrm{V}$)
and upper ($A^v_\mathrm{V}$) intersection points in that diagram are considered
as threshold values: all particles with associated Voronoi cell area 
lower than $A^c_\mathrm{V}$ are considered to reside in a cluster, those with a
volume larger than $A^v_\mathrm{V}$ are related to voids. 
%
We have computed for each particle
the number of times it crosses these threshold values and 
we have determined the duration of all temporal intervals during which
the Voronoi cell area is continuously below (above) the lower (upper)
threshold value.  
This type of study allows us to estimate the residence 
time of particles in a region 
with an `extreme' concentration 
(e.g.\ a cluster or a void). 
%
  It turns out that the average residence time in a cluster (void) is
  approximately 
  \revision{%
    $7.6T_\mathrm{b}$ ($6.6T_\mathrm{b}$) 
  }{%
    $7.8T_\mathrm{b}$ ($6.9T_\mathrm{b}$) 
  }
  while the average frequency of a particle entering a cluster (void)
  region is 
  \revision{%
    $0.06/T_\mathrm{b}$ ($0.02/T_\mathrm{b}$). 
    }{%
    $0.06/T_\mathrm{b}$ ($0.02/T_\mathrm{b}$). 
    }
  Concerning the Voronoi cell aspect ratio (which exhibits a single
  intersection point, cf.\ figure~\ref{fig:pdf_voronoi}$b$), one can
  define a threshold below which the Voronoi cell's shape can be
  considered as 
  \revision{streamwise}
  {spanwise}
  elongated. The analysis yields an average
  temporal interval of 
  \revision{%
    $5.4T_\mathrm{b}$ 
  }{%
    $5.5T_\mathrm{b}$     
  } 
  during which a particle has an
  associated 
  \revision{streamwise}
  {spanwise} 
  elongated Voronoi cell. 
  %

  The data from the Voronoi analysis can be further utilized for the
  purpose of conditionally averaging the Lagrangian particle
  quantities. To this end we have computed the difference between the
  streamwise component of the instantaneous particle velocity and the
  mean fluid velocity at the same wall-normal distance, i.e.\ 
  $u^{(i)}_\mathrm{p}(t) - \langle u_\mathrm{f}\rangle(y^{(i)}_\mathrm{p}(t))$ 
  for the $i$th particle. 
  The p.d.f.\ of this velocity difference is shown in
  figure~\ref{fig:u_pdf_conditioned_to_voronoi_area} for those
  particles whose center is located within one diameter of the wall
  plane.  
  The curve for the unconditioned quantity is clearly skewed towards
  positive values (i.e.\ large positive fluctuations are more probable
  than large negative ones). 
  Conditioning the same quantity upon the Voronoi cell area has a
  substantial effect, which is, however, mostly limited to a
  different probability for observing positive values. 
  In particular, particles located in clusters (voids) have a
  significantly lower (higher) probability to exhibit streamwise
  velocities which exceed the average fluid velocity. 
  One can conclude from this analysis that particles located in void
  regions have a higher tendency to be located in high-speed fluid
  regions than particles located in accumulation regions. 

  Finally, we consider the auto-correlation function of the particle
  velocity components, which is defined as 
  \begin{equation}
    \label{equ:def-lagrangian-auto-corr-partvel}
    R_{\mathrm{L},u_\mathrm{p},\alpha}(\tau_{\mathrm{sep}}) = 
      \frac{\langle u_{\mathrm{p},\alpha}^\prime(t)
      \,
      u_{\mathrm{p},\alpha}^\prime(t+\tau_{\mathrm{sep}})\rangle
    }
    {\langle u_{\mathrm{p},\alpha}^\prime(t)
      \,
      u_{\mathrm{p},\alpha}^\prime(t)
      \rangle
    }
    \qquad\forall\,\alpha=1,2,3
    \,,
  \end{equation}
  in analogy to (\ref{equ:def-lagrangian-auto-corr-voronoi-quant}). 
  %
  %
  Note that again only those data points are considered which
  correspond to particles remaining near the wall during the entire
  separation interval (i.e.\ with center $y_c^{(i)}(t^\prime)<D \
  \forall\,t^\prime \in [t,t+\tau_{sep}]$),
  consistent with the previous discussion. 
  Figure~\ref{fig:lagrangian_autocorrelation_velocities} shows that
  the wall-normal particle velocity component  
  decorrelates fastest, followed by the spanwise component and (with a
  much longer time scale) the streamwise component. 
  \revision{
    The corresponding Taylor microscales measure $1.38\,T_\mathrm{b}$, 
    $0.15\,T_\mathrm{b}$ and $0.75\,T_\mathrm{b}$ for the streamwise,
    wall-normal and spanwise components, respectively.
  }{%
    The corresponding Taylor microscales measure $1.41\,T_\mathrm{b}$,
    $0.15\,T_\mathrm{b}$ and $0.75\,T_\mathrm{b}$ for the streamwise,
    wall-normal and spanwise components, respectively.
  }
  The longer time scale of the streamwise particle velocity is owed to
  the anisotropy of the near-wall turbulence, i.e.\ the persistence of
  velocity streaks affecting the particle motion. A similar
  observation has been made in vertical particulate channel flow
  \citep{garcia-villalba:2012}. Contrary to that configuration,
  however, the wall-normal component behaves quite differently from
  the spanwise component in the present case. In particular, the curve
  for $R_{\mathrm{L},u_\mathrm{p},2}$ exhibits significant negative correlations at
  small separation times, with the first change of sign taking place
  at 
  $\tau_{\mathrm{sep}}=0.43\,T_\mathrm{b}$.  
  This feature is absent in the corresponding correlation curve in the
  case of the vertical particulate channel, and it must therefore be
  due to the action of gravity in the wall-normal direction. More
  specifically, gravity is counteracting any particle motion away from
  the wall, 
  \revision{leading to particle trajectories which often exhibit  a 
    reversal of sign after relatively short times.}
  {leading to wall-normal particle velocities which often exhibit  a 
    reversal of sign after relatively short times.}

\begin{figure}
   \centering
        \begin{minipage}{1ex}
        \rotatebox{90}{\small $\quad \tilde{z}/h$}
        \end{minipage}
        \begin{minipage}{.9\linewidth}
        \centerline
        {\small $\tilde{x}^+$}
        \includegraphics[width=\linewidth,clip=true,
        viewport=100 1 2320 580]
        {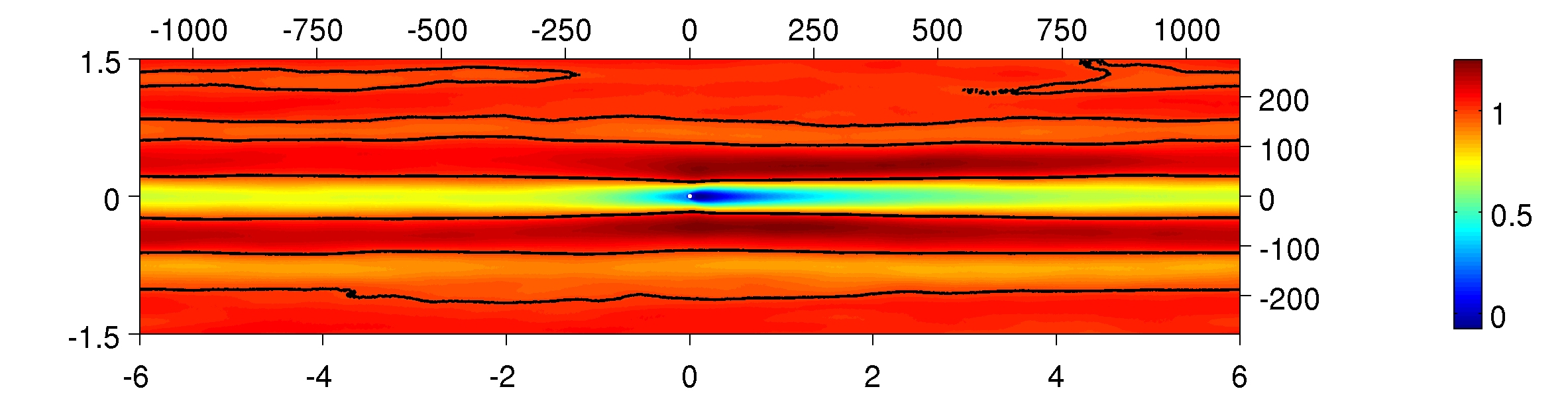}
        \\
        \centerline
        {\small $\tilde{x}/h$}
        \end{minipage}
        \hspace*{-10ex}
        \begin{minipage}{1ex}
        \rotatebox{90}{\small $\quad \tilde{z}^+$}
        \end{minipage}
        \caption{
          The average of the particle-conditioned
          relative streamwise velocity 
          $\tilde{u}_\mathrm{r}$ 
          (defined in equation
          \ref{equ-def-average-relative-velocity}), 
          normalized by $u_{lag}$ in the 
          near
          wall region. 
          The solid black lines correspond to a value of unity. 
        }
        \label{fig:ufield_conditioned2part_1}
\end{figure}
\begin{figure}
   \centering
        \begin{minipage}{2ex}
        \rotatebox{90}{\small 
         $\tilde{u}_\mathrm{r}/u_\mathrm{lag}$}
        \end{minipage}
        \begin{minipage}{.5\linewidth}
        \centerline{\small $(a)$}
        \centerline{\small $ \tilde{x}^+ $}
        \includegraphics[width=\linewidth]
        {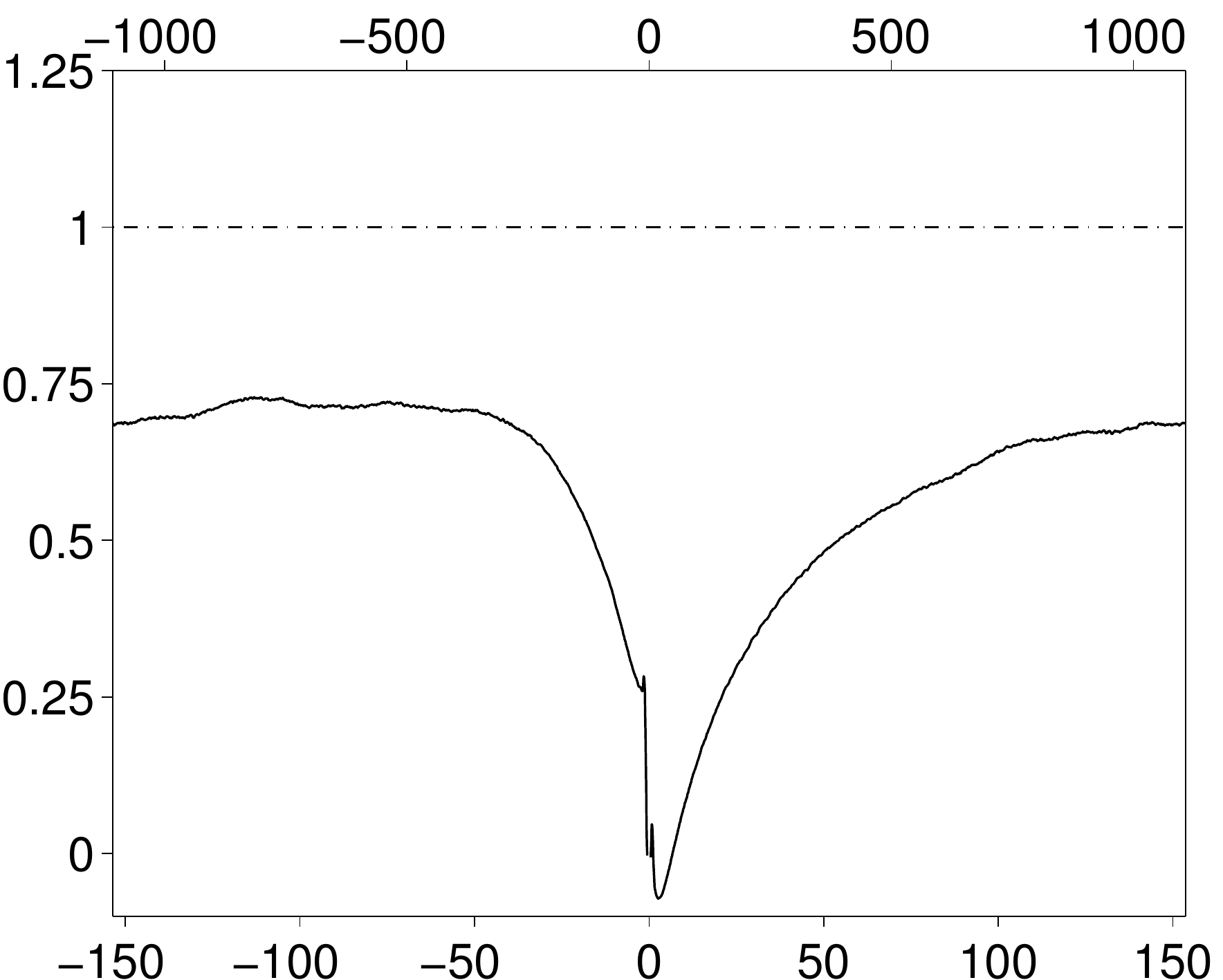}
        \hspace*{-.63\linewidth}
        \raisebox{.43\linewidth}{
          \includegraphics[width=.39\linewidth]
          {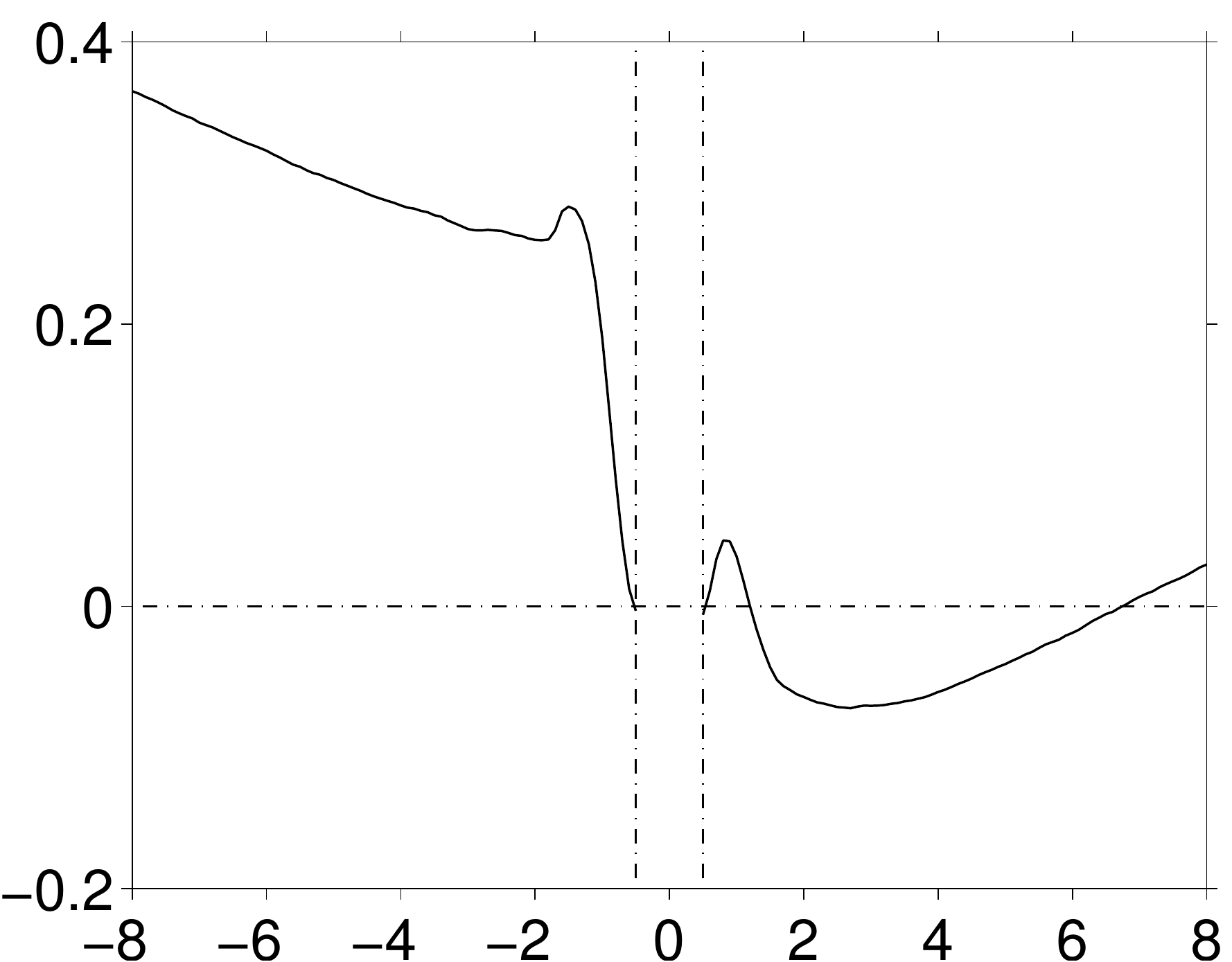}
        }
        \\
        \centerline
        {\small $ \tilde{x}/D $ }
        \end{minipage}\\[10pt]
        %
        \begin{minipage}{3ex}
        \rotatebox{90}
        {\small 
         $\tilde{u}_\mathrm{r}/u_\mathrm{lag}$}
        \end{minipage}
        \begin{minipage}{.5\linewidth}
        \centerline{\small $(b)$}
        \centerline{\small $ \tilde{z}^+ $}
        \includegraphics[width=\linewidth]
        {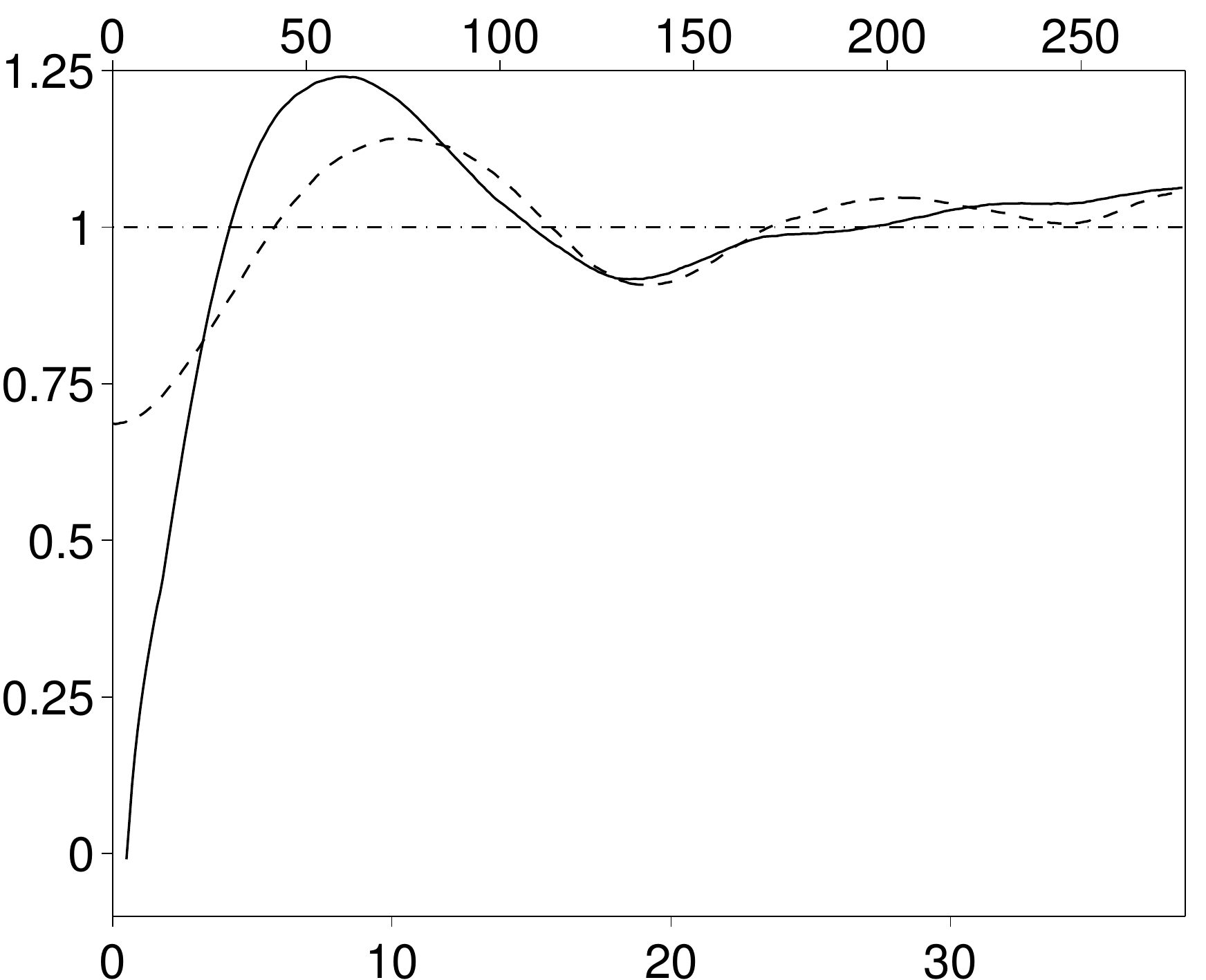}
        \\
        \centerline{\small $ \tilde{z}/D $}
        \end{minipage}  

        \caption{
          The same quantity $\tilde{u}_\mathrm{r}/u_\mathrm{lag}$ as in figure 
                 \ref{fig:ufield_conditioned2part_1}. 
                 (\textit{a}) shows a streamwise
                 profile 
                 through the particle center
                 ($\tilde{z}=0$), 
                 with the 
                 insert 
                 presenting a zoom around the particle
                 location. 
                 (\textit{b}) 
                 shows spanwise profiles through
                 the particle center (at $\tilde{x}=0$, solid line) and 
                 at $\tilde{x}=6h$ (dashed line). 
                 Note that the 
                 data in 
                 (\textit{b}) 
                 is symmetric with respect to $\tilde{z}=0$. 
                 %
                  }
        \label{fig:ufield_conditioned2part_2}
\end{figure}
%

\begin{figure}
   \centering
        \begin{minipage}{1ex}
           \rotatebox{90}{\hspace{5ex}\small$ z^+$}
        \end{minipage}
        \begin{minipage}{.7\linewidth}
        \includegraphics[width=\linewidth]
        {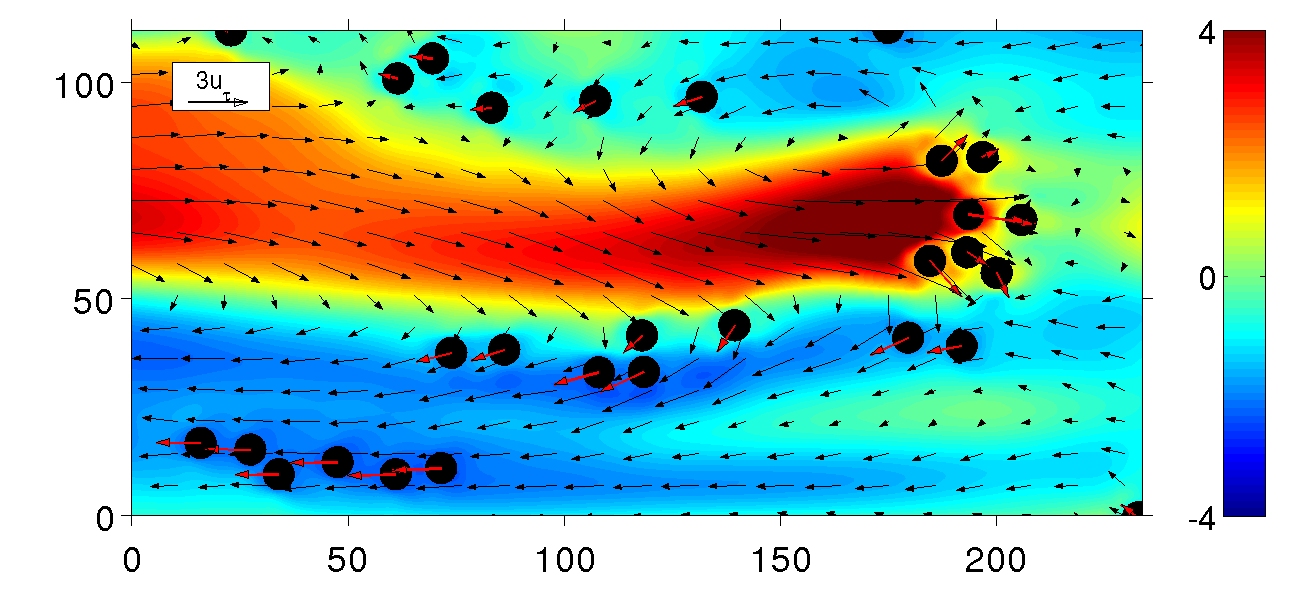}
        \\
        \centerline
        {\small $x^+$}
        \end{minipage}
        %
          \begin{minipage}{2ex}
            {\small$\displaystyle\frac{u_\mathrm{f}^\prime}{u_\tau}$}
          \end{minipage}
        %
        \caption{
          Instantaneous velocity fluctuations of the fluid and the
          particle phase in a wall-parallel plane at $y^+=5$ indicated
          by the black (fluid) and red (particles) arrows. 
          The colorplot represents the streamwise component
          of fluid velocity fluctuation, normalized by the friction
          velocity ($u_\mathrm{f}'/u_\tau$). 
          Note that backward pointing arrows do not correspond 
          to upstream motion.
          In fact, upstream particle motion has not been observed in
          the present case. 
          The graph shows only a small part of the entire plane, and
          the fluid velocity vectors have only been plotted at every 
          15th (10th) grid point in the $x$ ($z$) direction. 
        }
        \label{fig:velocity-vector-field-with-particles}
\end{figure}
\subsection{Particle-conditioned relative velocity field}
\label{subsec-particle-position-and-coherent-structures}
%
In this section we 
%
discuss the 
fluid 
velocity field conditioned 
upon 
the presence of
the particles.
This analysis provides a quantitative measure of the 
spatial 
distribution of the low- and high-speed fluid regions 
with respect to particle positions. 
%
First, we define 
the instantaneous relative velocity field with respect to the $l$th
particle as
\begin{equation}
\label{equ-def-relative-velocity}
\mathbf{u}^{(l)}_r(\tilde{\mathbf{x}}^{(l)},t) = 
    \mathbf{u}_f(\tilde{\mathbf{x}}^{(l)},t) - \mathbf{u}^{(l)}_p(t)
\end{equation}
where 
\begin{equation}
\label{equ-def-relative-position}
\tilde{\mathbf{x}}^{(l)}(t) = \mathbf{x} - \mathbf{x}^{(l)}_p(t)
\end{equation}
is an instantaneous coordinate relative to the particle's
center position.
The average of the relative velocity field
(with respect to the particle center)
over time
and over all particles located within a given wall-parallel slab 
$s$ is defined as follows:
\begin{equation}\label{equ-def-average-relative-velocity}
   \tilde{\mathbf{u}}_r(\tilde{\mathbf{x}},y^{(s)}) =
   \langle \mathbf{u}^{(l)}_r(\tilde{\mathbf{x}}^{(l)},t) \rangle_{p,t}
\end{equation}
where $\langle \cdot\rangle_{p,t}$ is the particle-centered 
field averaging operator defined in 
(\ref{equ-def-avg-operator-wake-time-particles-ybinned}).
%
We have carried out the averaging procedure  by considering 
all particles whose centers
are located at wall-normal distances within one particle diameter
from the bottom wall. That is, the considered slab has a wall normal
thickness of $\Delta h = D$ and its bottom edge coincides with 
the bottom wall.
Note that this averaging procedure is identical to the one
employed by \citet{garcia-villalba:2012} in a vertically
oriented particulate channel flow. 

If the particle distribution at the bottom were homogeneous, 
there would be no correlation between the spatial 
distribution of the coherent structures and the particle positions. 
In this case, the average of the 
particle-centered relative velocity component in the streamwise
direction 
$\tilde{u}_r(\tilde{\mathbf{x}},y^{(s)})$ 
is expected to tend to the value $u_{lag}(y^{(s)})$,
except in the immediate vicinity of the particle. 
Regardless of the particle distribution, the average relative 
velocity is zero at the fluid/solid interface $|\tilde{\mathbf{x}}| = D/2$, 
as a consequence of 
\revision{the the}{the} no-slip condition. 
%
If, however, particles are 
located at certain preferred regions with respect 
to the flow structures, or particle are considerably affecting the
flow field such that significant wakes exist behind them, then
$\tilde{u}_r$  is
expected to be significantly different from 
 $u_{lag}$ for 
appreciable regions of the averaging area, except at sufficiently
large distances 
where the averaged flow field
is expected to be uncorrelated with particle positions. 
We 
can rule
out the existence of significant
particle-induced wakes 
from the analysis in 
\S~\ref{subsubsec-mean-streamwise-relative-velocity}. 
%
Thus, locations where 
$\tilde{u}_r< u_{lag}$ correspond to regions of surrounding low-speed 
fluid and contrarily, locations where 
$\tilde{u}_r>u_{lag}$
correspond to high-speed fluid regions
with respect to particle positions.

Figure \ref{fig:ufield_conditioned2part_1} shows the  quantity 
$\tilde{u}_r /u_{lag}$. 
%
%
%
It is clearly seen
that on average particles are located in regions where 
$\tilde{u}_r/u_{lag}$ is below unity. 
The low-speed fluid region surrounding the particles on average
is found to be persistent in the streamwise direction, 
extending for more than 1000 wall units both upstream and downstream 
of the particle position. It reaches the extents of the averaging plane
(limited by the value of half the streamwise period of the 
computational domain $L_x/2$). 
Laterally however, the low-speed fluid region is seen to be correlated 
with the particle position for only up to about
$30$ wall units both in the positive and negative spanwise 
directions, beyond which $\tilde{u}_r/u_{lag}$ 
rises above unity. 
The general pattern in figure \ref{fig:ufield_conditioned2part_1}
is one of spanwise alternating low- and high-speed regions
correlated over the entire streamwise length of the domain.
Figure~\ref{fig:ufield_conditioned2part_2} shows
profiles of the same data on axes through the particle center 
both in the streamwise and spanwise directions. 
Note that 
here
the spanwise profile 
is averaged over the positive and negative directions exploiting 
the symmetry of the flow. At increasing streamwise separation 
$\tilde{x}$ from the reference particle
(figure~\ref{fig:ufield_conditioned2part_2}$a$), $\tilde{u}_r$ 
gradually increases from its smallest value 
(in the vicinity of the particle)
towards a finite value of approximately 
\revision{%
  $0.65u_{lag}$
  at a distance
  of approximately $\tilde{x} = 40D$ ($\tilde{x}^+ = 290$) in the upstream
  direction and approximately $\tilde{x} = 125D$ ($\tilde{x}^+ = 900$) 
  in the downstream direction. 
}{%
  $0.7\,u_{lag}$
  at a distance
  of approximately $\tilde{x} = 45\,D$ ($\tilde{x}^+ = 330$) in the upstream
  direction and approximately $\tilde{x} = 140\,D$ ($\tilde{x}^+ =1015$) 
  in the downstream direction. 
}
For larger distances, 
$\tilde{u}_r$ levels off at this value and does not approach unity
in the present computational domain.
%
  The 
  insert 
  in figure~\ref{fig:ufield_conditioned2part_2}$(a)$ reveals 
  that the average near-field within several particle diameters
  exhibits some additional structure. On the upstream side, a change
  of slope occurs at a distance of approximately $1.5D$ from the
  particle center, while the downstream side features a change of sign
  and an interval of negatively-valued $\tilde{u}_r$ up to a distance
  of roughly $7D$. 
  It should be noted that the asymmetry of the profile of the average
  relative velocity is not the signature of individual particle wakes,
  since the Reynolds number of the flow around the individual
  particles is much too low (cf.\
  \S~\ref{subsubsec-mean-streamwise-relative-velocity}) and the length
  scales of the asymmetric features are too large. 
  At this point it can only be stated that the observed shape of
  $\tilde{u}_r$ on the $\tilde{x}$-axis through the particle center 
  might be related to the preferential streamwise location of the
  particles with respect to the flow features of the
  near-wall layer.  
  As exemplified in one snapshot in 
  figure~\ref{fig:velocity-vector-field-with-particles}, high-speed
  particles are predominantly found at locations near the front of a 
  high-velocity fluid region, thereby contributing to the asymmetry of
  the streamwise profile of 
  $\tilde{u}_r$. 
  This point, however, merits further investigation in the future. 

In the spanwise direction
(figure~\ref{fig:ufield_conditioned2part_2}$b$), $\tilde{u}_r$
increases faster than in the streamwise direction, crossing the value
$\tilde{u}_r=u_{lag}$ for the first time
at a distance of approximately
\revision{%
  $\tilde{z} = 4.5D$ ($\tilde{z}^+ = 30$) 
}{%
  $\tilde{z} = 4.2D$ ($\tilde{z}^+ = 30$) 
}
indicating the spanwise
extent of the low-speed fluid regions. 
\revision{%
  $\tilde{u}_r$ attains a maximum value of $1.2u_{lag}$ at a distance of
  approximately   
  $\tilde{z} = 8.5D$ ($\tilde{z}^+ = 60$); 
}{%
  The particle-conditioned relative velocity $\tilde{u}_r$ attains a
  maximum value of $1.2u_{lag}$ at a distance of approximately   
  $\tilde{z} = 8.2D$ ($\tilde{z}^+ = 60$); 
}
%
this characterizes the lateral
distance at which the neighbouring high-speed fluid region 
is strongly correlated 
with the particle position. For larger spanwise distances 
$\tilde{z}$, $\tilde{u}_r/u_{lag}$ oscillates with a small
amplitude around unity reflecting the alternating nature of 
the low- and high-speed fluid regions.

%
It can be concluded that the 
spatial distribution of the coherent structures exhibits a strong 
correlation with particle positions almost throughout the 
entire streamwise and spanwise extent of the computational domain. 
A computational domain with 
dimensions much superior to those adopted in the present study
would be required if a full decorrelation were desired.

\subsection{Conditional average of the 
            local volumetric particle concentration}
\label{subsec-conditional-average-local-particle-concentration}
%
In \S~\ref{subsec-distribution-motion-of-particles} 
it has been quantitatively shown through Voronoi analysis that 
particles are inhomogeneously distributed and that they
tend to be streamwise aligned. Furthermore, in 
\S~\ref{subsec-particle-position-and-coherent-structures}, 
it has been shown that the most probable position of a near-wall
particle is inside a low-speed fluid region. 
Here, we investigate the spatial distribution of the local
volumetric particle concentration conditioned to particle
positions. This type of analysis provides additional information
on the spatial pattern of particle accumulations/voids. 

\begin{figure}
  \hspace*{.15\linewidth}
        \begin{minipage}[b]{2ex}
           {\small$(a)$}
        \end{minipage}
        \raisebox{-15ex}{
          \begin{minipage}[b]{1ex}
            \rotatebox{90}{\small$ \tilde{z}/D$}
          \end{minipage}
        }
        \begin{minipage}[t]{.7\linewidth}
          \centerline{\small \hspace{-5ex}$\tilde{x}^+$}
        \includegraphics[width=\linewidth,clip=true,
        viewport=105 1 1630 700]
        {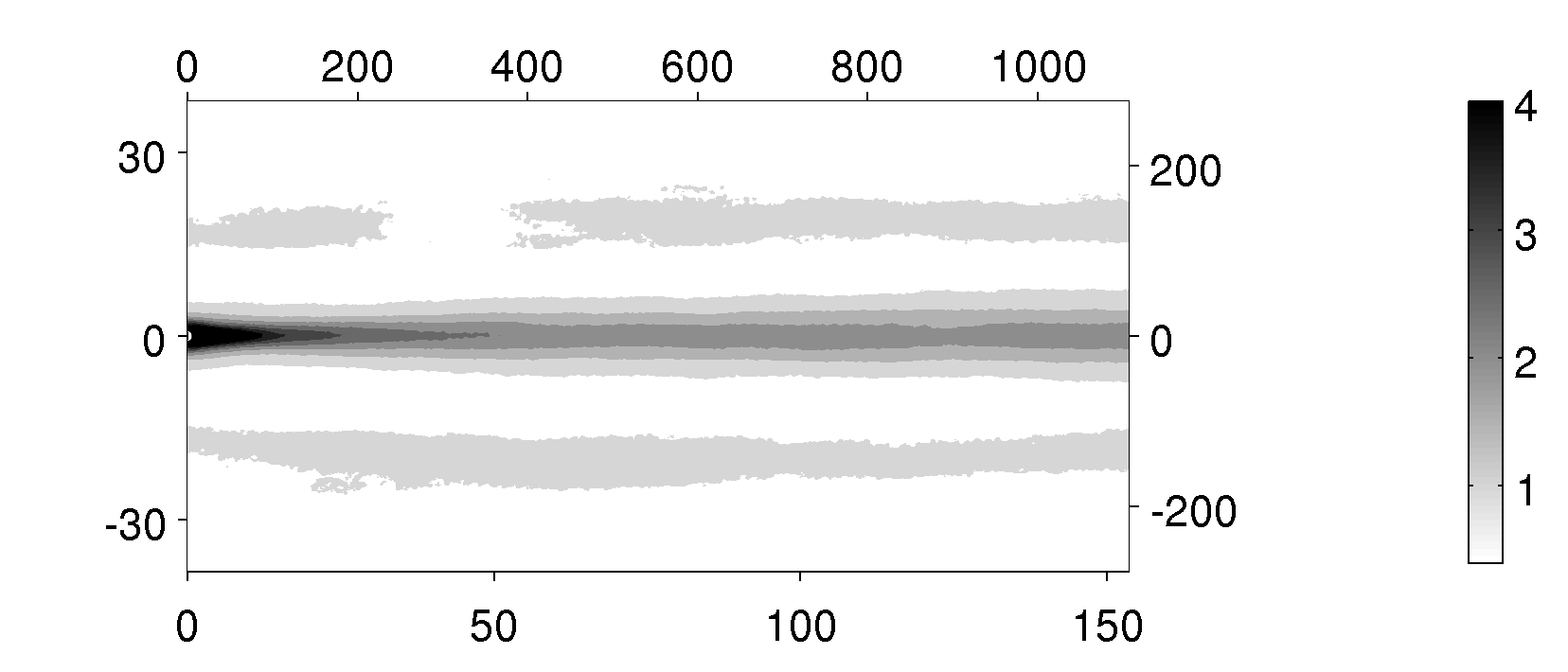}
        \\
        \centerline
        {\small\hspace{-5ex}$\tilde{x}/D$}
        \end{minipage}
        \hspace{-15ex}
        \raisebox{-15ex}{
          \begin{minipage}{1ex}
            \rotatebox{90}{\small $\tilde{z}^+$}
          \end{minipage}
        }
        \hspace{+10ex}
        \raisebox{-15ex}{
          \begin{minipage}{2ex}
            $\displaystyle \frac{\tilde{\psi}}{\langle \psi \rangle}$
          \end{minipage}
        }
        \\[3ex]
  \hspace*{.15\linewidth}
        \begin{minipage}[b]{2ex}
           {\small$(b)$}
        \end{minipage}
        \raisebox{-20ex}{
          \begin{minipage}[b]{1ex}
            \rotatebox{90}
            {\small $\tilde{\psi}/\langle \psi \rangle$}
          \end{minipage}
        }
        \begin{minipage}[t]{.5\linewidth}
          \centerline{\small $\tilde{x}^+,\tilde{z}^+$}
        \includegraphics[width=\linewidth]
        {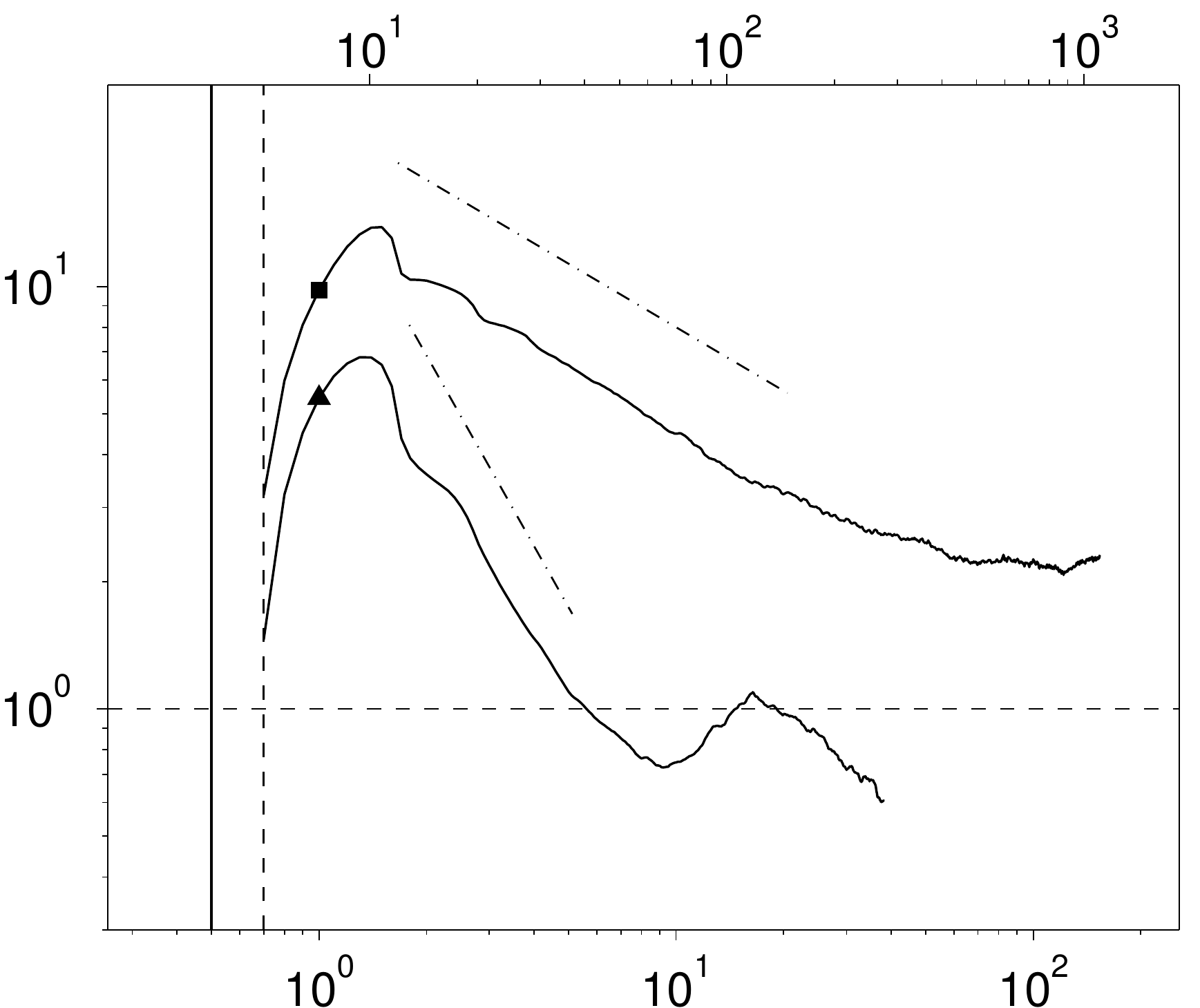}
        \\
        \centerline{\small $ \tilde{x}/D, \tilde{z}/D $}
        \end{minipage}
        \caption{ (\textit{a}) Contours of the
                 conditional local particle
                 concentration normalized by the mean solid volume
                 fraction ($\tilde{\psi}/\langle \psi \rangle$)
                 averaged over all particles which are located 
                 within the near-wall region ($y_\mathrm{p}<D$),
                 as defined in
                 (\ref{equ-conditionally_averaged_volume_fraction}).  
                 %
                 Only half of the averaging plane is shown 
                 due to symmetry. 
                 %
                 Contour 
                 levels
                 correspond to values 
                 (1:0.5:4). 
                 The graph in 
                 (\textit{b}) 
                 shows 
                 $\tilde{\psi}(\tilde{x},0,0,y^{(s)})/\langle \psi \rangle$ 
                 (square symbol) and  
                 $\tilde{\psi}(0,0,\tilde{z},y^{(s)})/\langle \psi \rangle$ 
                 (triangular symbol), both with $s=2$. 
                 The vertical solid line indicates the extent 
                 of the reference particle's surface,
                 and the vertical dashed line gives the distance 
                 $|\tilde{\mathbf{x}}| = D/2 + 2\Delta x$ 
                 from the center of the reference particle which marks
                 the limit of the range of action of the collisional
                 force. 
                 The chain-dotted lines indicate decay rates
                 proportional to 
                 $x^{-0.5}$ and $x^{-1.5}$.}
        \label{fig:spatial_probability}
\end{figure}
%
The indicator function of the solid phase, $\psi$, is simply related to
the fluid phase indicator function $\phi_\mathrm{f}$ defined in
(\ref{equ-def-fluid-indicator-fct}) by the relation 
\begin{equation}\label{local_volume_fraction}
\psi(\mathbf{x},t) = 1 - \phi_\mathrm{f}(\mathbf{x},t)\; .
\end{equation}
%
Its average over wall-parallel planes (using nodal values of the
discrete grid) and over time can be defined as
\begin{equation}\label{average_volume_fraction}
  \langle
  \psi
  \rangle(y_j)
  = 1 - \frac{n(y_j)}{N_\mathrm{t}^{\mathrm{(p)}}N_\mathrm{x}N_\mathrm{z}}
  \,,
\end{equation}
where $n(y_j)$ is the discrete counter given in 
(\ref{equ-def-sample-counter-plane-and-time-fluid-only}) with
$N_\mathrm{t}$ replaced by $N_\mathrm{t}^{\mathrm{(p)}}$ (the number
of available temporal records). 
The quantity defined in (\ref{average_volume_fraction}) is equivalent
to the average solid volume fraction at a given wall-normal location
$y_j$ of the discrete grid. Note that $\langle\psi\rangle(y_j)$ is 
alternative to the solid volume fraction
$\langle\phi_s\rangle(y^{(j)})$ defined in
(\ref{equ-def-avg-solid-volume-frac}) based on 
counting particles whose centers are located in wall-normal averaging
bins. In the limit of infinitesimally small particles and an infinitely
fine grid the difference vanishes. For a given finite grid and
particle size, however, the definition stated here (i.e.\ in
relation~\ref{average_volume_fraction}) yields smoother results,
especially when two-dimensional concentration fields are considered,
as will be done in the following.

Using the solid phase indicator function
(\ref{local_volume_fraction}), one can define an indicator field which
is relative to the location of the $l$th particle, i.e.\
$\psi^{(l)}(\tilde{\mathbf{x}}^{(l)},t)$, where the definition of the
relative coordinate given in (\ref{equ-def-relative-position}) is used. 
%
The average of this quantity over time and over all particles located
within a given wall-parallel slab $s$ corresponds to a
particle-conditioned field of the average solid volume fraction
relative to particle positions. 
%
%
The average is defined as follows
\begin{equation}
  \label{equ-conditionally_averaged_volume_fraction}
  \tilde{\psi}(\tilde{\mathbf{x}},y^{(s)}) 
  = 
  \langle 
  \psi^{(l)}(\tilde{\mathbf{x}}^{(l)},t)
  \rangle_{p,t}
  =
  1 - \frac{\tilde{n}^{(s)}(\tilde{\mathbf{x}})}
  {n_\mathrm{p}^{(s)}}
  \,,
\end{equation}
where $\tilde{n}^{(s)}(\tilde{\mathbf{x}})$ and $n_\mathrm{p}^{(s)}$ are the discrete counter
field and the sample counter given in
(\ref{equ-def-sample-counter}) and
(\ref{equ-def-sample-counter-binned-particles}) respectively 
(supposing that the temporal records in both definitions are
identical, i.e.\ $N_t^{(p)}=N_\mathrm{t}^\mathrm{(r)}$). 
For a finite number of samples ($n_\mathrm{p}^{(s)}>0$) we have by
definition that $\tilde{\psi}\in[0,1]$. 
%

The particle-conditioned average solid volume fraction $\tilde{\psi}$
can also be interpreted as the probability that a point
$\tilde{\mathbf{x}}$ (with respect to the center of a particle) is
located inside the solid phase domain $\Omega_\mathrm{p}$.  
Note that for monodispersed particles
$\tilde{\psi}$ is an even function, i.e.\ 
$\tilde{\psi}(\tilde{\mathbf{x}},y^{(s)}) = 
\tilde{\psi}(-\tilde{\mathbf{x}},y^{(s)})$.
For a random distribution of particles centered within
a given slab $s$, the
average local solid volume fraction is equal
to $\langle\psi\rangle(y_s)$ and has a uniform
distribution throughout the averaging volume, except for 
$|\tilde{\mathbf{x}}| \leq D/2$ (where $\tilde{\psi}=1$)
and in the immediate vicinity due to appreciable finite-size
effects. Thus
$\tilde{\psi}/\langle \psi \rangle$ is closely related to
the pairwise distribution function which gives 
a measure of the probability of 
finding a particle at a certain distance away from a given 
reference particle 
\citep[see e.g.][]{sundaram:1997,shotorban:2006, sardina:2012}. 

Figure \ref{fig:spatial_probability}\textit{a} shows
the spatial map of $\tilde{\psi}/\langle \psi \rangle$ 
for the immediate vicinity of the wall
(i.e.\ for all particles whose centers are located in the
region within $y_\mathrm{p} < D$).
The shaded part of the map shows the
regions where the conditional local particle
concentration is higher than that of a homogenous case
($\tilde{\psi}/\langle \psi \rangle > 1$). This means that
there is an increased probability of finding a particle in
this region when compared to that of a random particle
distribution. Contrarily,
the white region of the map corresponds to 
The figure clearly shows the 
elongated particle-conditioned preferred accumulation regions
which span the entire box in the streamwise direction
but which have only a small spanwise extent.
A repetitive pattern
of the regions is clearly observable in the spanwise direction
forming 
high conditional particle concentration
ridges and low conditional particle concentration troughs.

More quantitative information can be obtained by considering
profiles of $\tilde{\psi}$ through the particle
center both in the streamwise and spanwise directions, respectively
(figure \ref{fig:spatial_probability}\textit{b}).
Note that the reference particle's surface is marked by a 
vertical solid line in the figure,
and the distance $|\tilde{\mathbf{x}}| = D/2 + 2\Delta x$ 
below which the particle-particle repulsion force is active
(see \S~\ref{subsec-numerical-method}) is marked by a vertical 
dashed line. For increasing distances 
$|\tilde{\mathbf{x}}| > D/2 + 2\Delta x$, 
both in the streamwise
and spanwise directions, the quantity $\tilde{\psi}$ first rapidly
increases, takes a maximum value of 
\revision{%
  $14\langle \psi \rangle$ ($7\langle \psi \rangle$) 
}{%
  $13.8\langle \psi \rangle$ ($6.8\langle \psi \rangle$) 
}
in the streamwise (spanwise)
direction at a distance of approximately $1.3D$, and
then decays slowly. The decay depends strongly upon the coordinate
direction, reflecting the anisotropic nature of the pattern as seen in
figure \ref{fig:spatial_probability}\textit{a}.
The fact that
the highest values of $\tilde{\psi}$ are
located within the immediate vicinity of the reference 
particle is consistent with the results of the Voronoi area analysis,  
where we found that particles tend to locally cluster and 
result in much higher than average local particle concentration values.
Moreover, the anisotropic spatial 
distribution of the quantity $\tilde{\psi}$, i.e.\
the fact that the values of $\tilde{\psi}$ 
are larger in the streamwise direction than in the spanwise direction
for all values of $|\tilde{\mathbf{x}}|$, is again 
consistent with the picture deduced from the analysis of the
aspect ratio of Voronoi cells.
Furthermore, in the streamwise direction
the quantity $\tilde{\psi}$ persistently remains above the
average value $\langle \psi \rangle$ for extended streamwise 
distances exhibiting two regimes. First, a power law regime 
($\tilde{\psi} \propto x^{-\beta}$ with $\beta = 0.5$)
for up to a distance of about 
\revision{%
  $\tilde{x} = 50D$ ($\tilde{x}^+ = 370$), 
}{%
  $\tilde{x} = 50D$ ($\tilde{x}^+ = 370$), 
}
followed by a plateau regime extending for up to 
$\tilde{x} = L_\mathrm{x}/2$.
This result shows the large-scale nature of the particle 
clustering in the streamwise direction. 
Contrarily, the quantity $\tilde{\psi}$ decays
faster in the spanwise direction (with decay rate $\beta \approx 1.5$), 
first crosses the value $\tilde{\psi} = \langle \psi \rangle$ at
a lateral distace of approximately 
\revision{%
  $\tilde{z} = 5.4D$ ($\tilde{z}^+ \approx 40$) and attains a minimum
  at a distance of approximately $\tilde{z} = 9D$ ($\tilde{z}^+ \approx
  65$).
}{%
  $\tilde{z} = 5.6D$ ($\tilde{z}^+ \approx 40$) and attains a minimum
  at a distance of approximately $\tilde{z} = 9D$ ($\tilde{z}^+ \approx
  65$).
}
It slowly increases and attains a second local maximum which is slightly
above the value $\tilde{\psi} = \langle \psi \rangle$ at a distance of
roughly 
\revision{%
  $\tilde{z}^+ \approx 170$. 
}{%
  $\tilde{z}^+ \approx 120$. 
}
The existence of multiple local
maxima is a reflection of the 
ridge-trough pattern visible
in figure \ref{fig:spatial_probability}\textit{a}.

The decay rate 
of the radial distribution function with increasing
distance from the reference particle has been
previously taken as a measure of the small-scale clustering
intensity of point-particles: the higher the value of the 
decay rate the stronger is the small-scale clustering
\citep[see e.g.][]{sardina:2012,gualtieri:2009}. Incidentaly, 
the decay rate of $\tilde{\psi}$ in the streamwise direction is
identical to that of the radial distribution function
of point-particles in a channel flow configuration 
(without gravity), at a comparable value of the 
Stokes number \citep{sardina:2012}

\begin{figure}
   \centering
        %
        \begin{minipage}{3ex}
          \rotatebox{90}{\small $\quad \tilde{y}^+$}
        \end{minipage}
        \begin{minipage}{.5\linewidth}
          \centerline{\small $(a)$}
          \includegraphics[width=\linewidth]
          {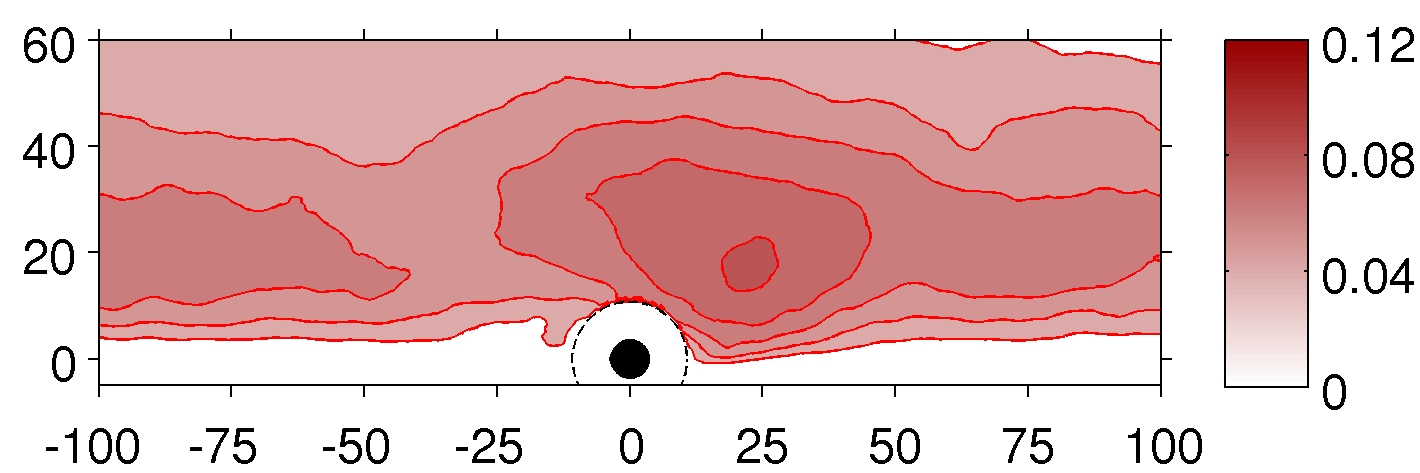}
          \\[-1.5ex]
          \includegraphics[width=\linewidth]
          {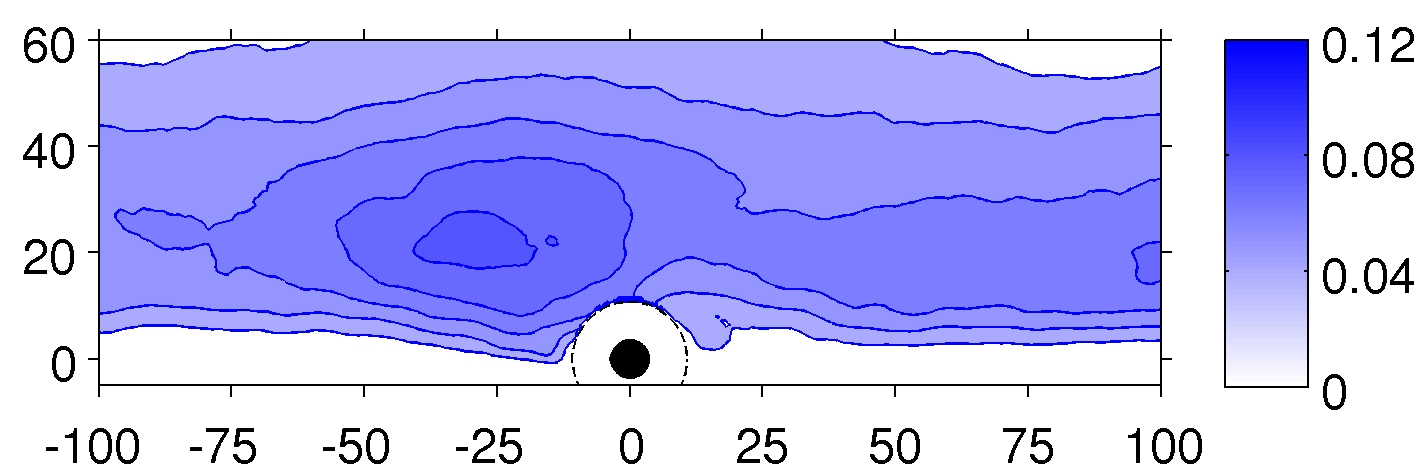}
          \centerline
          {\small $ \tilde{z}^+ $ }
        \end{minipage}\\[2ex]
        %
        \begin{minipage}{3ex}
          \rotatebox{90}{\small $\quad \tilde{y}^+$}
        \end{minipage}
        \begin{minipage}{.5\linewidth}
          \centerline{\small $(b)$}
          \includegraphics[width=\linewidth]
          {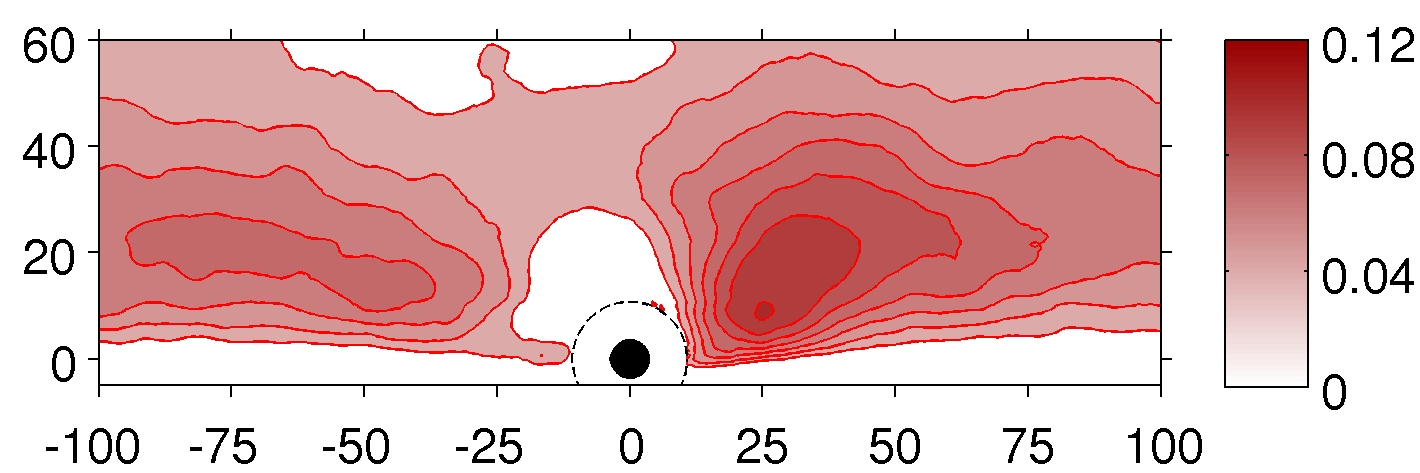}
          \\[-1.5ex]
          \includegraphics[width=\linewidth]
          {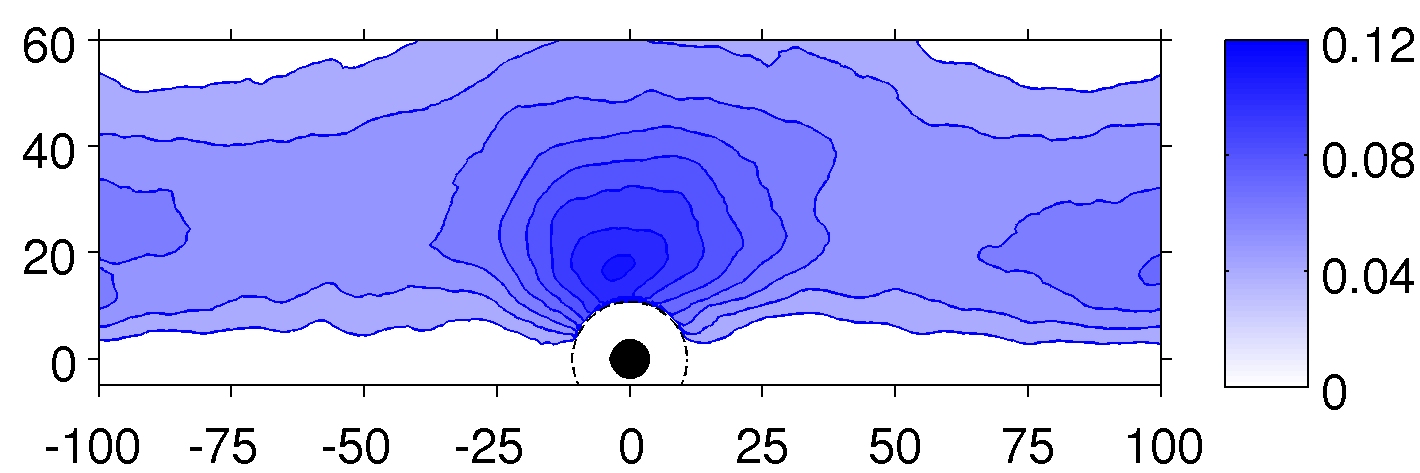}
          \centerline
          {\small $ \tilde{z}^+ $ }
        \end{minipage}\\[2ex]
        %
        \begin{minipage}{3ex}
          \rotatebox{90}{\small $\quad \tilde{y}^+$}
        \end{minipage}
        \begin{minipage}{.5\linewidth}
          \centerline{\small $(c)$}
          \includegraphics[width=\linewidth]
          {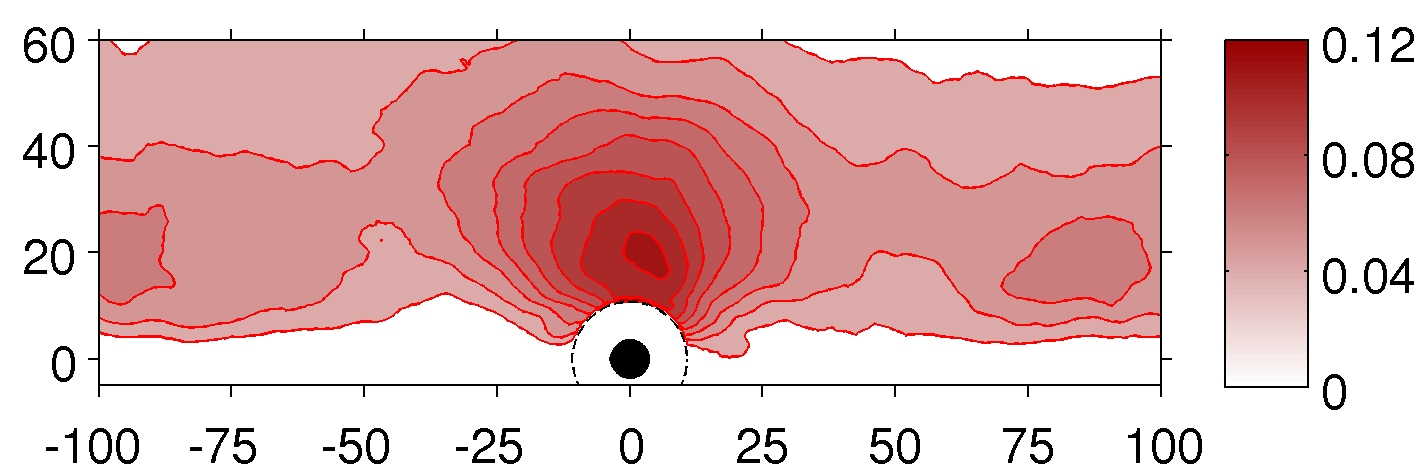}
          \\[-1.5ex]
          \includegraphics[width=\linewidth]
          {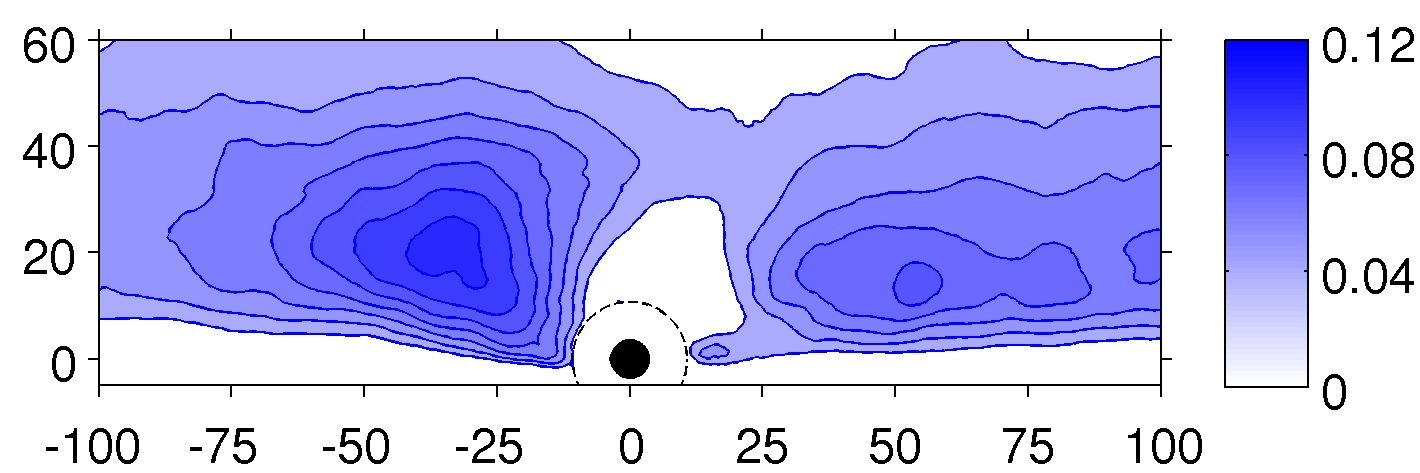}
          \centerline
          {\small $ \tilde{z}^+ $ }
        \end{minipage}
        \caption{
          The spatial distribution 
          of clockwise and counterclockwise rotating vortical regions 
          with respect to the position of a particle in a
          cross-sectional plane passing through the particle 
          center. 
          The vortical regions are educed from
          $\tilde{\lambda}_2^\mathrm{\omega_x+}$ (red-colored
          contours) and $\tilde{\lambda}_2^\mathrm{\omega_x-}$
          (blue-colored contours) as defined in (\ref{equ-def-averaged-vortices}). 
          (\textit{a}) shows the 
          probability distribution conditioned to 
          all particles  whose centers are located within one particle 
          diameter from the bottom wall.
          (\textit{b}) shows the same quantity 
          as in (\textit{a}) but additionally selecting only particles
          with positive 
          spanwise velocity $w_\mathrm{p} > 0$.
          (\textit{c}) shows the same quantity as in 
          (\textit{a}) but  additionally selecting only particles
          with negative spanwise 
          velocity $w_\mathrm{p} < 0$.
          Note that the positive $x$-direction points into the paper. 
        }
        \label{fig:conditional_lambda2_field}
\end{figure}
\subsection{The role of the streamwise vortices
            on the distribution of particles}
\label{subsec-role-of-streamwise-vortices-on-distribution-of-particles}
Visual and quantitative evidence provided so far shows that, 
consistent with previous experimental findings, 
particles are inhomogeneously distributed at the bottom wall.
This results on the one hand in regions of 
high 
particle concentrations where particles are observed to form 
streamwise elongated clusters 
and on the other hand in void regions of very low particle 
concentration. It has been shown that particles tend to avoid
the high-speed fluid regions and are on average
residing in the low-speed fluid regions.
As mentioned in the introduction, the behaviour of 
particle segregation at the bottom wall
is 
believed 
to be linked to the dynamics of the
quasi-streamwise vortices in the near-wall region. 
These coherent vortical structures are
well known to dominate the buffer region 
in boundary-layer-type flows and are responsible for most 
of the turbulence activity therein 
\citep{robinson:1991,jimenez:1999,adrian:2000,schoppa:2000,adrian:2007}.

With the aim to 
further analyze the role of the streamwise
vortices with respect to particle segregation,
we determine the correlation between 
particle positions and the position of vortical regions in their vicinity. 
For the identification of vortical regions we have adopted the 
definition of vortical structures proposed by 
\citet[][]{jeong:1995}. A vortex is defined
as a region where the second largest eigenvalue 
$\lambda_2$ of the tensor 
$\mathbf{S}^2 + \boldsymbol{\Omega}^2$ 
is negative, $\mathbf{S}$ and $\boldsymbol{\Omega}$ being the symmetric
and antisymmetric parts of the fluid velocity gradient tensor
$\nabla \mathbf{u}_\mathrm{f}$.
It is well known that in the buffer region of wall-bounded flows,
where the quasi-streamwise
vortices are dominant, the streamwise vorticity field
$\omega_\mathrm{f,x} = \partial_\mathrm{y}w_\mathrm{f} - \partial_\mathrm{z}v_\mathrm{f}$ 
is highly correlated with the
vortical regions, 
\citep[][]{kim:1987}. 
Here the sign of $\omega_\mathrm{f,x}$
(positive/negative) is used to determine the sense of rotation 
(clockwise/counterclockwise) of the vortices educed by means of $\lambda_2$.
Therefore, we define two real-valued fields 
$\lambda_2^\mathrm{\omega_x+}$ and $\lambda_2^\mathrm{\omega_x-}$  which indicate
whether a given point lies inside a vortical region with, respectively, clockwise 
or counterclockwise rotation around the streamwise axis, having the amplitude 
$|\lambda_2(\mathbf{x},t)|$, viz.  
\begin{subequations}   \label{eq:lambda2-indicator}
  \begin{eqnarray}
    \lambda_2^\mathrm{\omega_x+}(\mathbf{x},t) = 
    \left\{
      \begin{array}{ll}
        \phi_\mathrm{f}(\mathbf{x},t)|\lambda_2(\mathbf{x},t)| &
        \mbox{if}\, \lambda_2(\mathbf{x},t) < 0,\;
        \omega_\mathrm{f,x}(\mathbf{x},t) > 0\\
        0 & \mbox{else}\;,\\
      \end{array}\right.
    \label{eq:positive-lambda2-indicator}\\[2ex]
    \lambda_2^\mathrm{\omega_x-}(\mathbf{x},t) = 
    \left\{
      \begin{array}{ll}
        \phi_\mathrm{f}(\mathbf{x},t)|\lambda_2(\mathbf{x},t)| &
        \mbox{if}\, \lambda_2(\mathbf{x},t) < 0,\;,
        \omega_\mathrm{f,x}(\mathbf{x},t) < 0 \\
        0 & \mbox{else}\;.
      \end{array}\right.
    \label{eq:negative-lambda2-indicator}
  \end{eqnarray}
\end{subequations}
%
\begin{figure}
   \centering
        %
        \begin{minipage}{3ex}
          \rotatebox{90}{\small $\quad y^+$}
        \end{minipage}
        \begin{minipage}{.7\linewidth}
          \includegraphics[width=\linewidth]
          {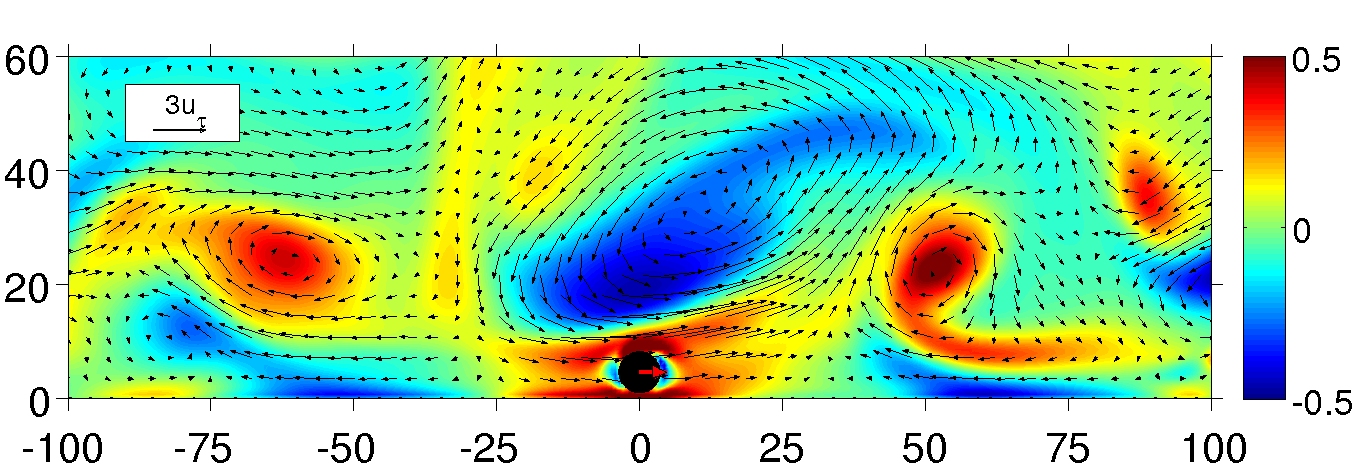}
          \centerline
          {\small $ \tilde{z}^+ $ }
        \end{minipage}
        \caption{%
          An instantaneous fluid velocity field (indicated by black
          arrows) around a particle in spanwise motion (red arrow) in
          a cross-sectional plane. 
          The background color indicates the streamwise vorticity
          normalized in wall units, i.e.\ $\omega_\mathrm{f,x}^+$.
          The graph shows only a small part of the entire plane, and
          the fluid velocity vectors have only been plotted at every 
          5th grid point in both coordinate directions.  
        }
        \label{fig:cross-stream-velocity-vector-field-with-a-particle}
\end{figure}
The presence of the fluid indicator function $\phi_\mathrm{f}$ in the above definition 
guarantees that only points in the fluid region are selected. Note
that we have additionally excluded a spherical region with radius $3R$ around each
particle center in order to consider only the part of the fluid domain
$\Omega_\mathrm{f}(t)$ which is not directly affected by the particle's own near-field 
(cf.\ \S~\ref{subsec-fluid-seen-by-particles}). 
Analogous to the notation in previous sections, the 
fields defined in (\ref{eq:lambda2-indicator}) conditioned upon the $l$th
particle are denoted as 
$\lambda_2^{\mathrm{\omega_x+},(l)}(\tilde{\mathbf{x}}^{(l)},t)$ and
$\lambda_2^{\mathrm{\omega_x-},(l)}(\tilde{\mathbf{x}}^{(l)},t)$. 
The average of the quantities $\lambda_2^{\mathrm{\omega_x+},(l)}$ and
$\lambda_2^{\mathrm{\omega_x-},(l)}$ over all particles and time 
(as in \ref{equ-def-average-relative-velocity})
is denoted as
\begin{subequations} \label{equ-def-averaged-vortices}
\begin{eqnarray}
   \tilde{\lambda}_2^\mathrm{\omega_x+}(\tilde{\mathbf{x}},y^{(s)}) =
   \langle \lambda_2^{\mathrm{\omega_x+},(l)} (\tilde{\mathbf{x}}^{(l)},t) \rangle_\mathrm{p,t}\;,
   \label{equ-def-averaged-clockwise-vortices}\\
   \tilde{\lambda}_2^\mathrm{\omega_x-}(\tilde{\mathbf{x}},y^{(s)}) =
   \langle \lambda_2^{\mathrm{\omega_x-},(l)} (\tilde{\mathbf{x}}^{(l)},t) \rangle_\mathrm{p,t}\;.
   \label{equ-def-averaged-counterclockwise-vortices}
\end{eqnarray}
\end{subequations}
The fields $\tilde{\lambda}_2^\mathrm{\omega_x+}$ and
$\tilde{\lambda}_2^\mathrm{\omega_x-}$ correspond to 
averages of 
the magnitude of
(negative-valued) $\lambda_2$, 
conditioned to particle centers located in the wall-normal slab $s$, 
further distinguishing between positive and negative streamwise
rotation of the fluid. 
Therefore, they give an indication of the intensity of vortical
structures (with the respective sense of streamwise rotation) 
on average found at a given position with respect to a particle.

Figure~\ref{fig:conditional_lambda2_field}$(a)$ shows the two fields
defined in (\ref{equ-def-averaged-vortices}) in a cross-sectional 
plane through the center of the reference particle, selecting by the
choice of the slab $s$ those particles with $y_\mathrm{p}\leq D$. 
The two graphs clearly show that those particles are on average found at
spanwise positions in between strong vortical regions of opposite sign. 
The signatures of these prominent counter-rotating vortical regions
visible in figure~\ref{fig:conditional_lambda2_field}$(a)$ are located
at wall-normal distances of roughly $\tilde{y}^+=20$ 
to $30$
measured 
from the reference particle center. 
The sense of streamwise rotation of the two intense vortical regions
is such that a 
region of low streamwise velocity 
is induced at around the particle
location, 
in agreement 
with the results of
\S~\ref{subsec-particle-position-and-coherent-structures}. 
This result is also consistent with the common occurrence of
streamwise vortices flanking low-speed velocity streaks 
with a positive or negative spanwise shift according to their sign. 

In order to analyze the correlation 
between 
lateral particle motion 
and the presence of 
coherent structures, we have further conditioned the fields defined in
(\ref{eq:lambda2-indicator}) with respect to the sign of the spanwise
particle velocity $w_\mathrm{p}$. 
The result is shown in figure~\ref{fig:conditional_lambda2_field}$(b)$
and \ref{fig:conditional_lambda2_field}$(c)$ for positive and negative
values of 
$w_\mathrm{p}$, respectively. 
The first observation from these figures is that the condition on
the sign of spanwise particle velocity is statistically relevant since
the resulting fields are clearly distinct from those in
figure~\ref{fig:conditional_lambda2_field}$(a)$. 
Most prominently, we find regions of large values of
$\tilde{\lambda}_2^\mathrm{\omega_x-}$ ($\tilde{\lambda}_2^\mathrm{\omega_x+}$) 
directly above particles for which $w_\mathrm{p}>0$ ($w_\mathrm{p}<0$). 
The oppositely signed fields ($\tilde{\lambda}_2^\mathrm{\omega_x+}$ 
in case of 
$w_\mathrm{p}>0$, and $\tilde{\lambda}_2^\mathrm{\omega_x-}$ for $w_\mathrm{p}<0$) do not
exhibit large values in the direct vicinity of the reference particle.  
%
These results confirm that the spanwise motion of near-wall particles
is statistically correlated with coherent vortices found in the buffer
layer. 
The data suggests that the lateral particle migration towards the
low-speed region is indeed a result of the velocity field induced by
the quasi-streamwise vortices. 

An example of a particle located below a counter-clockwise rotating
vortex and migrating in the positive spanwise direction can be found
in figure~\ref{fig:cross-stream-velocity-vector-field-with-a-particle}. 
%
This particle is immersed in the boundary layer induced by the
streamwise vortex immediately located above the particle (around 
$\tilde{y}^+\approx20$, $\tilde{z}^+\approx0$).
%
The direction of the depicted particle's spanwise motion is clearly
towards the location where an ejection of low-momentum fluid takes
place at that instant ($\tilde{z}^+\approx30$ in the figure), i.e.\
a low-velocity region. 
The scenario which is captured in 
figure~\ref{fig:cross-stream-velocity-vector-field-with-a-particle}
is representative of many events involving near-bottom particles and
quasi-streamwise vortices which we have observed in our data-base. 

\section{Conclusion}
\label{sec-conclusion}
  In the present study we have investigated the motion of heavy
  spherical particles in horizontal open channel flow by means of
  DNS. 
  The particles are larger than the viscous length scale of the
  near-wall region, requiring full resolution of the flow in their
  near-field. 
  The gravitational acceleration was chosen such that the overwhelming
  part of the particles is residing near the wall-plane. 
  Since the global solid volume fraction in the considered case is
  very low ($5\cdot10^{-4}$), the maximum of the average particle
  concentration obtained near the wall is still comparatively low. 
  %
  The particle-to-fluid density ratio is moderate, leading to
  a value of the Stokes number based on the viscous time scale which
  is larger than (but of the same order as) unity. 

  We have found that the basic statistics of the present fluid
  velocity field are essentially the same as those of single-phase
  flow at the same Reynolds number. The two-point correlation of the
  fluid velocity field, however, is somewhat modified, exhibiting
  slightly larger correlation lengths in the near-wall region where
  the bulk of the particles are located. 

  Concerning the dispersed phase, the mean particle velocity 
  is found to be smaller than the mean fluid velocity at all
  wall distances. 
  In previous investigations \citep{kaftori:1995b,kiger:2002} 
  an explanation of this apparent velocity lag based upon preferential
  particle concentration has already been proposed. 
  However, its confirmation hinges on technical
  points such as the definition of a relative inter-phase velocity and
  the performance of particle-conditioned averaging. 
  Therefore, we have investigated the origin of this apparent velocity
  lag in more detail. 
  For this purpose we have first proposed a method to
  determine a characteristic fluid velocity in the vicinity of a
  particle (i.e.\ the fluid velocity ``seen'' by the
  particle) for the case of finite particle sizes. Our definition is
  based on an average of the fluid velocity over a spherical surface
  (with radius $R_s$) centered at the particle center. 
  The validity of this definition was tested in the case of a single
  fixed sphere in uniform flow which also allowed us to calibrate the
  value of $R_s$, henceforth set to three times the particle radius. 
  Subsequently, our definition of fluid velocity ``seen'' by the
  particles was applied to the present horizontal channel flow
  data. It was found that the two phases are instantaneously much 
  closer to equilibrium than indicated by the apparent velocity lag,
  as already observed by previous authors in similar cases
  \citep{kaftori:1995b,kiger:2002}, albeit using less precise 
  measures of the relative velocity.  
  
  %
  We have analyzed the spatial distribution of near-wall particles by
  means of Voronoi tesselation in a wall-parallel plane. 
  The analysis shows that particles near the
  wall are strongly accumulating into streamwise elongated structures.  
  From the particle conditioned local volumetric concentration field
  (closely related to the pairwise distribution function) 
  we deduce that these accumulation regions are of a very large scale,
  with their streamwise extent of the order of the current domain
  size and a spanwise repeating pattern over distances of
  approximately 
  \revision{%
    $170$ wall units. 
  }{%
    $120$ wall units. 
  }
  The auto-correlation of Voronoi cell areas was found to decay 
  very slowly (Taylor microscale of 
  \revision{%
    $5.7$
  }{%
    $5.1$
  }
  bulk time units), which
  indicates that the particle accumulation regions are extremely
  stable in time. In fact, their time scale is much larger than the
  time scale deduced from the auto-correlation of particle velocities 
  (the Taylor microscale of the streamwise component measures $1.4$
  bulk time units).  

  %
  %
  %
  Furthermore, we have computed the particle-conditioned fluid
  velocity field (for near-wall particles) in a coordinate system
  relative to the particle centers. The average relative velocity is
  found to exhibit spanwise alternating ridges and troughs
  (approximately $50$ wall units apart) which are also essentially
  spanning the entire streamwise domain size. 
  The principal minimum of the average relative velocity is centered
  upon the reference particle, confirming previous observations that
  the particles are preferentially residing 
  inside the low-speed 
  streaks of the buffer layer.

  %
  Finally, we have performed a vortex eduction study \citep[based upon the
  $\lambda_2$ criterion of][]{jeong:1995} with the purpose of elucidating
  the role of coherent structures upon the spanwise motion of
  particles located near the wall.  
  The particle-conditioned average of the negative part of the
  $\lambda_2$ field (distinguishing between positive and
  negative values of the streamwise vorticity) clearly shows that
  near-wall particles are preferentially located at spanwise positions
  in between and beneath counter-rotating vortical regions, their sense
  of rotation being consistent with the quasi-streamwise vortices
  which are known to flank the typical low-speed streaks in the buffer
  region. 
  When the above particle-conditioned $\lambda_2$ field is
  additionally conditioned upon the spanwise particle velocity, 
  it is observed that the direction of spanwise particle motion is
  strongly correlated with the presence of a strong vortical region
  found directly above the reference particle location.
  The sign of the streamwise vorticity associated with the vortical
  region found on average above the particles is such that the induced
  velocity at the particle location is consistent with the sign of the
  spanwise particle velocity. 
  This result suggests that the spanwise particle motion near the wall
  is indeed caused by the quasi-streamwise vortices in the buffer layer. 

  The present study opens several new perspectives. 
  The proposed method to determine a fluid velocity ``seen'' by the
  particles appears potentially suitable for the investigation of
  fluid-particle interaction in 
  a wide range of
  flow configurations. 
  It has the advantage of being free from an a priori bias towards a
  specific spatial direction. 
  At the same time the definition can be readily evaluated with the
  aid of flow data generated by means of interface-resolved direct
  numerical simulation, or, possibly, 
  data obtained through 
  high-resolution volumetric
  measurements at the particle scale. 

  Concerning the interaction of near-wall turbulence and particles in
  the present case, the dynamical mechanism leading to the formation
  (and sustenance) of particle accumulation regions in low-speed
  streaks merits further investigation. 
  In order to provide a ``cleaner'' and more tractable approach to 
  the numerical simulation of such interaction mechanisms it appears 
  beneficial to consider instead of full turbulence in relatively
  large domains a smaller ``laboratory flow''. 
  This can be achieved either by turning to turbulent flow in smaller
  domain sizes, i.e.\ the minimal flow unit of \citet{jimenez:91}, 
  or, alternatively, to invariant solutions of the Navier-Stokes
  equations \citep{kawahara:12a}.  
  These alternatives will be considered in future studies. 

The data presented in this work can be found at the following URL: 
\url{http://www.ifh.kit.edu/dns_data}. 
\section*{Acknowledgments}
The simulations were performed at SCC Karlsruhe.
The computer resources, technical expertise and assistance provided by
the staff are thankfully acknowledged.
AGK has received financial support through a FYS grant from KIT within
the framework of the German Excellence Initiative.
Support from the German Research Foundation (DFG) through grants 
JI~18/19-1 and UH~242/1-1 is gratefully acknowledged. 
Thanks is also due to Manuel Garc\'ia-Villalba for many
fruitful discussions.

%
\addcontentsline{toc}{section}{Bibliography}
%

\begin{thebibliography}{56}
\providecommand{\natexlab}[1]{#1}
\providecommand{\url}[1]{\texttt{#1}}
\expandafter\ifx\csname urlstyle\endcsname\relax
  \providecommand{\doi}[1]{doi: #1}\else
  \providecommand{\doi}{doi: \begingroup \urlstyle{rm}\Url}\fi

\bibitem[Sumer and O\u{g}uz(1978)]{sumer:1978}
B.~Mutlu Sumer and Beyhan O\u{g}uz.
\newblock Particle motions near the bottom in turbulent flow in an open
  channel.
\newblock \emph{J.~Fluid Mech.}, 86:\penalty0 109--127, 1978.

\bibitem[Yung et~al.(1989)Yung, Merry, and Bott]{yung:1989}
B.~P.~K. Yung, H.~Merry, and T.~R. Bott.
\newblock The role of turbulent bursts in particle re-entrainment in aqueous
  systems.
\newblock \emph{Chem.~Eng.~Sci.}, 44:\penalty0 873--882, 1989.

\bibitem[Rashidi et~al.(1990)Rashidi, Hetsroni, and Banerjee]{rashidi:1990}
M.~Rashidi, G.~Hetsroni, and S.~Banerjee.
\newblock Particle-turbulence interaction in a boundary layer.
\newblock \emph{Int.~J.~Multiphase Flow}, 16:\penalty0 935 -- 949, 1990.

\bibitem[Kaftori et~al.(1995{\natexlab{a}})Kaftori, Hetsroni, and
  Banerjee]{kaftori:1995a}
D.~Kaftori, G.~Hetsroni, and S.~Banerjee.
\newblock Particle behaviour in the turbulent boundary layer. {I}. {M}otion,
  deposition, and entrainment.
\newblock \emph{Phys.~Fluids}, 7\penalty0 (5):\penalty0 1095 -- 1107,
  1995{\natexlab{a}}.

\bibitem[Kaftori et~al.(1995{\natexlab{b}})Kaftori, Hetsroni, and
  Banerjee]{kaftori:1995b}
D.~Kaftori, G.~Hetsroni, and S.~Banerjee.
\newblock Particle behaviour in the turbulent boundary layer. {II}. {V}elocity
  and distribution profiles.
\newblock \emph{Phys.~Fluids}, 7\penalty0 (5):\penalty0 1107 -- 1121,
  1995{\natexlab{b}}.

\bibitem[Ni{\~{n}}o and Garc\'ia(1996)]{nino:1996}
Y.~Ni{\~{n}}o and M.~H. Garc\'ia.
\newblock Experiments on particle-turbulence interactions in the near-wall
  region of an open channel flow: implications for sediment transport.
\newblock \emph{J.~Fluid Mech.}, 326:\penalty0 285--319, 1996.

\bibitem[Kiger and Pan(2002)]{kiger:2002}
K~T Kiger and C~Pan.
\newblock Suspension and turbulence modification effects of solid particulates
  on a horizontal turbulent channel flow.
\newblock \emph{J.~Turb.}, 3\penalty0 (019):\penalty0 1--17, 2002.

\bibitem[Righetti and Romano(2004)]{righetti:2004}
M.~Righetti and G.~P. Romano.
\newblock Particle-fluid interactions in a plane near-wall turbulent flow.
\newblock \emph{J.~Fluid Mech.}, 505:\penalty0 93 -- 121, 2004.

\bibitem[Hetsroni and Rozenblit(1994)]{hetsroni:1994}
G.~Hetsroni and R.~Rozenblit.
\newblock Heat transfer to a liquid-solid mixture in a flume.
\newblock \emph{Int.~J.~Multiphase Flow}, 20:\penalty0 671--689, 1994.

\bibitem[Rouson and Eaton(2001)]{rouson:2001}
Damian W.~I. Rouson and John~K. Eaton.
\newblock On the preferential concentration of solid particles in turbulent
  channel flow.
\newblock \emph{J.~Fluid Mech.}, 428:\penalty0 149--169, 2001.

\bibitem[Marchioli and Soldati(2002)]{marchioli:2002}
Cristian Marchioli and Alfredo Soldati.
\newblock Mechanisms for particle transfer and segregation in a turbulent
  boundary layer.
\newblock \emph{J.~Fluid Mech.}, 468:\penalty0 283 -- 315, 2002.

\bibitem[Narayanan et~al.(2003)Narayanan, Lakehal, Botto, and
  Soldati]{narayanan:2003}
Chidambaram Narayanan, Djamel Lakehal, Lorenzo Botto, and Alfredo Soldati.
\newblock Mechanisms of particle deposition in a fully developed turbulent open
  channel flow.
\newblock \emph{Phys. Fluids}, 15:\penalty0 1545473, 2003.

\bibitem[Picciotto et~al.(2005)Picciotto, Marchioli, and
  Soldati]{picciotto:2005}
Maurizio Picciotto, Cristian Marchioli, and Alfredo Soldati.
\newblock Characterization of near-wall accumulation regions for inertial
  particles in turbulent boundary layers.
\newblock \emph{Phys. Fluids}, 17:\penalty0 098101, 2005.

\bibitem[Soldati and Marchioli(2009)]{soldati:2009}
Alfredo Soldati and Cristian Marchioli.
\newblock Physics and modelling of turbulent particle deposition and
  entrainment: {R}eview of a systematic study.
\newblock \emph{Int.~J.~Multiphase Flow}, 35:\penalty0 827--839, 2009.

\bibitem[Pan and Banerjee(1997)]{pan:1997}
Y.~Pan and S.~Banerjee.
\newblock Numerical investigation of the effects of large particles on
  wall-turbulence.
\newblock \emph{Phys.~Fluids}, 9\penalty0 (12):\penalty0 3786 -- 3807, 1997.

\bibitem[Uhlmann(2008)]{uhlmann:2008}
Markus Uhlmann.
\newblock Interface-resolved direct numerical simulation of vertical
  particulate channel flow in the turbulent regime.
\newblock \emph{Phys.~Fluids}, 20\penalty0 (5):\penalty0 053305, 2008.

\bibitem[Shao et~al.(2012)Shao, Wu, and Yu]{shao:2012}
Xueming Shao, Tenghu Wu, and Zhaosheng Yu.
\newblock Fully resolved numerical simulation of particle-laden turbulent flow
  in a horizontal channel at a low reynolds number.
\newblock \emph{J.~Fluid Mech.}, 693:\penalty0 319--344, 2012.

\bibitem[Garc\'ia-Villalba et~al.(2012)Garc\'ia-Villalba, Kidanemariam, and
  Uhlmann]{garcia-villalba:2012}
M.\ Garc\'ia-Villalba, A.~G.\ Kidanemariam, and M.\ Uhlmann.
\newblock D{N}{S} of vertical plane channel flow with finite-size particles:
  Voronoi analysis, acceleration statistics and particle-conditioned averaging.
\newblock \emph{Int.\ J.\ Multiphase Flow}, 46:\penalty0 54--74, 2012.

\bibitem[Kajishima and Takiguchi(2002)]{kajishima:2002}
Takeo Kajishima and Saotshi Takiguchi.
\newblock Interaction between particle clusters and particle-induced
  turbulence.
\newblock \emph{Int.~J.~Heat Fluid Fl.}, 23:\penalty0 639--646, 2002.

\bibitem[{Ten Cate} et~al.(2004){Ten Cate}, Derksen, Portela, and {Van Den
  Akker}]{tencate:2004}
Andreas {Ten Cate}, Jos~J. Derksen, Luis~M. Portela, and Harry E.~A. {Van Den
  Akker}.
\newblock Fully resolved simulations of colliding monodisperse spheres in
  forced isotropic turbulence.
\newblock \emph{J.~Fluid Mech.}, 519:\penalty0 233--271, 2004.

\bibitem[Lucci et~al.(2010)Lucci, Ferrante, and Elghobashi]{lucci:2010}
Francesco Lucci, Antonino Ferrante, and Said Elghobashi.
\newblock Modulation of isotropic turbulence by particles of {T}aylor
  length-scale size.
\newblock \emph{J.~Fluid Mech.}, 650:\penalty0 5--55, 2010.

\bibitem[Lucci et~al.(2011)Lucci, Ferrante, and Elghobashi]{lucci:2011}
Francesco Lucci, Antonino Ferrante, and Said Elghobashi.
\newblock Is {S}tokes number an appropriate indicator for turbulence modulation
  by particles of {T}aylor length-scale size.
\newblock \emph{Phys.~Fluids}, 23:\penalty0 025101, 2011.

\bibitem[Monchaux et~al.(2010)Monchaux, Bourgoin, and
  Cartellier]{monchaux:2010b}
R.~Monchaux, M.~Bourgoin, and A.~Cartellier.
\newblock Preferential concentration of heavy particles: A vorono analysis.
\newblock \emph{Phys.~Fluids}, 22:\penalty0 103304, 2010.

\bibitem[Monchaux et~al.(2012)Monchaux, Bourgoin, and
  Cartellier]{monchaux:2012}
R.~Monchaux, M.~Bourgoin, and A.~Cartellier.
\newblock Analyzing preferential concentration and clustering of intertial
  particles in turbulence.
\newblock \emph{Int.\ J.\ Multiphase Flow}, 40:\penalty0 1--18, 2012.

\bibitem[Bagchi and Balachandar(2003)]{bagchi:2003}
P.~Bagchi and S.~Balachandar.
\newblock Effect of turbulence on the drag and lift of a particle.
\newblock \emph{Phys. Fluids}, 15:\penalty0 3496, 2003.

\bibitem[Merle et~al.(2005)Merle, Legendre, and Magnaudet]{merle:2005}
Axel Merle, Dominique Legendre, and Jacques Magnaudet.
\newblock Forces on a high-{R}eynolds-number spherical bubble in a turbulent
  flow.
\newblock \emph{J.~Fluid Mech.}, 532:\penalty0 53--62, 2005.

\bibitem[Zeng et~al.(2008)Zeng, Balachandar, Fischer, and Najjar]{zeng:2008}
Lanying Zeng, S.~Balachandar, Paul Fischer, and Fady Najjar.
\newblock Interactions of a stationary finite-sized particle with wall
  turbulence.
\newblock \emph{J.~Fluid Mech.}, 594:\penalty0 271--305, 2008.

\bibitem[Uhlmann(2005{\natexlab{a}})]{uhlmann:2005a}
Markus Uhlmann.
\newblock An immersed boundary method with direct forcing for the simulation of
  particulate flows.
\newblock \emph{J.~Comput.~Phys.}, 209:\penalty0 448 -- 476,
  2005{\natexlab{a}}.

\bibitem[Verzicco and Orlandi(1996)]{verzicco:1996}
R.~Verzicco and P.~Orlandi.
\newblock A finite-difference scheme for three-dimensional incompressible flows
  in cylindrical coordinates.
\newblock \emph{J.~Comput.~Phys.}, 123:\penalty0 402--414, 1996.

\bibitem[Peskin(2002)]{peskin:2002}
Charles~S. Peskin.
\newblock The immersed boundary method.
\newblock In \emph{Acta Numerica}, volume~11, pages 479 -- 517. Cambridge
  University Press, 2002.

\bibitem[Glowinski et~al.(1999)Glowinski, Pan, Hesla, and
  Joseph]{glowinski:1999}
R.~Glowinski, T.~W. Pan, T.~I. Hesla, and D.~D. Joseph.
\newblock A distributed lagrange multiplier/fictitious domain method for
  particulate flows.
\newblock \emph{Int.~J.~Multiphase Flow}, 25:\penalty0 755, 1999.

\bibitem[Uhlmann(2010)]{uhlmann:2010}
M.\ Uhlmann.
\newblock Hydrodynamic turbulence induced by sedimenting particles.
\newblock In B.\ Mohr and W.\ Frings, editors, \emph{J\"ulich Blue Gene/P
  Extreme Scaling Workshop 2010}, Technical Report FZJ-JSC-IB-2010-03, pages
  35--37, 2010.

\bibitem[Uhlmann(2004)]{uhlmann:2004}
Markus Uhlmann.
\newblock New results on the simulation of particulate flows.
\newblock Technical Report 1038, CIEMAT, Madrid, Spain, 2004.

\bibitem[Uhlmann(2005{\natexlab{b}})]{uhlmann:2005b}
M.\ Uhlmann.
\newblock An improved fluid-solid coupling method for {D}{N}{S} of particulate
  flow on a fixed mesh.
\newblock In M.\ Sommerfeld, editor, \emph{Proc.\ 11th Workshop Two-Phase Flow
  Predictions}, Merseburg, Germany, 2005{\natexlab{b}}. Universit\"at Halle.
\newblock ISBN 3-86010-767-4.

\bibitem[Uhlmann(2006)]{uhlmann:2006}
M.\ Uhlmann.
\newblock Experience with {D}{N}{S} of particulate flow using a variant of the
  immersed boundary method.
\newblock In P.\ Wesseling, E.\ O{\~n}ate, and J.\ P\'eriaux, editors,
  \emph{Proc.\ ECCOMAS CFD 2006}, Egmond aan Zee, The Netherlands, 2006. TU
  Delft.
\newblock ISBN 90-9020970-0.

\bibitem[Chan-Braun et~al.(2011)Chan-Braun, Garc\'{\i}a-Villalba, and
  Uhlmann]{chan-braun:2011}
C.\ Chan-Braun, M.\ Garc\'{\i}a-Villalba, and M.\ Uhlmann.
\newblock Force and torque acting on particles in a transitionally rough open
  channel flow.
\newblock \emph{J.~Fluid Mech.}, 684:\penalty0 441--474, 2011.
\newblock \doi{10.1017/jfm.2011.311}.

\bibitem[Kim et~al.(1987)Kim, Moin, and Moser]{kim:1987}
John Kim, Parviz Moin, and Robert Moser.
\newblock Turbulence statistics in fully developed channel flow at low reynolds
  number.
\newblock \emph{J.~Fluid Mech.}, 177:\penalty0 133 -- 166, 1987.

\bibitem[Taniere et~al.(1997)Taniere, Oesterle, and
  Monnier]{taniere_oesterle_monnier_EIF_97}
A~Taniere, B~Oesterle, and JC~Monnier.
\newblock On the behaviour of solid particles in a horizontal boundary layer
  with turbulence and saltation effects.
\newblock \emph{Exp.~Fluids}, 23\penalty0 (6):\penalty0 463--471, DEC 1997.
\newblock ISSN 0723-4864.

\bibitem[Muste et~al.(2009)Muste, Yu, Fujita, and Ettema]{muste:09}
M.\ Muste, K.\ Yu, I.\ Fujita, and R.\ Ettema.
\newblock Two-phase flow insights into open-channel flows with suspended
  particles of different densities.
\newblock \emph{Environ.\ Fluid Mech.}, 9:\penalty0 161--186, 2009.

\bibitem[Noguchi and Nezu(2009)]{noguchi:09}
K.\ Noguchi and I.\ Nezu.
\newblock Particle-turbulence interaction and local particle concentration in
  sediment-laden open-channel flows.
\newblock \emph{J.\ Hydro-env.\ Res.}, 3:\penalty0 54--68, 2009.

\bibitem[Johnson and Patel(1999)]{johnson:1999}
T.~A. Johnson and V.~C. Patel.
\newblock Flow past a sphere up to a reynolds number of 300.
\newblock \emph{J.~Fluid Mech.}, 378:\penalty0 19--70, 1999.

\bibitem[Naso and Prosperetti(2010)]{naso:10}
A.\ Naso and A.\ Prosperetti.
\newblock The interaction between a solid particle and a turbulent flow.
\newblock \emph{New J.\ Phys.}, 12\penalty0 (033040), 2010.

\bibitem[Okabe et~al.(1992)Okabe, Boots, and Sugihara]{okabe:1992}
Atsuyuki Okabe, Barry Boots, and Kokich Sugihara.
\newblock \emph{Spatial {T}essellations: {C}oncepts and {A}pplications of
  Voronoi {D}iagrams}.
\newblock Wiley, Chichester, 2nd edition, 1992.

\bibitem[Ferenc and N\'{e}da(2007)]{Ferenc:2007}
J\'{a}rai-Szab\'{o} Ferenc and Zolt\'{a}n N\'{e}da.
\newblock On the size distribution of poisson voronoi cells.
\newblock \emph{Physica A}, 385\penalty0 (2):\penalty0 518 -- 526, 2007.
\newblock ISSN 0378-4371.
\newblock \doi{DOI: 10.1016/j.physa.2007.07.063}.

\bibitem[Sundaram and Collins(1997)]{sundaram:1997}
Shivshankar Sundaram and Lance~R. Collins.
\newblock Collision statistics in an isotropic particle-laden turbulent
  suspension. {P}art 1. {D}irect numerical simulations.
\newblock \emph{J.~Fluid Mech.}, 335:\penalty0 75--109, 1997.

\bibitem[Shotorban and Balachandar(2006)]{shotorban:2006}
Babak Shotorban and S.~Balachandar.
\newblock Particle concentration in homogeneous shear turbulence simulated via
  lagrangian and equilibrium eulerian approaches.
\newblock \emph{Phys. Fluids}, 18:\penalty0 065105, 2006.

\bibitem[Sardina et~al.(2012)Sardina, Schlatter, Brandt, Picano, and
  Casciola]{sardina:2012}
G.~Sardina, P.~Schlatter, L.~Brandt, F.~Picano, and C.~M. Casciola.
\newblock Wall accumulation and spatial localization in particle-laden wall
  flows.
\newblock \emph{J.~Fluid Mech.}, 699:\penalty0 50--78, 2012.

\bibitem[Gualtieri et~al.(2009)Gualtieri, Picano, and Casciola]{gualtieri:2009}
P.~Gualtieri, F.~Picano, and C.~M. Casciola.
\newblock Anisotropic clustering of inertial particles in homogeneous shear
  flow.
\newblock \emph{J.~Fluid Mech.}, 629:\penalty0 25--39, 2009.

\bibitem[Robinson(1991)]{robinson:1991}
S.~K. Robinson.
\newblock Coherent motions in the turbulent boundary layer.
\newblock \emph{Annu.~Rev.~Fluid Mech.}, 23:\penalty0 601 -- 639, 1991.

\bibitem[Jim\'enez and Pinelli(1999)]{jimenez:1999}
Javier Jim\'enez and Alfredo Pinelli.
\newblock The autonomous cycle of near-wall turbulence.
\newblock \emph{J.~Fluid Mech.}, 389:\penalty0 335--359, 1999.

\bibitem[Adrian et~al.(2000)Adrian, Meinhart, and Tomkins]{adrian:2000}
R.~J. Adrian, C.~D. Meinhart, and C.~D. Tomkins.
\newblock Vortex organization in the outer region of the turbulent boundary
  layer.
\newblock \emph{J.~Fluid Mech.}, 422:\penalty0 1--54, 2000.

\bibitem[Schoppa and Hussain(2000)]{schoppa:2000}
Wade Schoppa and Fazle Hussain.
\newblock Coherent structure dynamics in near-wall turbulence.
\newblock \emph{J.~Fluid Dyn. Res.}, 26:\penalty0 119--139, 2000.

\bibitem[Adrian(2007)]{adrian:2007}
Ronald~J. Adrian.
\newblock Hairpin vortex organization in wall turbulence.
\newblock \emph{Phys. Fluids}, 19:\penalty0 041301, 2007.

\bibitem[Jeong and Hussain(1995)]{jeong:1995}
Jinhee Jeong and Fazle Hussain.
\newblock On the identification of a vortex.
\newblock \emph{J.~Fluid Mech.}, 285:\penalty0 69--94, 1995.

\bibitem[Jim\'enez and Moin(1991)]{jimenez:91}
J.\ Jim\'enez and P.\ Moin.
\newblock The minimal flow unit in near-wall turbulence.
\newblock \emph{J.\ Fluid Mech.}, 225:\penalty0 213--240, 1991.

\bibitem[Kawahara et~al.(2012)Kawahara, Uhlmann, and van Veen]{kawahara:12a}
G.\ Kawahara, M.\ Uhlmann, and L.\ van Veen.
\newblock The significance of simple invariant solutions in turbulent flows.
\newblock \emph{Ann.\ Rev.\ Fluid Mech.}, 44:\penalty0 203--225, 2012.

\end{thebibliography}

%
\clearpage
\begin{appendices}
\section{Statistical convergence of the fluid phase data}
\label{sec-statistical-convergence-fluid-phase}
\begin{figure}[b]
   \centering
        \begin{minipage}{2ex}
        \rotatebox{90}{\small $\quad y/h$}
        \end{minipage}
        \begin{minipage}{.5\linewidth}
        \includegraphics[width=\linewidth]
        {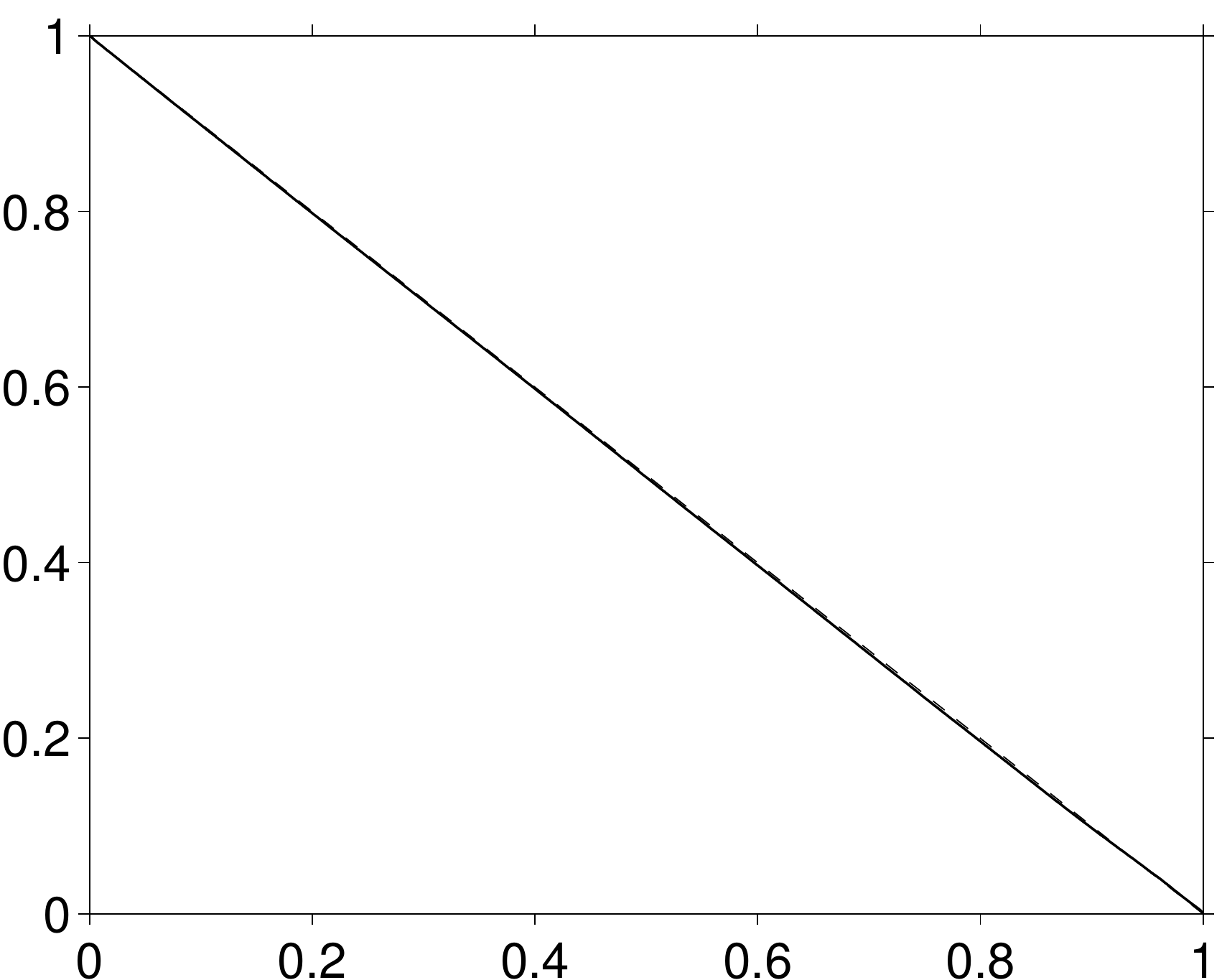}
        \\
        \centerline{\small $\langle\tau\rangle^*/(\rho_\mathrm{f}u_{\tau}^2)$ }
        \end{minipage}      
        \caption{Wall-normal profile of the plane- and 
                 time-averaged 
                 total shear
                 stress as defined in (\ref{eq:momentumbalance}). 
                 The dashed line 
                 indicates the function $(1-y/h)$. 
               }
        \label{fig:momentum_balance}
\end{figure}
Fully developed single phase channel flows are characterized by a 
linear variation as a function of the wall-distance 
of the plane- and time-averaged total shear stress 
\revision{
  $\langle \tau \rangle/\rho_\mathrm{f}
  \equiv\nu \partial_\mathrm{y}\langle u \rangle 
  -\rho\langle u'v' \rangle$.}
{  
  $\langle \tau \rangle/\rho_\mathrm{f}
  \equiv\nu \partial_\mathrm{y}\langle u \rangle 
  -\langle u'v' \rangle$.}
Traditionally this linearity critereon is 
used to check the convergence of statistics. However, in particulate 
channel flows, particles contribute additional terms to the
streamwise momentum balance modifying the total shear stress 
$\langle \tau \rangle$. 
In the context of the immersed boundary method, the 
modified Navier-Stokes equations are solved throughout 
the entire domain 
while adding an appropriate volume force term 
$\mathbf{f}$ to the equations, 
which imposes the no-slip condition at the location of the particles
viz.
\begin{equation}\label{navier-stokes-equations}
  \left.
    \begin{array}{lll}
      \partial_\mathrm{t}\mathbf{u}+(\mathbf{u}\cdot \nabla)\mathbf{u}
      + \nabla p &=& \nu\nabla^2\mathbf{u} + \mathbf{f} \\[5pt]
      \nabla \cdot \mathbf{u}&=&0
    \end{array}
  \right]
  \quad \mathbf{x}\in\Omega
  \,,
\end{equation}
where $\mathbf{u} = (u,v,w)^T$ 
is the vector of fluid velocities (without distinction between the
regions occupied by the fluid and the solid), $p$ is the
pressure normalized with the fluid density and 
$\mathbf{f} = (f_\mathrm{x},f_\mathrm{y},f_\mathrm{z})^T $
is the localized volume force term.
Integrating the streamwise momentum equation 
over wall-parallel planes and in time (supposing statistical
stationarity), and additionally integrating in the wall-normal
direction (from the wall plane $y^\prime=0$ up to an undetermined
distance $y^\prime=y$) 
yields the following equation:
\begin{equation}
 \underbrace{ 
  \langle \tau \rangle/\rho_\mathrm{f} + \int_0^y{\langle f_\mathrm{x}
    \rangle}dy^\prime  
 }_{\langle \tau \rangle^*}
   = 
 u_\tau^2 \Big(1 - \frac{y}{h}\Big).
\label{eq:momentumbalance}
\end{equation}
It can be seen from (\ref{eq:momentumbalance}) that 
there is indeed an additional contribution to the momentum balance 
due to the particle-related forcing term. Thus 
the modified total shear stress $\langle\tau \rangle^*$ should vary 
linearly for a fully developed particle laden channel flow. 
Figure \ref{fig:momentum_balance} shows the wall-normal profile of
$\langle \tau \rangle^*$ for the fluid statistics accumulated 
excluding the transient period. The near perfect match with the
linear profile indicates that the flow is in a statistically steady
state in the considered interval.
%

\section{Averaging operations}
\label{sec-ensemble-averaging}
\subsection{Wall-parallel plane and time averaging}
\label{subsec-wall-parallel-plane-and-time-averaging}
Let us first define an indicator function $\phi_\mathrm{f}(\mathbf{x},t)$ for the
fluid phase which tells us whether a given point with a position vector
$\mathbf{x}$ lies inside $\Omega_\mathrm{f}(t)$, the part of the 
computational domain $\Omega$ which is occupied by fluid at time $t$, as
follows:
\begin{equation}\label{equ-def-fluid-indicator-fct}
  \phi_\mathrm{f}(\mathbf{x},t)
  =
  \left\{
    \begin{array}{lll}
      1&\mbox{if}&
      \mathbf{x}\in\Omega_\mathrm{f}(t)
      \\
      0&\mbox{else}&
    \end{array}
  \right.
  \,.
\end{equation}
Furthermore, a discrete counter of fluid sample points in a wall-parallel 
plane at a given wall-distance $y_j$ is defined as:
\begin{equation}\label{equ-def-sample-counter-plane-and-time-fluid-only}
  n(y_j)
  =
  \sum_{m=1}^{N_\mathrm{t}}
  \sum_{i=1}^{N_\mathrm{x}}
  \sum_{k=1}^{N_\mathrm{z}}
  \phi_\mathrm{f}(\mathbf{x}_{ijk},t^m)
  \,,
\end{equation}
where $N_\mathrm{x}$ and $N_\mathrm{z}$ are the number of grid nodes
in the streamwise and spanwise directions, $N_\mathrm{t}$ is the
number of time records in the data set, and a discrete grid position
is denoted as $\mathbf{x}_{ijk}=(x_i,y_j,z_k)^T$.  
Consequently, 
the ensemble average of an Eulerian quantity $\boldsymbol{\xi}_\mathrm{f}$ 
of the fluid 
phase over wall-parallel planes and in time, while considering only
grid points located in the fluid domain, is defined as:
\begin{equation}\label{equ-def-avg-operator-plane-and-time-fluid-only}
  \langle
  \boldsymbol{\xi}_\mathrm{f}
  \rangle(y_j)
  =
  \frac{1}{n(y_j)}
  \sum_{m=1}^{N_\mathrm{t}}
  \sum_{i=1}^{N_\mathrm{x}}
  \sum_{k=1}^{N_\mathrm{z}}
  \phi_\mathrm{f}(\mathbf{x}_{ijk},t^m)\,
  \boldsymbol{\xi}_\mathrm{f}(\mathbf{x}_{ijk},t^m)
  \,.
\end{equation}
In the present study wall-parallel plane averaging of 
Eulerian quantities 
was performed during runtime from a 
total of 
\revision{%
  $N_\mathrm{t} = 20900$ 
}{%
  $N_\mathrm{t} = 26700$ 
}
flow field snapshots.

\subsection{Instantaneous box averaging}
\label{subsec-box-averaging}
A global counter of samples in the fluid domain can be defined
analogously to
(\ref{equ-def-sample-counter-plane-and-time-fluid-only}), viz.  
\begin{equation}\label{equ-def-sample-counter-box-fluid-only}
  n_{xyz}(t)
  =
  \sum_{i=1}^{N_\mathrm{x}}
  \sum_{j=1}^{N_\mathrm{y}}
  \sum_{k=1}^{N_\mathrm{z}}
  \phi_\mathrm{f}(\mathbf{x}_{ijk},t)
  \,,
\end{equation}
where $N_\mathrm{y}$ is the number of grid nodes in the wall-normal
direction. The instantaneous average of a quantity
$\boldsymbol{\xi}_\mathrm{f}$ over the entire volume $\Omega_f$
occupied by the fluid is then defined as follows:
\begin{equation}\label{equ-def-avg-operator-box-fluid-only}
  \langle
  \boldsymbol{\xi}_\mathrm{f}
  \rangle_\mathrm{xyz}(t)
  =
  \frac{1}{n_{xyz}(t)}
  \sum_{i=1}^{N_\mathrm{x}}
  \sum_{j=1}^{N_\mathrm{y}}
  \sum_{k=1}^{N_\mathrm{z}}
  \phi_\mathrm{f}(\mathbf{x}_{ijk},t)\,
  \boldsymbol{\xi}_\mathrm{f}(\mathbf{x}_{ijk},t)
  \,.
\end{equation}
%

\subsection{Binned averages over particle-related quantities}
\label{subsec-binned-averages-over-particle-related-quantities}
Concerning Lagrangian quantities related to particles, the domain
was decomposed into discrete wall-parallel bins of thickness $\Delta h$ 
and averaging was performed over all those particles within each bin.
Similar to (\ref{equ-def-fluid-indicator-fct}), 
we  define an indicator function $\phi_\mathrm{bin}^{(j)}(y)$ which
tells us whether a given wall-normal position $y$ is located inside or
outside a particular bin with index $j$, viz.
\begin{equation}\label{equ-def-ybin-indicator-fct}
  \phi_\mathrm{bin}^{(j)}(y)=
  \left\{
    \begin{array}{lll}
      1&\mbox{if}&
      (j-1)\Delta h\leq y<j\Delta h
      \\
      0&\mbox{else}&
    \end{array}
  \right.
\end{equation}
as well as a sample counter for each bin viz.
\begin{equation}\label{equ-def-sample-counter-binned-particles}
  n_\mathrm{p}^{(j)}
  =
  \sum_{m=1}^{N_\mathrm{t}^\mathrm{(p)}}
  \sum_{l=1}^{N_\mathrm{p}}
  \phi_\mathrm{bin}^{(j)}(y_\mathrm{p}^{(l)}(t^m))\,
  \,,
\end{equation}
where $N_\mathrm{t}^\mathrm{(p)}$ is the number of available 
snapshots of the solid phase.
From the sample counter we can deduce the average solid volume
fraction in each bin, viz.
\begin{equation}\label{equ-def-avg-solid-volume-frac}
  \langle\phi_s\rangle(y^{(j)})
  =
  \frac{n_p^{(j)}}{N_\mathrm{t}^\mathrm{(p)}}\,\frac{\pi D^3}{6L_xL_z\Delta h}
  \,.
\end{equation}
Finally, the binned average (over time and the number of particles) of a
Lagrangian quantity $\boldsymbol{\xi}_\mathrm{p}$ is defined as follows:
\begin{equation}\label{equ-def-avg-operator-binned-particles}
  \langle
  \boldsymbol{\xi}_\mathrm{p}
  \rangle
  (y^{(j)})
  =
  \frac{1}{n_\mathrm{p}^{(j)}}
  \sum_{m=1}^{N_\mathrm{t}^\mathrm{(p)}}
  \sum_{l=1}^{N_\mathrm{p}}
  \phi_\mathrm{bin}^{(j)}(y_\mathrm{p}^{(l)}(t^m))\,
  \boldsymbol{\xi}_\mathrm{p}^{(l)}(t^m)
  \,,
\end{equation}
supposing that a finite number of samples has been encountered
($n_\mathrm{p}^{(j)}>0$).
Accumulated data related to Lagrangian quantities
amounts to a total of 
\revision{%
  $N_\mathrm{t}^\mathrm{(p)} = 12900$
  particle fields (approximately $6.7$ million particle samples). 
}{%
  $N_\mathrm{t}^\mathrm{(p)} = 20314$
  particle fields (more than $10^7$ particle samples). 
}
A bin thickness of $\Delta h = D/2$ was adopted to evaluate
the bin-averages unless otherwise stated.
%
The distribution of samples can be deduced from
figure~\ref{fig:particle_solid_volume_fraction} which shows the
wall-normal profile of the average solid volume fraction. 
\subsection{Particle-centered averaging}
\label{subsec-particle-centered-averaging}
For performing the particle-conditioned averaging of a 
field $\boldsymbol{\xi}_\mathrm{f}(\mathbf{x},t)$, 
an averaging volume is considered such that
the coordinate system within the volume is defined relative to
the center of each of the considered particles. 
Applying the indicator functions defined above,
we define a discrete counter field
$\tilde{n}^{(s)}(\tilde{\mathbf{x}}_{ijk})$ which holds 
the number of samples obtained through averaging at a given discrete
location with indices $i,j,k$ for a given 
$y$-slab with index $s$, viz.
\begin{equation}\label{equ-def-sample-counter}
  \tilde{n}^{(s)}(\tilde{\mathbf{x}}_{ijk})
  =
  \sum_{m=1}^{N_\mathrm{t}^\mathrm{(r)}}
  \sum_{l=1}^{N_\mathrm{p}}
  \phi_\mathrm{bin}^{(s)}(y_\mathrm{p}^{(l)}(t^m))\,
  \phi_\mathrm{f}(\tilde{\mathbf{x}}_{ijk}^{(l)}(t^m),t^m)
  \,,
\end{equation}
In equation (\ref{equ-def-sample-counter}) the symbol $t^m$ indicates
the time corresponding to the $m$th snapshot in the database 
(comprising a total of $N_\mathrm{t}^\mathrm{(r)}$ snapshots) and
$\tilde{\mathbf{x}}_{ijk}^{(l)}(t^m)$ is a discrete coordinate
relative to the 
$l$th particle's center position at time $t^m$, as defined in
(\ref{equ-def-relative-position}).
The  average of the quantity $\boldsymbol{\xi}_\mathrm{f}$
over time and over all particles located within the given 
wall-parallel slab is then defined as follows:
\begin{eqnarray}
  \nonumber
  \langle \boldsymbol{\xi}_\mathrm{f} \rangle_\mathrm{p,t}
  (\tilde{\mathbf{x}}_{ijk},y^{(s)}) =
  \frac{1}{\tilde{n}^{(s)}(\tilde{\mathbf{x}}_{ijk})}
  \sum_{m=1}^{N_\mathrm{t}^\mathrm{(r)}}
  \sum_{l=1}^{N_\mathrm{p}}
  &&
  \phi_\mathrm{bin}^{(s)}(y_\mathrm{p}^{(l)}(t^m))\,
  \phi_\mathrm{f}(\tilde{\mathbf{x}}_{ijk}^{(l)}(t^m),t^m)
  \\
  \label{equ-def-avg-operator-wake-time-particles-ybinned}
  &&
  \times\,
  \boldsymbol{\xi}_\mathrm{f}(\tilde{\mathbf{x}}_{ijk}^{(l)}(t^m),t^m)
  \,
\end{eqnarray}
Note that the coordinates  $\tilde{\mathbf{x}}_{ijk}^{(l)}(t^m)$ do
not 
necessarily 
coincide with the fixed grid used in the direct numerical 
simulation. Thus, 
values of $\boldsymbol{\xi}_\mathrm{f}$ at $\tilde{\mathbf{x}}_{ijk}^{(l)}(t^m)$  
are determined by a tri-linear interpolation from those values 
at the neighbouring fixed grid locations.

For evaluating the average defined by 
(\ref{equ-def-avg-operator-wake-time-particles-ybinned})
a total amount of 
\revision{%
  $N_\mathrm{t}^\mathrm{(r)}= 45$ 
}{%
  $N_\mathrm{t}^\mathrm{(r)}= 70$ 
}
flow field snapshots evenly
distributed in the considered observation interval were used.

\end{appendices}
\end{document}